\documentstyle[12pt,aaspp4,psfig]{article}
\newcommand{\kms}          {\mbox{${\rm km~s^{-1}}$}}
\newcommand{\cc}           {\mbox{${\rm cm^{-3}}$}}

\def\cm2{\mbox{${\rm cm^{-2}}$}}
\def\R21{\mbox{$W_{2-1}/W_{1-0}$}}

\def\13co{\mbox{$^{13}$CO}}
\def\W13{\mbox{$W_{^{13}{\rm CO}}$}}
\def\N13{\mbox{$N_{^{13}{\rm CO}}$}}
\def\c18o{\mbox{C$^{18}$O}}
\def\h2{\mbox{${\rm H}_2$}}
\def\nh2{\mbox{$n_{\rm H_2}$}}
\def\Nh2{\mbox{$N_{{\rm H}_2}$}}
\def\Mh2{\mbox{$M_{{\rm H}_2}$}}

\def\cm{{\rm\,cm}}

\def\msun{{\rm\,M-_\odot}}

\def\kmse{{km\thinspace s$^{-1}$}\ }              %kms -1%
\def\kms{{km\thinspace s$^{-1}$}}
\def\13co{$^{13}$CO}
\def\h2{H$_2$}

\def\le{{$l$}\ }

\def\be{{$b$}\ }
\def\deg{\ifmmode^\circ\else$^\circ$\fi}
\def\dege{{\ifmmode^\circ\else$^\circ$\fi}\ }
\def\solar{\ifmmode _{\mathord\odot}\else$_{\mathord\odot}$\fi}
\def\sun{$_\odot$}
\def\sune{{$_\odot$}\ }

\def\msun{M\sun}

\def\msune{{M\sun}\ }

\def\asec{$^{\prime \prime}$}

\def\amin{$^{\prime}$}
\def\microne{{$\mu$m}\ }

\def\hper{\ifmmode \rlap.^{h} \else $\rlap{.}^h$\fi}
\def\mper{\ifmmode \rlap.^{m} \else $\rlap{.}^m$\fi}
\def\sper{\ifmmode \rlap.^{s} \else $\rlap{.}^s$ \fi}
\def\degper{\ifmmode \rlap.^{\circ} \else $\rlap{.}^{\circ} $\fi}
\def\arcmper{\ifmmode \rlap.{' } \else $\rlap{.}' $\fi}
\def\arcsper{\ifmmode \rlap.{'' } \else $\rlap{.}'' $\fi}
\def\cc{cm$^{-3}$}

\def\c2{cm$^{-2}$}
\def\m12{\magnification=1200}

\def\deg{\ifmmode^\circ\else$^\circ$\fi}              %Degree sign%
\def\arcs{\ifmmode {'' }\else $'' $\fi}           %Arc seconds%
\def\arcm{\ifmmode {' }\else $' $\fi}             %Arc minutes%
\def\spose#1{\hbox to 0pt{#1\hss}}
\def\lta{\mathrel{\spose{\lower 3pt\hbox{$\mathchar"218$}}
     \raise 2.0pt\hbox{$\mathchar"13C$}}}
\def\gta{\mathrel{\spose{\lower 3pt\hbox{$\mathchar"218$}}
     \raise 2.0pt\hbox{$\mathchar"13E$}}}
\def\clock{\count0=\time \divide\count0 by 60
     \count1=\count0 \multiply\count1 by -60 \advance\count1 by \time
\number\count0:\ifnum\count1<10{0\number\count1}\else\number\count1\fi}

\begin{document}
\title{High--Velocity Clouds: Building Blocks of the Local Group}
\author{Leo Blitz\footnote{blitz@gmc.berkeley.edu}}
\affil{University of California, Berkeley, CA 94720}
\affil{University of Maryland, College Park, MD 20742}
\author{David N.\ Spergel\footnote{dns@astro.princeton.edu}}
\affil{Princeton University Observatory, Princeton, NJ 08544}
\affil{University of Maryland, College Park, MD 20742}
\author{Peter J. Teuben\footnote{teuben@astro.umd.edu}}
\affil{University of Maryland, College Park, MD 20742}
\author{Dap Hartmann\footnote{dap@abitibi.harvard.edu}}
\affil {Harvard-Smithsonian Center for Astrophysics, Cambridge, MA 02138}\author{W. Butler 
Burton\footnote{burton@strw.leidenuniv.nl}}
\affil{Leiden University Observatory, Leiden, Netherlands}

\dates

\begin{abstract}
\rightskip = 0pt
\noindent
We suggest that the high--velocity clouds (HVCs) are large clouds, with
typical diameters of 25 kpc and containing $5 \times 10^7$ solar masses of
neutral gas and $3 \times 10^8$ solar masses of dark matter, falling
onto the Local Group; altogether the HVCs contain 10$^{11}$ solar
masses of neutral gas.  Our reexamination of the Local--Group hypothesis
for the HVCs connects their properties to the hierarchical structure
formation scenario and to the gas seen in absorbtion towards quasars.

We begin by showing that at least one HVC complex (besides the 
Magellanic Stream) must be extragalactic at a
distance ${>} 40$ kpc from the Galactic center, with a diameter ${>}
20$ kpc and a mass ${>} 10^8$ solar masses. We then discuss a number
of other clouds that are positionally associated with the Local Group
galaxies. The kinematics of the entire ensemble of HVCs is inconsistent
with a Galactic origin, but implies that the HVCs are falling towards
the Local Group barycenter. The HVCs obey an angular--size/velocity
relation consistent with the Local Group infall model.

We simulate the dynamical evolution of the Local Group. The simulated
properties of material falling into the Local Group reproduce
the location of two of the three most significant groupings of clouds
and the kinematics of the entire cloud ensemble (excluding the
Magellanic Stream). We interpret the third grouping (the A, C, and M
complexes) as tidally unstable nearby material falling onto the Galactic
disk. We interpret the more distant HVCs as dark matter ``mini--halos''
moving along filaments towards the Local Group.  Most poor galaxy
groups should contain HI structures to large distances bound to the
group.  We suggest that the HVCs are local analogues of the Lyman--limit
absorbing clouds observed against distant quasars.

We argue that there is a Galactic fountain in the Milky Way, but that
the fountain does not explain the origin of the HVCs. Our analysis of
the HI data leads to the detection of a vertical infall of low--velocity
gas towards the plane.  We suggest that the fountain is a local
phenomenon involving neutral gas with characteristic velocities of 6
\kmse rather than 100 \kms. This implies that the chemical evolution of
the Galactic disk is governed by episodic infall of metal-poor HVC gas
that only slowly mixes with the rest of the interstellar medium.

The Local--Group infall hypothesis makes a number of testable
predictions.  The HVCs should have sub-solar metallicities. Their
H$\alpha $ emission should be less than that seen from the Magellanic
Stream. The clouds should not be seen in absorption to nearby stars.
The clouds should be detectable in both emission and absorption around
other groups. We show that current observations are consistent with
these predictions and discuss future tests.

\end{abstract}

\keywords{Galaxy: general, formation, evolution, and structure --- Local Group 
--- intergalactic medium --- ISM: clouds, high--velocity clouds, structure, 
kinematics and dynamics --- quasars: absorption lines} 

\clearpage \rightskip = 0pt

\section{Introduction} \label{sec:intro}

Since their discovery in 1963 by \markcite{Muller63}Muller, Oort,
\& Raimond, the
nature of the high--velocity hydrogen clouds (HVCs) has remained a
mystery.  HVCs are clouds that deviate from Galactic circular rotation
by as much as several hundred kilometers per second.  Although
astronomers have speculated about the origin of HVCs since their detection,
no single explanation has encompassed the vast quantity of data that
has been collected on the clouds (see \markcite{WvW97} Wakker \& van
Woerden 1997, hereafter WvW97, and references therein). Particularly
important is the lack of agreement on a characteristic distance for the
clouds, because most of the relevant physical parameters depend on
distance to one order or another.  In the 1970s, a well--defined subset
of the clouds was identified as a tidal stream associated with the
Magellanic Clouds (\markcite{Mathewson74}Mathewson, Cleary, \& Murray 1974),
but since then no consensus has arisen regarding the nature of the
remaining clouds which constitute the majority of HVCs.

In this paper, we suggest that the HVCs represent infall of the
intergalactic medium onto the Local Group.  
%we are the first to consider this suggestion.  Other models consider 
%the HVCs to be a quasi-static phenomenon within the local group 
%WBB isn't sure he agrees ...
Previous authors have
explored the possibility that the HVCs are infalling primordial gas
(\markcite{Oort66,Oort70}Oort 1966, 1970) and have associated the HVCs
with the Local Group (\markcite{Verschuur69}Verschuur 1969;
\markcite{Kerr69}Kerr \& Sullivan 1969; \markcite{Arp85}Arp 1985;
\markcite{Bajaja87}Bajaja, Morras, \& P\"oppel 1987; \markcite{Arp91}Arp \& 
Sulentic
1991), but subsequent observations always produced fundamental
difficulties for the particular models considered.  Here, we assemble
evidence based on new general--purpose surveys of atomic hydrogen gas
by \markcite{Stark92}Stark et al. (1992) and by
\markcite{Hartmann97}Hartmann \& Burton (1997), and on the HVC surveys by
\markcite{Hulsbosch88}Hulsbosch \& Wakker (1988) and by
\markcite{Bajaja85}Bajaja et al. (1985), and consider theoretical arguments in
the context of modern cosmology. We argue that the clouds are matter
accreting onto the Local Group of galaxies. Their velocities would thus
largely reflect the motion of the clouds in the gravitational potential of the
Local Group and the motion of the LSR about the Galactic center.  We
suggest that the clouds represent the building blocks from which the
Local Group was assembled and that they continue to fuel star formation
in the disk of the Milky Way.

The evidence is presented as follows. In Section 2, we review some of
the observed properties of the high--velocity clouds, and indicate
those which are not consistent with a Galactic origin for the HVCs. In
Section 3, we detail the observations entering our analysis. In Section
4, we discuss the stability of the HVCs against gravitational collapse
and against Galactic tidal forces, and suggest that these
considerations imply  that the most of the clouds are extragalactic at
distances typical of the Local Group. The stability criteria imply,
however, that the largest clouds are nearby and possibly are interacting
with the Milky Way.  In Section 5, we discuss three individual clouds,
one of which must be beyond the disk of the Milky Way, and two others
that appear to be associated with M31 and M33, suggesting that at least
some of the HVCs may be extragalactic. We identify a subset of the HVCs
centered near the barycenter of the Local Group, and show that its
kinematics as well as that of the entire HVC ensemble are well
described as being at rest with respect to the Local--Group Standard of
Rest (LGSR); the kinematics are thus inconsistent with a Galactic
origin.  The entire HVC ensemble is also shown to exhibit an
angular--size/velocity relation consistent with membership in the Local
Group.  In Section 6, we simulate the accretion history of the Local
Group and show that the simulation accounts for the observed
distribution and kinematics of the HVC ensemble. The agreement between
the simulation and the observations supports inferences about
similarities between the Local Group HVCs and the Ly--$\alpha$ absorbing
clouds observed toward quasars. We show that the velocity extrema
observed for the HVCs are consistent with their membership in the Local
Group.  In Section 7, we discuss the distances and abundances of the
HVCs in the context of the Local Group HVC hypothesis, and show that
the hypothesis is consistent with all of the observations made to
date.
We review extragalactic HI searches for HVCs which have revealed 
clouds with properties similar to those we derive, in about the
expected numbers.  In Section 8, we discuss the implied mass accretion
rate, and implications for the chemical evolution of the disk of the the
Milky Way.  We also present evidence for the Galactic fountain in
low--velocity HI which suggests that the HI disk of the Milky Way is
not in hydrostatic equilibrium.  In Section 9, we conclude by discussing
predictions and future tests of the model, and summarize the principal
arguments made in this paper.

\section{High--Velocity--Cloud Properties}

\label{sec:hvcprops}

HVCs are HI clouds with radial velocities inconsistent with circular or 
near--circular rotation about the Galactic center, and in this and
other regards unlike most of the HI making up the Galactic 
disk.  Perhaps the most useful kinematic definition is that of the
``deviation velocity'' introduced by \markcite{Wakker91} Wakker (1991),
indicating the degree to which a cloud deviates from a reasonable
circular--rotation model. We note the following general properties of
HVCs other than those which are part of the Magellanic Stream (see
\markcite{WvW97}WvW97 for a thorough review):

(1) HVCs have LSR radial velocities as extreme as $-464$ \kms.  The most
extreme radial velocity consistent with circular rotation is about $\pm
220$ \kms.

(2) HVCs have both positive and negative radial velocities, but clouds
with positive radial velocities are distributed differently on the sky
than clouds with negative velocities.

(3) Two HVC complexes (The Magellanic Stream and Complex C) are contiguous
% wouldn't Complex C be intended here, as it is much larger than A?
structures covering thousands of square degrees on the sky.  Most HVCs,
however, have angular extents of a few square degrees or less.
Individual clouds separated by several or tens of degrees are often
separated in velocity by hundreds of kilometers per second.  Except for
the Magellanic Stream, the largest cloud complexes tend to lie at
positive latitudes in the first and second Galactic quadrants; these
include the well known complexes A, C, and M, which have received the
most observational attention.  We refer to these clouds as the Northern
Hemisphere Clouds and find them to be the nearest examples of the HVC
phenomenon (see \S\ref{sec:bighvcs}. 

(4) HVCs have a narrow range of internal velocity dispersions, centered
near 13 \kmse (see Section 4).

(5) HVCs have low dust--to--gas ratios, at least a factor of three below
that of normal Galactic clouds\markcite{Wakker86} (Wakker \& Boulanger
1986).

(6) Measured heavy--element abundances are all well below solar values.
Although few metallicities are available, those measured are also
significantly subsolar.

(7) Distance measurements to the clouds have been largely indeterminate,
although several recent observations in the direction of Northern
Hemisphere Clouds suggest a distance of about 5 kpc for Complex
A\markcite{vanWoerden98} (van Woerden et al. 1998) as well as for Complex
M\markcite{Danly93} (Danly, Albert, \& Kuntz 1993).

Clearly, any explanation of the nature and origin of the HVCs must
account for these observed properties.  Previous explanations fall into
two broad categories: Galactic and extragalactic.  In the first general
discussion of the nature of the HVCs,\markcite{Oort66} Oort (1966)
realized that if the clouds are self--gravitating then they must have
distances on the order of a Mpc. In Galactic models, of which the most
popular is the Galactic fountain (\markcite{Shapiro76}Shapiro \& Field
1976; \markcite{Bregman80} Bregman 1980), the HVCs would be too near to
be bound by their own gravity and, because of their large internal
velocity dispersions, would be transient objects with typical lifetimes
$\lta 10^6$ y.  It is difficult to understand in the Galactic fountain
context why the clouds would not be metal rich or how their vertical
velocities would be greater than 70 to 100 \kms.  Extragalactic models,
on the other hand, generally require HVCs to be gravitationally stable
for periods comparable to a Hubble time, and to be metal poor (though
not necessarily with zero metallicity).

\section{HI Observations} 
\label{sec:obs}

Our analysis relies principally on the catalogue of HVCs compiled by
\markcite{WvW91} Wakker \& van Woerden (1991, hereafter WvW91) and on
the new Leiden/Dwingeloo survey of HI of\markcite{Hartmann97} Hartmann
\& Burton (1997, hereafter LD).  The WvW91 compilation is based largely
on data from the HI survey of \markcite{Hulsbosch88} Hulsbosch \&
Wakker (1988), made with the Dwingeloo 25--m telescope over the
northern sky at 1\dege intervals covering LSR velocities from  $-900$
to $+800$ \kmse with a velocity resolution of 8.25 \kms, together with
data from the HI survey of \markcite{Bajaja85} Bajaja et al. (1985),
made with the Villa Elisa 100--foot telescope over declinations only
accessible from the southern hemisphere, at somewhat coarser resolution
and somewhat lower sensitivity.  The Wakker \& van Woerden catalogue
also includes data from a number of other observational programs
covering specific HVCs in enhanced detail.

The Leiden/Dwingeloo HI survey of \markcite{Hartmann97}Hartmann \&
Burton (1997) utilized the Dwingeloo 25--m telescope to observe the sky
at declinations north of $-30\deg$, at the relatively high resolutions
of 0\degper5 in angle and 1 \kmse in velocity, covering the velocity
range $-450 <  v_{\rm LSR} < +400 $ \kms.  Although the velocity
coverage of the LD survey is less than that of WvW91, few HVCs have
been found beyond the negative--velocity range of the LD survey, and
none beyond the positive--velocity edge.  The sidelobe response has
been removed from the LD data by subtracting, from each spectrum, the
simulated sidelobe response of a modeled telescope antenna to a
full--sky HI map (\markcite{Hartmann94}Hartmann 1994;
\markcite{Hartmann96}Hartmann et al. 1996).  Although the survey was
motivated by problems pertaining to the Milky Way, it is useful for
some HVC studies too.  When smoothed to the resolution of the Hulsbosch
\& Wakker data, the sensitivity of the LD data is similar; its superior
spatial and velocity resolution makes it particularly useful for
obtaining observational parameters of some of the HVCs.  The LD survey
has not, however, been analyzed for high--velocity emission in the
detail of WvW91.

We also utilize the Bell Telephone Laboratories HI survey of
\markcite{Stark92}Stark et al. (1992, hereafter BTL), which was the first
northern hemisphere HI sky survey with full beamwidth sampling and low
sidelobe response. Although the rather coarse (2\deg) angular
resolution is not particularly well--suited to investigations of most
HVCs, the velocity resolution of 5 \kmse is not a disadvantage in
global HVC studies because most HVC lines are considerably broader than
this.  The BTL and LD surveys have similar noise levels when convolved
to the same resolution.  Although possible sidelobe contamination is
not particularly worrisome for HI features with velocities beyond those
of conventional Galactic emission, low sidelobe response is useful for
studying those HVCs seen at high Galactic latitudes within the velocity
range of Galactic plane HI. We checked the reality of weak features by
making cross comparisons between the BTL and LD surveys.

\section{Stability Criteria for the HVCs}
\label{sec:stability}
We begin our analysis by calculating the characteristic
self-gravitating distances for the HVCs.  As mentioned above,
\markcite{Oort66}Oort (1966) noted that self-gravitating HVCs imply
that the HVCs are at a distance of $\geq$ Mpc.
\markcite{Giovanelli79}Giovanelli (1979), using better HI data, 
argued that the virial
distances would place many HVCs outside the Local Group.  Here, we
revisit the calculation with new data and include the presence of dark
matter in our analysis.

If the HVCs are objects nearly as old as the Universe, they must be
gravitationally and tidally stable for a Hubble time. A
self--gravitating cloud obeys the relationship

\begin{equation} 
{v_{3{\rm D}}}^2~~<~~\frac{2GM_{\rm HI}}{fR}_{}
\end{equation} 
where $M_{\rm HI}$ is the mass in neutral hydrogen, $f$
is the hydrogen/total mass ratio, and $R$ is the gravitational radius of
the cloud.  Because HVCs are not sites of star formation nor has CO
ever been found associated with them \markcite{Wakker97}(Wakker et al.
1997),  the inequality in equation (1) is reversed for HVCs.  The equation can
therefore be used to estimate an upper limit to the distance of an HVC,
assuming that it is self--gravitating:

\begin{equation}
r_{\rm g}~~<~~88.6~~{{f\Delta v} \over {T_{\rm B}\Omega ^{1/2}}}~~{\rm kpc},
\end{equation}

\noindent where  $\Delta v$ is the observed FWHM of the HI emission
from a cloud, measured in \kms, $T_{\rm B}$ is the peak HI brightness
temperature, measured in K, and $\Omega $ is the solid angle projected
by the cloud in square degrees; the coefficient includes a 40\%
correction by mass for helium.

Stable clouds must also be able to withstand the shear of the Galactic
tidal field; this consideration yields a lower limit to the distance of
an HVC:

\begin{equation}
-{\frac {\rm d}{{\rm d}r}}({{\Theta ^2}\over r})R~~<~~{{GM}\over {R^2}}.
\end{equation}
This implies, in terms of observables:

\begin{equation}
r_{\rm t}~~>~~24.8~{{{\Theta ^2\Omega
^{1/2}f}\over {T_{\rm B}\Delta v}}~~{\rm pc,}} 
\end{equation}
where $r$ is the distance to the cloud and $\Theta$ is the circular
speed of the Galaxy at the distance of an HVC. If one assumes that
$\Theta =\Theta_{\odot}$ for $r_{\rm t}<r_{\rm c}$, where $r_{\rm c}$ is the distance
where the rotation curve becomes Keplerian, and that $\Theta
^2=r_{\rm c}{\Theta _{\odot}}^2/r_{\rm t}$ for $r_{\rm t}>r_{\rm c}$, then for $r_{\rm c}$ = 100 kpc,
equation (4) becomes

\begin{equation}
r_{\rm t}~~>~~1.34\times 10^3~~{{\Omega ^{1/4}\Theta _\odot f^{1/2}}\over
{\left( T_{\rm B}\Delta v\right) ^{1/2}}}~~{\rm pc,}
\end{equation}
for $r>$ 100 kpc, and where equation (4) holds for $r<$ 100 kpc.  For
$r_{\rm t} < r_{\rm g}$, only the gravitational distance is important,
because clouds that are not gravitationally stable are also not tidally
stable.

If $f$ is sufficiently small that either dark matter or ionized gas
dominates the mass of an individual HVC, that is, if the mass of the
cloud is given by $R{v_{\rm 3D}}^2/G$ independent of its distance, then
the tidal criterion becomes a criterion on the angular size of an HVC:

\begin{equation}
\Omega~~<~~ 7.13\times 10^3~{\Delta v\over \Theta _{\odot}^2}~~{\rm sq~deg,}\ 
\end{equation} 
for $R <$ 100 kpc, and 

\begin{equation}
\Omega~~<~~ 7.13\times 10^3~{\Delta v\over \Theta _{\odot}^2} {R\over 100~{\rm 
kpc}}~~{\rm sq~deg,}\
\end{equation}
for $R >$ 100 kpc, assuming a rotation curve flat to 100 kpc.

An additional useful relation is the crossing time, $t_{\rm c}$, for
the cloud, defined as the timescale for which a cloud would double in
size if it were not self gravitating:

\begin{equation}
t_{\rm c} ~~ = ~~ 17.1 ~ {{\Omega^{1/2} r_{\rm kpc}}\over {\Delta v}}~~ {\rm 
My,} 
\end{equation} 
where $r_{\rm kpc}$ is the distance to the cloud in kpc.

\subsection{HVC Statistics}
\label{sec:statistics}

\begin{figure}[htpb]
\vskip -0.25in
\psfig{figure=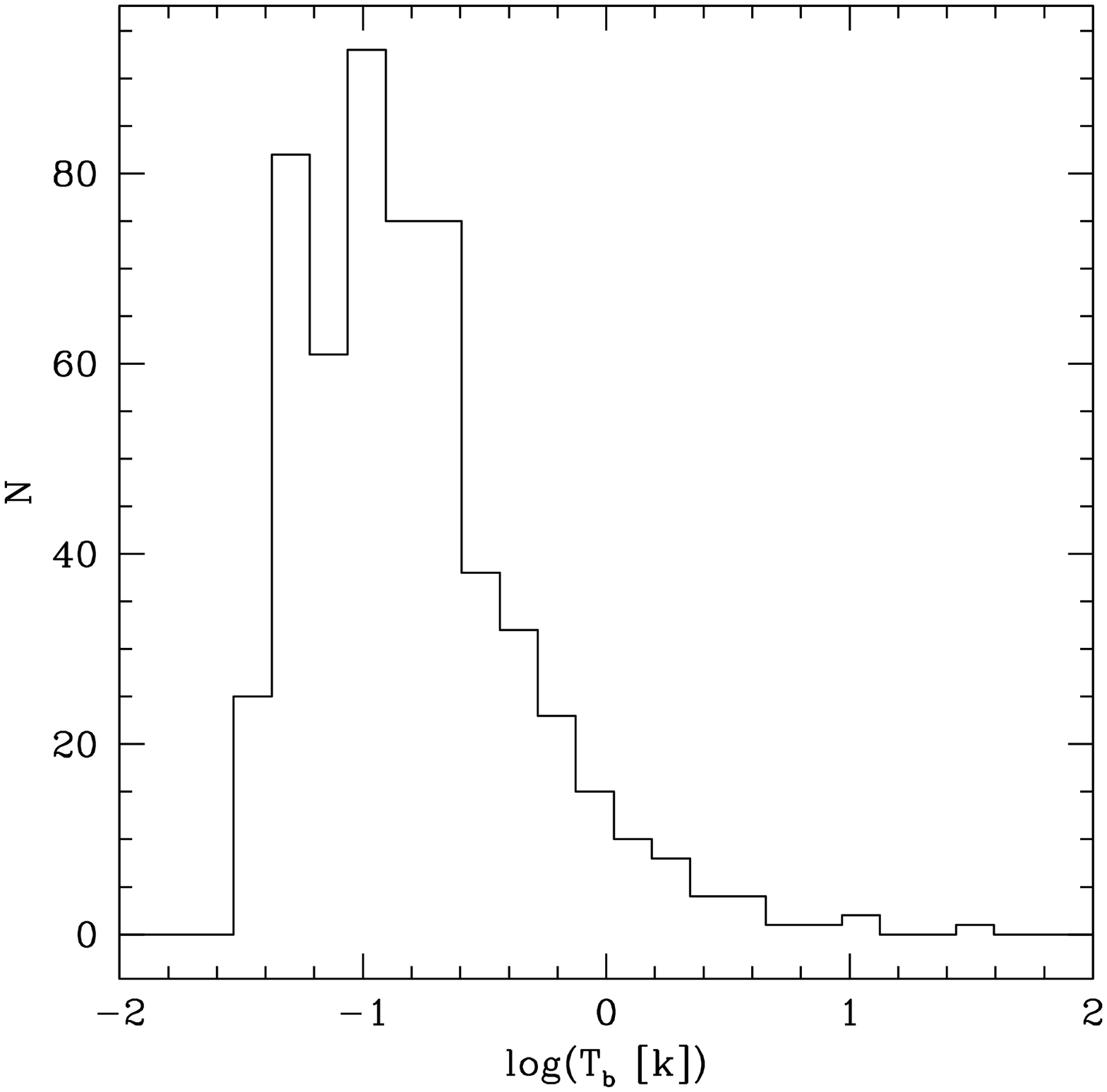,width=2.0in,angle=0,silent=1} 
\vskip -2.0in
\hskip 2.0in
\psfig{figure=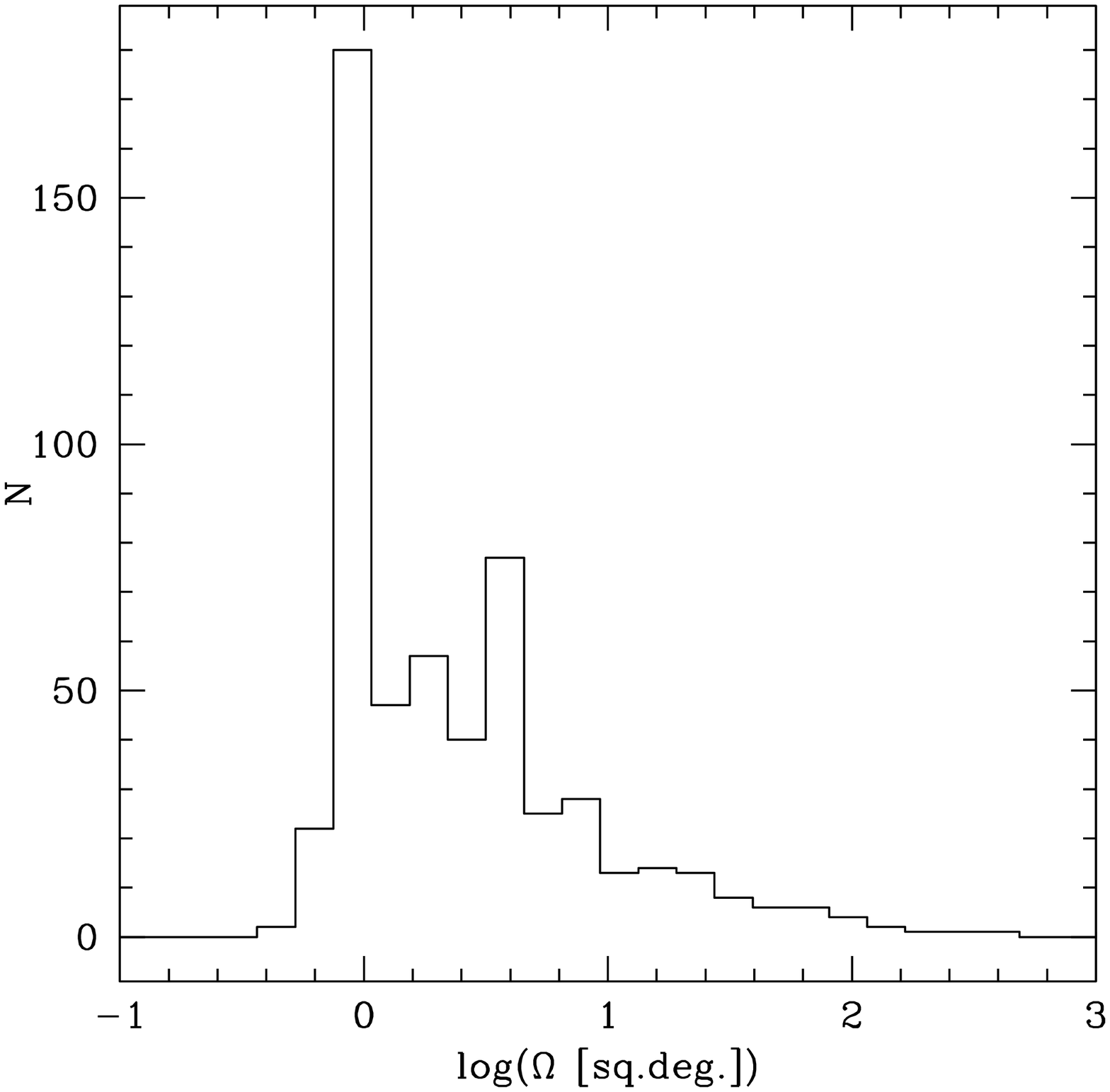,width=2.0in,angle=0,silent=1}
\vskip -2.0in
\hskip 4.0in
\psfig{figure=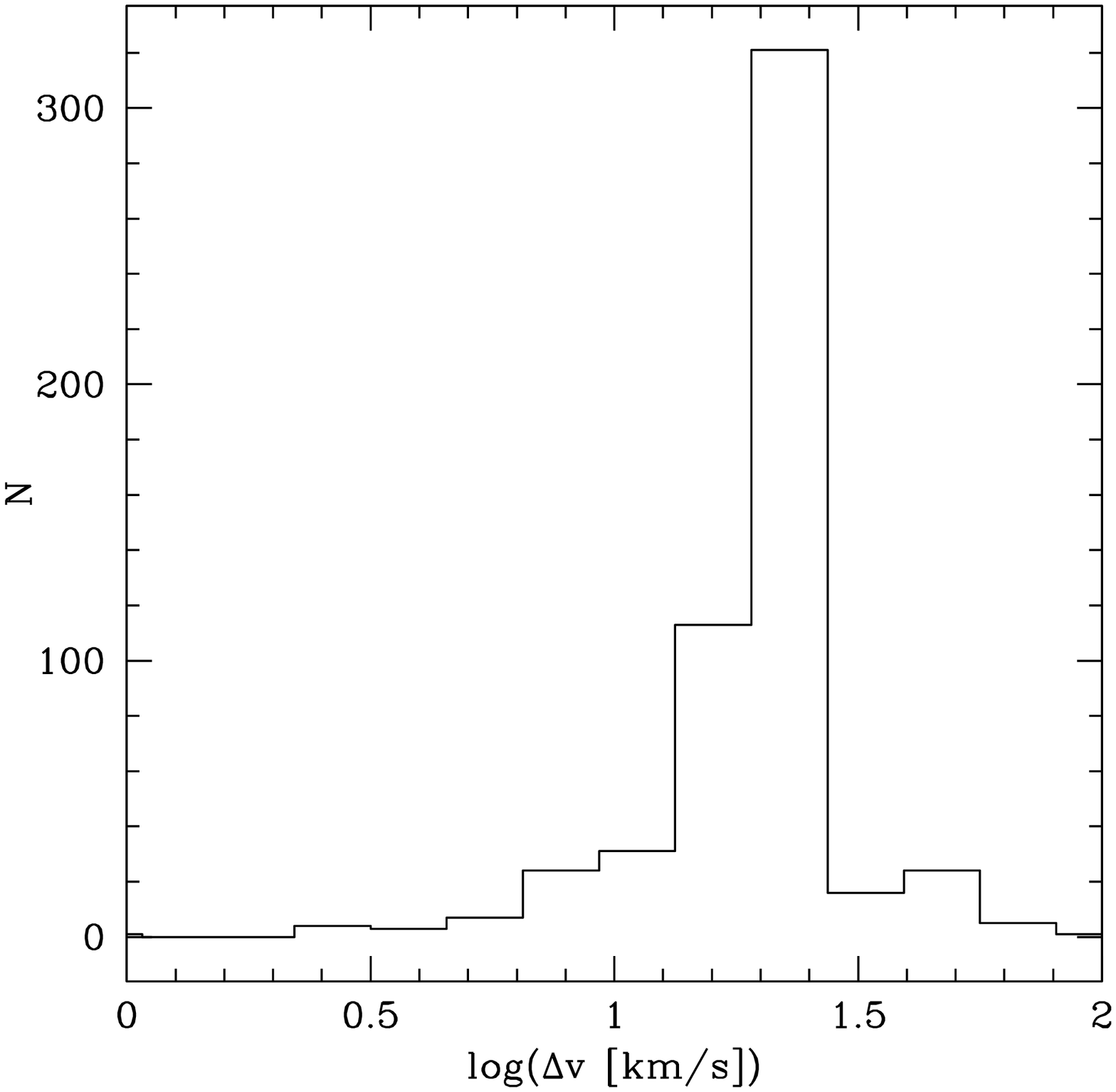,width=2.0in,angle=0,silent=1}
\caption{{\it Left:} Histogram of brightness temperature (in K) for the HVCs 
in the WvW91 compilation. {\it Center:} Histogram of solid angle (in square 
degrees) subtended by the HVCs in the WvW91 compilation. {\it Right:} 
Distribution of average linewidths (FWHM) (in \kms) for the HVCs in the WvW91 
compilation.}
%\caption{Histogram of the distribution of brightness temperature (in K) for 
%the WvW91 compilation of HVCs.}
\label{fig:3wakker}
\end{figure}

We first consider the statistics of the clouds to determine mean
observational quantities, $T_{\rm B}$, $\Omega$, and $\Delta v$.  We
use the WvW91 compilation because of the better spatial resolution
compared to the BTL catalogue.  However, the line width is not explicitly
given and
must be estimated from the total flux and $T_{\rm max}$.  Histograms of
these quantities are given in Figure~\ref{fig:3wakker}.  The
statistical properties are given in Table 1.  We have checked the
distribution of these quantities with the BTL compilation and find good
agreement between the two catalogues given their differences between them.

\begin{table}
\begin{center}
TABLE 1\\Mean Observed HVC properties\\\vskip 2mm 
\begin{tabular}{lrrr}\hline\\[-2mm]
Quantity     & mean  & dispersion & median 
\\[2mm]\hline \hline \\[-3mm]$T_{\rm B}$ (K)       &   0.35   & $\pm$ 0.88 &  
0.14   \nl$\Omega$~~(sq deg)   &   16.1    & $\pm$ 133.9  &  1.9    \nl 
$\Delta v$~(\kms)     &   20.6   & $\pm$ 9.2  &  20.9   \\[2mm]\hline\\[-2mm] 
\end{tabular}
\vskip 2mm Mean of log HVC properties\\\vskip 2mm
\begin{tabular}{lrrr} \hline\\[-2mm]$T_{\rm B}$ (K)       &   0.16   & $\pm$ 
0.28 &  0.14   \nl$\Omega$~~(sq deg)   &   2.5    & $\pm$ 6.8  &  1.9    \nl 
$\Delta v$~(\kms)     &   18.9   & $\pm$ 10.5  &  20.9   \\[2mm]
\hline\\[-2mm] 
\end{tabular}
\end{center}
\label{tab:meanobs}
\end{table}

From Figure~\ref{fig:3wakker}, as well as from
Table 1, it is clear that the mean and median of both $T_{\rm B}$ and
$\Omega$ are quite different, which is reflected in the dispersion of
these quantities.  The means are dominated by the skewed distribution
and a few very large values.  We obtain representative values of the
observed quantities from the means of log $T_{\rm B}$, log $\Omega$ and
log $\Delta$v,\\ which give values close to the medians of the observed
quantities.  Note the narrowness of the distribution of velocity
dispersion; this small dispersion occurs in both the BTL and WvW91 data
and suggests that the HVCs as a whole are a single collection of objects
with similar intrinsic properties.

\begin{figure}[htpb]
\psfig{figure=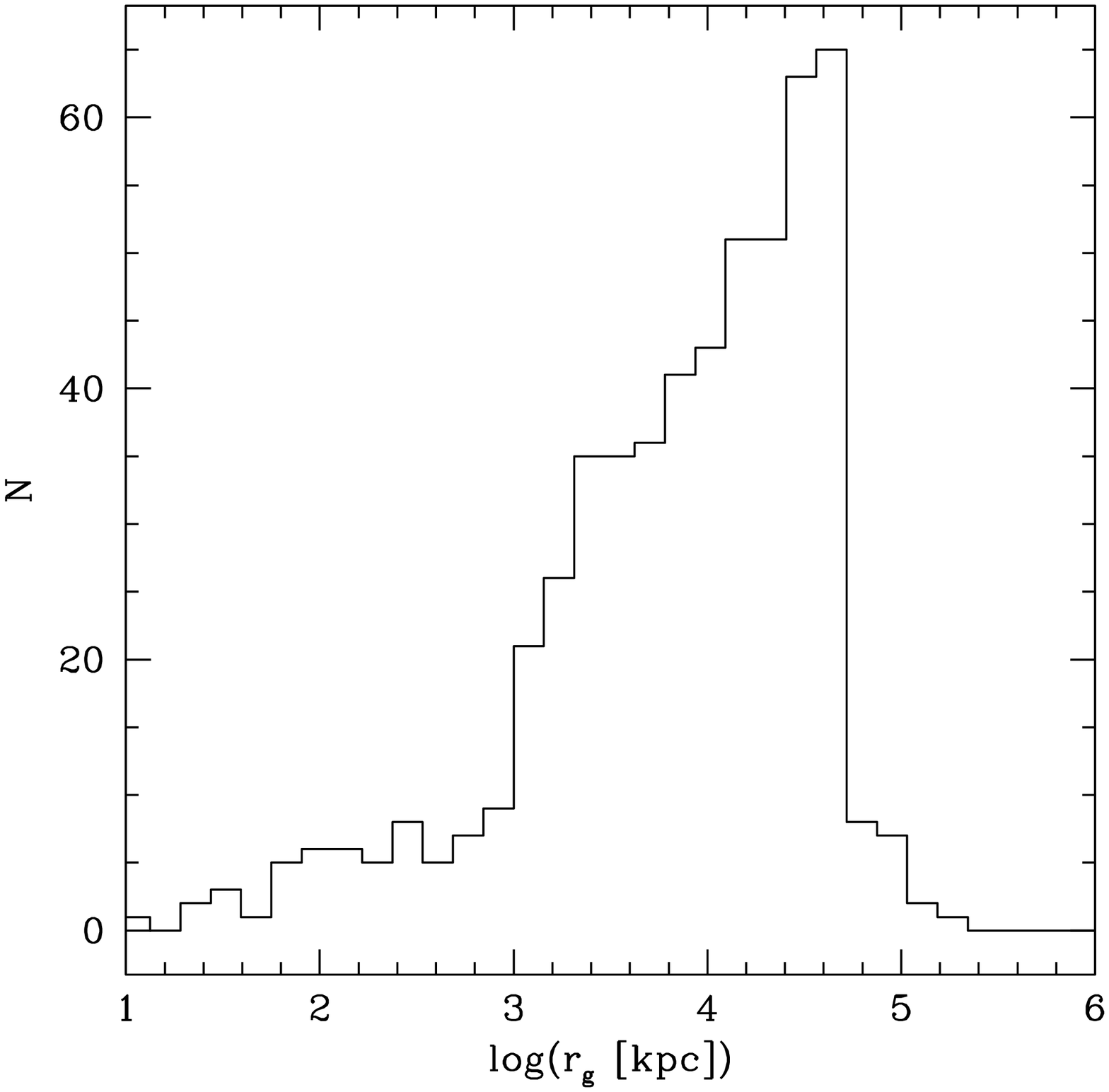,width=3.0in,angle=0,silent=1}
\vskip -3.0in
\hskip 3.0in
\psfig{figure=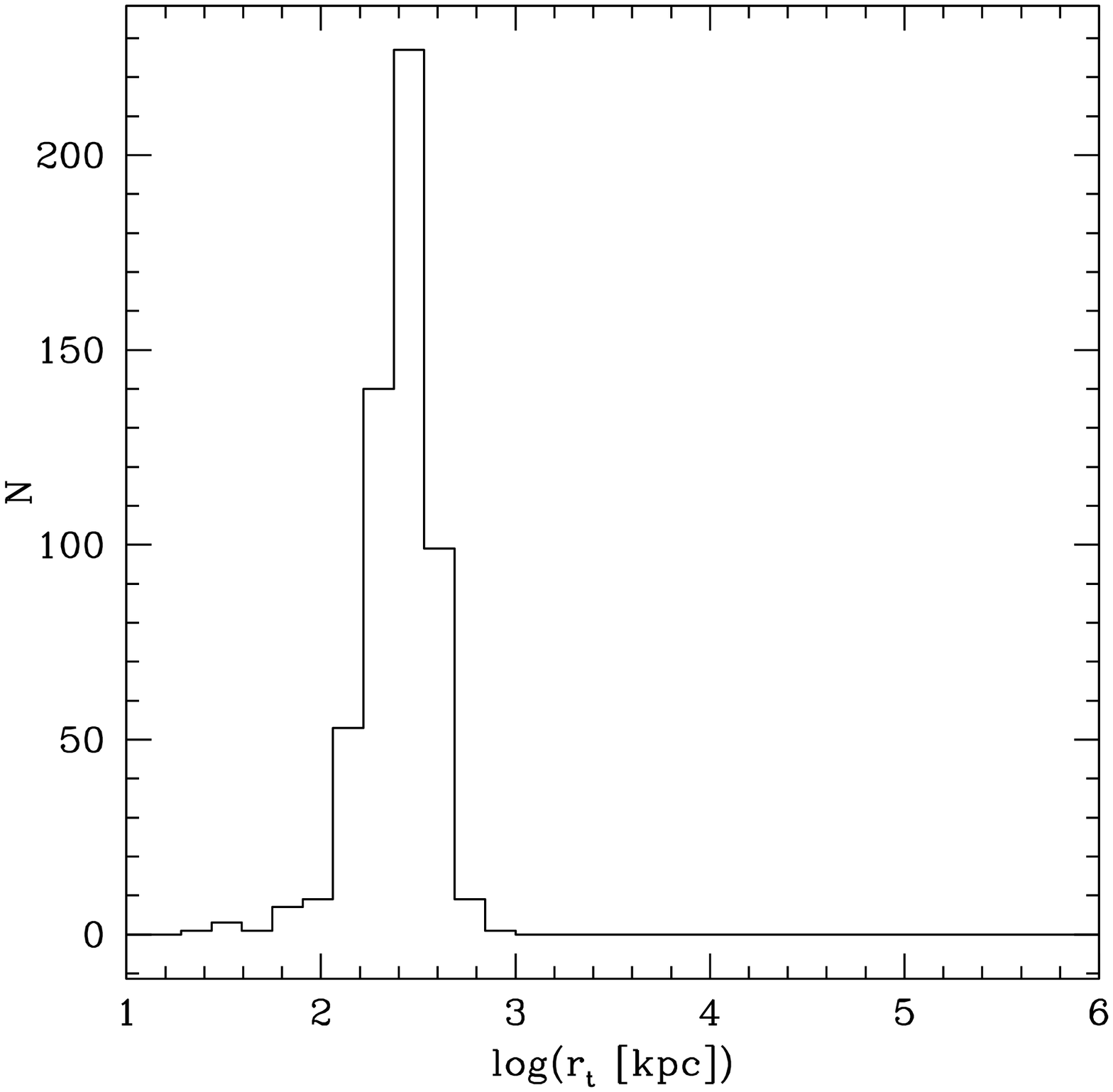,width=3.0in,angle=0,silent=1}
\vskip -0.2in
\caption{{\it Left:} Histogram of the distribution of distances in kpc
at which the neutral hydrogen in an HVC in the WvW91 compilation
would make the cloud self gravitating. {\it Right:} Distribution of
distances (in kpc) of the clouds in the WvW91 compilation within which
the clouds would have to be if they are tidally stable.
}
\label{fig:r_gwakker}
\end{figure}

We now wish to determine the mean values of $r_{\rm g}$ and $r_{\rm t}$.
In the
WvW91 compilation we eliminate the clouds for which we do not have a
good estimate of $\Delta$v: clouds with $\Omega >100$ sq deg (where
which the assumption that T$_{max}$ is a representative temperature
over most of the cloud probably breaks down). This introduces only a
small bias because there are only nine clouds with such large areas.
Histograms of the results are shown in Figure~\ref{fig:r_gwakker}.  In
the BTL sample, we eliminate unresolved clouds, which leaves 444 clouds
from the sample of 1312. As a check we also calculate $r_{\rm g}$, and $r_{\rm t}$
from the mean of the log cloud properties in Table 1.  A summary 
of the results is
given in Table 2.  All three methods give consistent determinations of
these quantities.

\begin{table}[htpb]
\begin{center}
TABLE 2\\Stability distances\\\vskip 2mm 
\begin{tabular}{llll}
\hline\\[-2mm]
Quantity     & mean  & dispersion & median 
\\[2mm]
\hline \hline \\[-3mm]
%{Wakker - Hulsbo-sch Sample} \nl
{Wakker--Hulsbosch sample} & kpc & kpc & kpc \nl
\hline\\[-2mm]
$r_{\rm g}$~(kpc) & 6.5 $\times 10^3$~$f$  & $\pm$ 6.92  &  10.0 
$\times 10^3$~$f$ \nl
$r_{\rm t}$~(kpc) & 2.5 $\times 10^2$~$f^{1/2}$ & $\pm$ 1.51  &  
2.7 $\times 10^2$~$f^{1/2}$  \\[2mm]
\hline\\[-2mm]
{BTL sample} \nl 
\hline\\[-2mm]
$r_{\rm g}$~(kpc) & 4.9 $\times 10^3$~$f$  & $\pm$ 3.5  &  5.9 $\times 
10^3$~$f$ \nl
$r_{\rm t}$~(kpc) & 3.0 $\times 10^2$~$f^{1/2}$  & $\pm$ 2.0  &  3.0 
$\times 10^2$~$f^{1/2}$ \\[2mm]
\hline\\[-2mm]
{From mean cloud properties} \nl 
\hline\\[-2mm]
$r_{\rm g}$~(kpc) & 6.7 $\times 10^3$~$f$    &   &    \nl 
$r_{\rm t}$~(kpc) & 
2.1 $\times 10^2$~$f^{1/2}$   &   &    \\[2mm]
\hline\\[-2mm] \hline\\[-2mm] 
\end{tabular}
\end{center}
\label{tab:stabdistances}
\end{table} 

\begin{table}
\begin{center}
TABLE 3\\
Mean derived HVC parameters\\
\vskip 2mm 
\begin{tabular}{lr}
\hline\\[-2mm]
Quantity     & value \\[2mm]
\hline \hline 
\\[-3mm]
Mass    &  3.2 $\times 10^8$ \msun \nl
HI mass &  3.4 $\times 10^7$ 
\msun \nl
Diameter   &   28  kpc \nl
Distance  &   1 Mpc  \nl
$<n_{\rm HI}>$ &    
1.2 $\times 10^{-4}$ \cc  \nl
$f$ &           0.15 \\[2mm]
\hline\\[-2mm] 
\end{tabular}
\end{center}
\label{tab:meanprops}\end{table}

We note first that $r_{\rm t} < r_{\rm g}$ for the great majority of clouds;
thus the tidal distance is generally unimportant.  The typical value of
$r_{\rm g}$ is of the order of 6$f$ Mpc.  If $f = 1$ and the clouds were
self-gravitating at this distance, they would, on average, be part of
the Hubble flow, they would not exhibit the velocity cutoff described
in \S\ref{sec:extrema}, and they would also not exhibit the
overwhelmingly negative observed velocities relative to the LGSR. A
mean distance of 6$f$ Mpc is also an order of magnitude larger than the
distance to M31 for values of $f$ near unity, and much larger than the
1.5 Mpc radius at which the clouds are turned around from the Hubble
flow in the simulations (see Figure~\ref{fig:arrowc}).  However, if the
HVCs are extragalactic, we expect that they contain copious amounts of
dark matter.  Observations of galaxies and cluters suggest that the
baryon/dark matter ratio is roughly 0.1 \markcite{Fukugita97}(Fukugita,
Hogan, \& Peebles 1997).  If HVCs are fair samples of material from
which galaxies are made, then we expect that
$f \sim 0.1$.  This implies that their true distance is $\sim $ 0.5 -
1.0 Mpc.

Using the mean cloud values, at a typical distance of 1 Mpc the typical
diameter of an HVC is about 28 kpc, a large value, but comparable to
the value we obtain for Complex H in \S\ref{sec:hulsbosch}.  At a
distance of 1 Mpc, the HVCs have a typical neutral hydrogen mass of
$\sim 3.4 \times 10^7$ \msun, a value close to the HI mass of Complex H,
a total neutral gas mass of $\sim 4.7 \times 10^7$ \msun, and a total
mass of $\sim 3.2 \times 10^8$ \msun.  The mean HI density ($n_{\rm
HI}$) of a typical cloud is then $1.2 \times 10^{-4}$ \cc.  A mean
density this small might require some clumpiness for the cloud to
remain neutral even in the metagalactic radiation field (see
\S~\ref{sec:ionization}).  The typical derived HVC properties are
summarized in Table 3.

\subsection{Stability of the Largest HVCs}
\label{sec:bighvcs}

Although for most HVCs the tidal--stability distance is less than the 
self--gravitating distance, the opposite is the case for the largest
HVCs.  Furthermore, for these largest HVCs, the conditions imposed by
equations (6) and (7) are so severe that all clouds with angular sizes
greater than about 100 sq deg cannot be tidally stable at any
reasonable distance. Equation (7) can be rewritten, for example, as
\begin{equation} 0.34~\Omega~\Delta v_{20}~~<~~  {R\over 100~ {\rm
kpc}} \end{equation} where $\Delta v_{20}$ is the linewidth measured in
units of 20 \kms; Figure~\ref{fig:3wakker} shows that this quantity is
generally of order unity.  Thus any cloud with an angular size $>$ 60
sq deg will be tidally unstable for any distance $<$ 2 Mpc.  Complexes
A, C, and M, the three main structures comprising the Northern
Hemisphere grouping, are all larger than this: Complex C, for example,
covers 1814 sq deg.  All three of these complexes in the Northern
Hemisphere grouping therefore are tidally unstable for any reasonable
distance if they are self-gravitating.  The gravitational--binding
distance, $r_{\rm g}$, is moreover quite small for the largest clouds.
For  Complex A, which has a mean $\Delta v$ of $\sim$40 \kms, and a
mean $T_{\rm B}$ of 1.0 K, $r_{\rm g} < 85f$ kpc.  That is, this cloud
cannot be more than 85 kpc distant even if it is self--gravitating by
its gas content alone ($f = 1$); if it contains substantial amounts 
of ionized gas or dark matter, it is
probably substantially closer than 85 kpc.  For values of $f$ in the
range 0.1 to  0.2, Complex A would have a maximum distance of  $\sim
10$ to 15 kpc.  If Complexes C and M, together with A, are all part of
a single general grouping, then this distance would be consistent with
the two measured absorption--line distances to the HVCs by
\markcite{Danly93}Danly et al. (1993) and by \markcite{van Woerden97}van
Woerden et al. (1997).  This distance is also consistent with the large
total angular extent of this structure, nearly 180\deg.  If the mean
diameter typical of HVCs is about 28 kpc, and if this structure is a
typical HVC, then its distance would be $< 15$ kpc, since it subtends
more than 2 radians.  We conclude then that the Northern Hemisphere
Complex probably has a mean distance of $\sim$ 10 to 15 kpc, an extent
of $\sim$ 25 kpc, is being tidally sheared, 
and traverses a range of distances from the Sun,
including some which are evidently as near as several kpc.

These considerations suggest that the HVC Complexes A, C, and M, which
have been the most studied of the HVCs because of their large angular
extent and location at declinations accessible from northern hemisphere
telescopes, are atypical because they are so close;  moreover they appear to be
interacting with the Milky Way.  The proximity to these clouds might
also explain why distance determinations to the Northern Hemisphere
grouping have been so discrepant, with most absorption--line
measurements giving only upper limits:  the material is probably strung
out over a large range of distances because of its intrinsic size and
the tidal shearing.

There are two other complexes with angular sizes in excess of 1000 sq
deg: the Outer Arm Complex and the Magellanic Stream. The tidal and
gravitational binding considerations discussed above apply equally well
to these  complexes if they are self--gravitating.  In the case of the
Outer Arm, it is not clear whether it is really a collection of
HVCs, or part of the normal structure and dynamics of the
outermost Milky Way.  In any event, it must be tidally unstable if it
is self--gravitating,  and both it and the Magellanic Stream have
self-gravitating distances $<$ 100 kpc, possibly considerably less than
this.  It might even be that the Magellanic Stream is not gas tidally
disrupted from either the Milky Way or the LMC, but rather an HVC that
has become entrained in the tidal field of the LMC/Milky Way  system.

For clouds with angular sizes between $\sim$ 100 and 1000 sq deg, the
gravitational binding distance is greater than that for the three
largest clouds complexes, and such clouds would be expected to be more
distant than Complex A.  Complex H, which we discuss in detail in
\S\ref{sec:features} has $\Omega$ = 250 sq deg, $r_{\rm g}$ = 580 $f$
kpc, and $r_{\rm t}$ = 190 $f^{1/2}$ kpc.  If Complex H is typical of
other HVCs, and $f$ is about 0.1, then $r_{\rm g} \simeq r_{\rm t}$ =
60 kpc, and its distance would be close to the value of 50 kpc value
determined from kinematic considerations in \S\ref{sec:hulsbosch}.  It
seems, therefore, that the stability argument yields some information
about distances, and that this information is consistent both with
direct distance measurements, in the case of the nearby Northern Hemisphere
clouds, with inferences from the kinematics in the case of Complex H,
and with cloud properties statistically inferred from numerical
simulations.

\section{Evidence for Extragalactic Nature of HVCs}
\label{sec:evidence}
\subsection{Individual HI features}\label{sec:features}

We consider here three individual clouds seen in the second 
quadrant of 
Galactic longitude which suggest that at least some of the HVCs must be 
extragalactic.

\subsubsection{HVC Complex H}
\label{sec:hulsbosch}

\begin{figure} [htpb]
\centerline{\psfig{figure=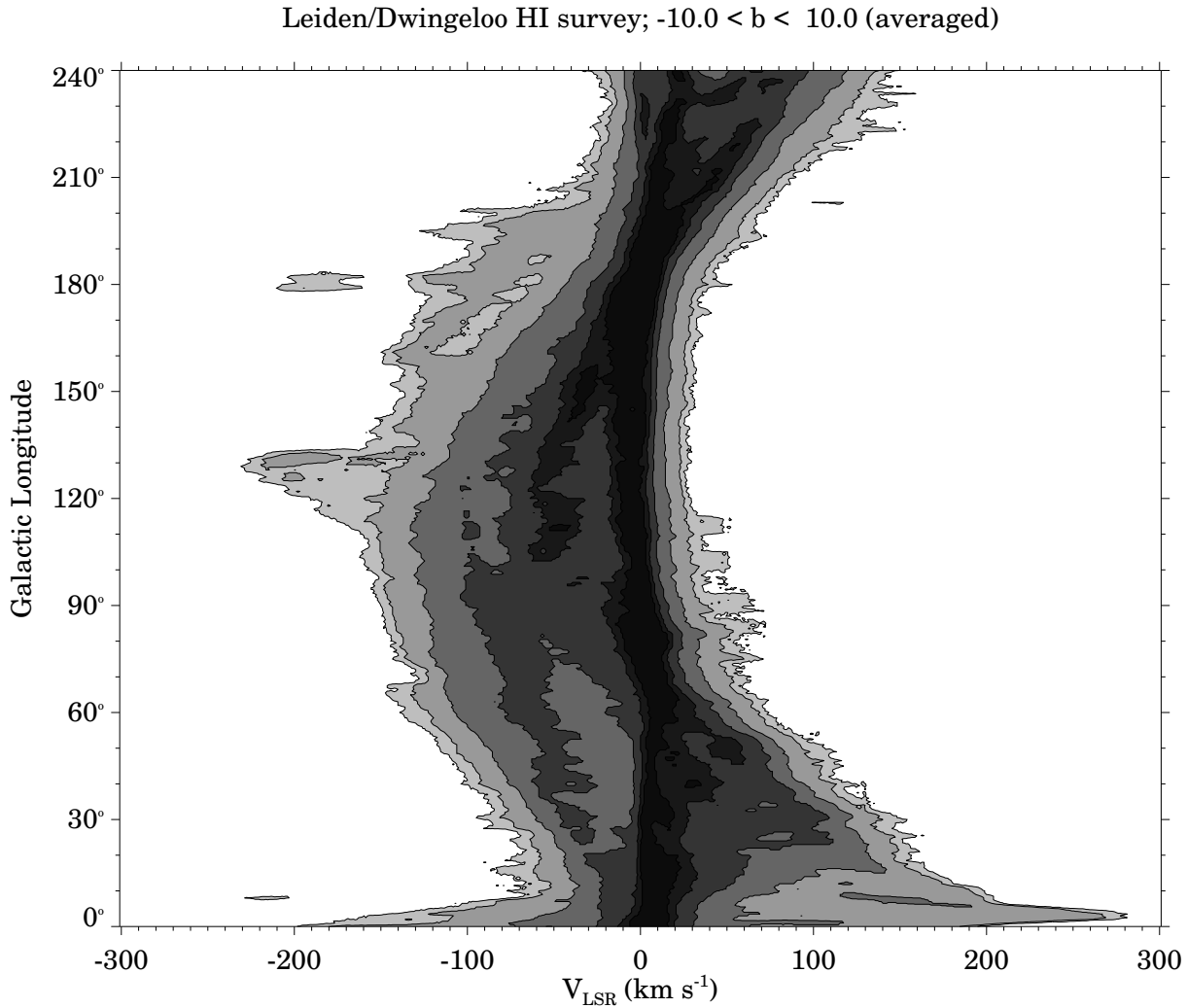,height=6in,angle=+90,silent=1}}
\vskip -0.2in
\caption{Longitude--velocity plot of HI emission from the LD survey averaged in 
latitude over $|b| \leq 10\deg$.  The contour intervals are spaced 
logarithmically.  Gas at negative velocities for $l < 180\deg$ corresponds to 
gas beyond the Sun's radius in a circular--rotation model.  The lowest 
negative--velocity contours are approximately sinusoidal, indicating nearly 
circular rotation to a distance of about $R = 40$ kpc for a flat rotation 
curve.  Note, however the emission at $110\deg < l < 133\deg$, with velocities 
up to about $-230$ \kms.  This is Complex H; higher--sensitivity maps show
that the velocity of Complex H extends to $-240$ \kms.}
\label{fig:lvhulsbos}
\end{figure}

Longitude--velocity plots of HI lying near the Galactic equator show
that most of the gas in the Milky Way is in nearly--circular orbits
about the Galactic center.  Figure~\ref{fig:lvhulsbos} shows such a
plot for LD--survey data over the longitude range $0\deg \leq l \leq
250\deg$, averaged over $|b| \leq$ 10\deg.  Negative velocities in the
longitude range \le = 0\deg -- 180\dege correspond to gas at distances
greater than $R_\odot$, under the assumption of circular rotation;
near--circular orbits are suggested by the approximately sinusoidal
contours of the HI, especially at the lower contour levels.  The most
distant gas would contribute the lowest contours, at velocities near
170~sin\thinspace$l$  \kms, corresponding to a distance of about 37 kpc from the
Galactic center, if $\Theta(R\geq R_\odot) = \Theta_\odot =$ 220 \kmse
(or to a distance of about 27 kpc if the rotation curve were to become
Keplerian at the last measured point about 20 kpc from the center).
Gas with rotation velocities in excess of $\Theta_\odot$ cannot be in
circular rotation anywhere in the Milky Way.  For $\Theta_\odot <$ 220
\kms, kinematic distances are increased.

\begin{figure} [htpb]
\centerline{\psfig{figure=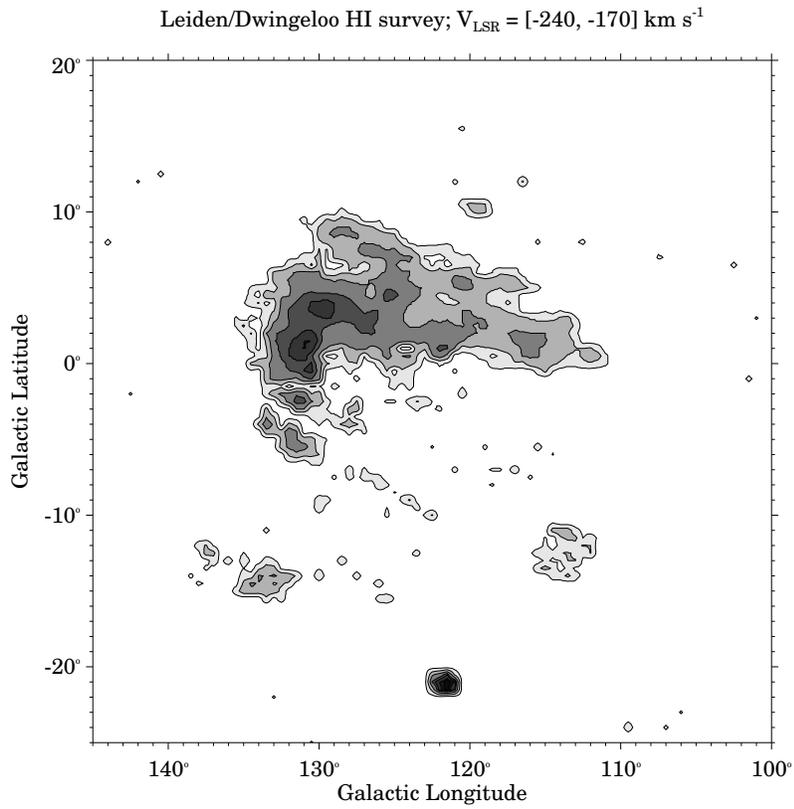,height=6in,angle=0,silent=1}}
\vskip -0.2in
\caption{HI emission from Complex H integrated over the velocity range $-240 
\leq~ v_{\rm LSR} \leq -170$ \kms.  This range was chosen to exclude the 
conventional Galactic gaseous disk; parts of Complex H may in fact extend 
to more extreme velocities.  The contour intervals are spaced linearly.  The 
bright object at $l = 122\deg$, $b = -21\deg$ represents the portion of M31 
emitting within the chosen velocity range.}
\label{fig:hulsbos}
\end{figure}

Figure~\ref{fig:lvhulsbos} also shows emission, however, from an HI
structure, in the longitude range $l =$ 110\dege to 135\deg, extending
to velocities as high as --240 \kms, well beyond the most extreme
circular--rotation speeds permitted.  A map of this HVC material is
shown in Figure~\ref{fig:hulsbos}, which integrates all of the HI
emission beyond the range of normal circular velocities, from $-240 <
v_{\rm LSR} < -170$ \kms. This structure was first detailed by
\markcite{Hulsbosch75}Hulsbosch (1975) and therefore named Complex H by
WvW91. Complex H is quite large, with an angular extent of $\simeq$
25\deg. (The complex has a somewhat larger extent in the compilation of
WvW91 than that shown in Figure~\ref{fig:hulsbos} because WvW include
velocities that are still part of the outermost gas in the Galactic
disk.) The radial--velocity centroid of the brightest part of the
emission is --200 \kms; if this characterizes the systemic velocity of
Complex H, and if the higher velocities are due to the velocity
dispersion of the HI, the distance from the Galactic center would be
about 85 kpc for a flat rotation curve (43 kpc assuming that the
rotation curve becomes Keplerian at $R = 20$ kpc). In any event, Complex H
is quite large, with an angular extent of about 25\dege and a radial
velocity that places it beyond $\sim$ 50 kpc from the center, under any
reasonable Galactic rotation curve.

Because Complex H lies directly in the plane, it is at a minimum distance
of 40 kpc from the Galactic center, whether or not its radial velocity
corresponds to circular motion.  If the complex were at distances
smaller than 40 kpc (that is, if the velocity of the cloud has a
substantial non--circular component), then the velocity difference
between the complex and the ambient Galactic disk gas would range from
30 to 200 \kms,  producing a huge region of highly shocked
gas.  No major disturbance in the HI images of this part of the disk is
apparent, however.  One would also expect strong radio--continuum and X--ray
emission, and other indicators of strong shocks, but no such shock
tracer is evident over the region. H$\alpha$ or other optical emission
would be expected at least at the higher latitudes where the extinction
is relatively low, because the high velocity of Complex H relative to
the ambient gas would be comparable to that of the jets from young
stars, which are all strong optical emitters \markcite{Lada85}(Lada
1985).  There is no indication that the emission from the disk is
anomalous, supporting the conclusion that the complex lies beyond the
gaseous disk of the Milky Way.

The large angular size of Complex H implies that, if it is at a distance
of 50 kpc from the Sun, then it has a diameter of about 20 kpc and an HI
mass of about 9$\times 10^7$ \msun, using the total HI flux seen in
Figure~\ref{fig:hulsbos}, an enormous gas structure by Galactic 
standards; the mass
would scale with the square of the distance. Using equation  (2) of
\S4, the distance of the cloud would be 1.0~$f$ Mpc if it were
self--gravitating, where $f$ is the ratio of neutral gas/total mass of
the structure. Thus $f$ would have to be $\sim$ 0.1 if Complex H were
self gravitating, a value similar to that deduced in \S4.1 for the 
ensemble of HVCs.

\subsubsection{Kinematic Distances to Other HVCs}
\label{sec:galactic}

The argument supporting a large distance for Complex H is more
difficult to apply to HVCs located out of the Galactic plane.  The
constraint that Complex H lies beyond the outer edge of the gaseous
disk is not applicable to most HVCs because they are largely identified
out of the Galactic plane.  For example, another HVC with a substantial
deviation velocity is seen in Figure~\ref{fig:lvhulsbos} near $l =
180\deg$, $v =-180$ \kms; this is the ACI knot in the Anticenter
Complex, most of which extends beyond the borders of 
Figure~\ref{fig:lvhulsbos}.  The ACI concentration is centered only
about 10\dege below the Galactic plane and, as Complex H, shows no 
association with
any disturbance in the conventional gaseous disk.  However, because it
does not lie directly in the Galactic plane, the constraint on the
distance to the ACI knot is weaker than for Complex H.

There have been several suggestions of evidence for interactions of HVC
material with the conventional gaseous disk in some of the
Northern--Hemisphere Clouds which were argued in \S\ref{sec:bighvcs} 
to be relatively
local, but none for the HVCs which we argue are dispersed throughout
the Local Group (\S\ref{sec:lgkinematics}).  \markcite{Burton97}Burton (1997) 
discussed two
general observational constraints on the distances of high-- and
intermediate--velocity clouds. The ``scale--height constraint"
recognizes the measured half--thickness of the HI gaseous disk, which
is well--established by the observations to be about 120 pc; if HVCs
were a general property of the disk--halo transition region, they would
have to be confined to $|z| \leq 100$ pc to be consistent with the 
measured thickness;
otherwise, HVCs would have to occur beyond the disk--halo interface,
typically at large enough distances that their existence does not
contaminate the scale--height measurement.  

\subsubsection{HVCs in the Direction of M31} 
\label{sec:m-31}

\begin{figure}[htpb] 
\centerline{\psfig{figure=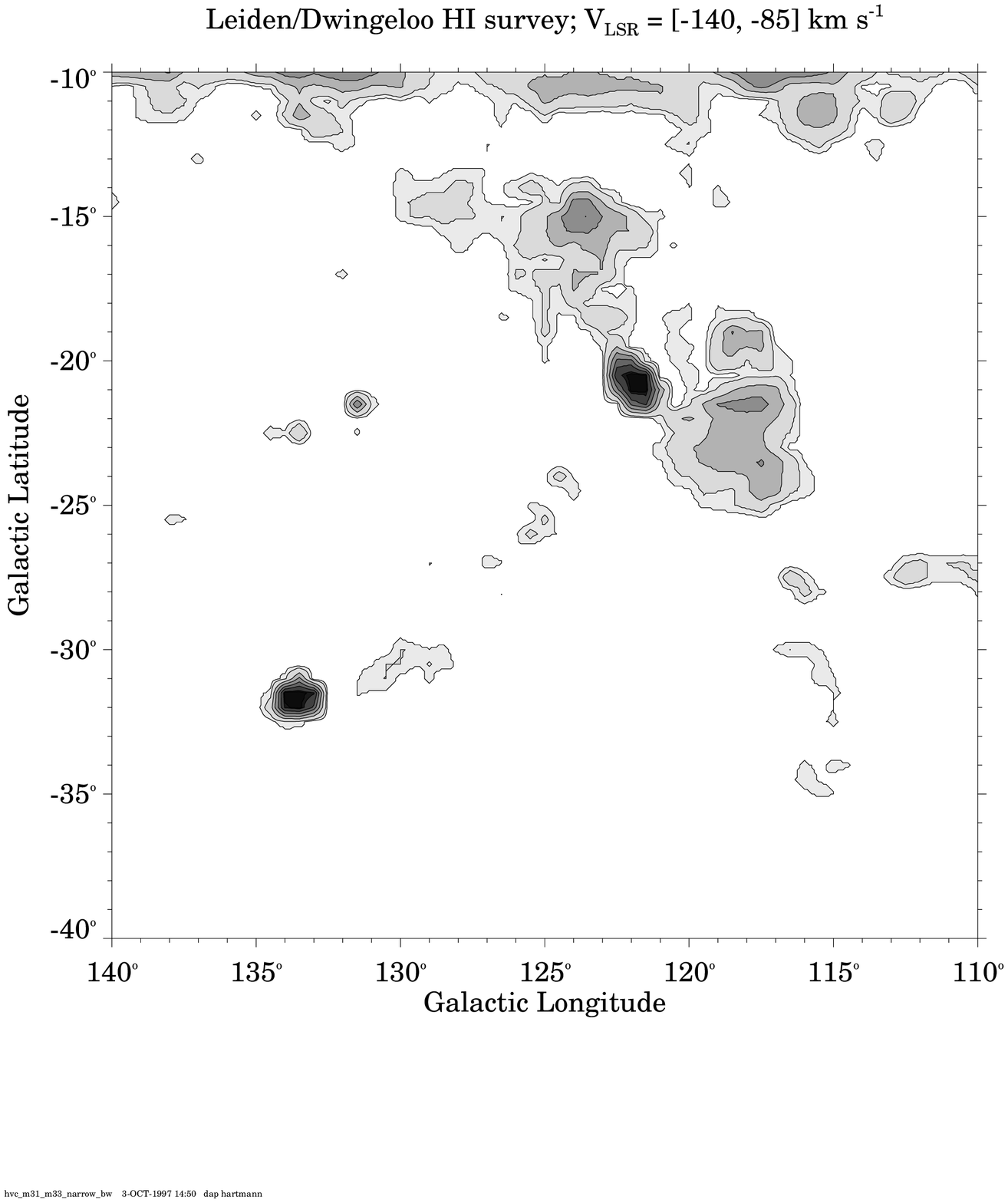,height=5.5in,angle=0,silent=1}}
\vskip -0.2in 
\caption{HI emission in the vicinity of M31 integrated over the
velocity range $-140 \leq~ v_{\rm LSR} \leq -85$ \kms. HI emission from
M31 in this velocity range yields the bright knot of emission at $l
=122\deg$, $b = -21\deg$.  The extended halo of emission around M31 is
the M31 Cloud.  The bright knot at $l = 133\degper5$, $b =
-31\degper5$, is the HI emission from M33 in the velocity range
plotted.  Other smaller HVCs are also evident in the figure; several of
them give the appearance of a broken chain extending between M31 and
M33.  The band of emission across the top of the image represents HI in
the gaseous disk of the Milky Way.}
\label{fig:m31}
\end{figure}

The bright emission feature evident in Figure~\ref{fig:hulsbos} at $l
= 122 \deg, b = -21\deg$ represents HI in the disk of M31.  Could it be
that Complex H is part of a tidal tail extending toward Andromeda?  We
used the BTL and LD surveys to map the HI emission in the general
region around M31, but found no HVC feature continuous or nearly
continuous in position and velocity comparable to the Magellanic
Stream.  We did, however, find the remarkable HI halo seen in
Figure~\ref{fig:m31} as it appears in the LD survey, approximately 
centered on M31,
which we discuss below as the M31 Cloud.  In this figure, M31 itself is
the bright spot in the central region of the cloud, some 2\dege in
extent; M33 is the bright feature at $l =133\degper5$, $b = -32\deg$.
The gaseous disk of the Milky Way contributes the emission along the
top edge of the figure. The M31 Cloud is also seen in the BTL survey
and a small portion of it occurs in the WvW91 catalogue.

The extent of the emission surrounding the direction to M31 is about
14\deg.  If the M31 cloud were at the distance of M31, its diameter
would be about 170 kpc and its mass about $9 \times 10^8$ \msun.  The
velocity extent of the cloud is --145 $< v_{\rm LSR} < -80 $ \kms; it
is therefore not centered on the systemic velocity of M31 ($v_{\rm LSR}
= -300$ \kms), but emission from the cloud does blend with
the lower--absolute--velocity side of M31.  The apparent blending may,
however, be fortuitous: there is no direct evidence for an interaction
between the HI cloud and M31, nor is there direct evidence that the cloud
and M31 are at the same distance.  Because the M31 Cloud does not connect
smoothly with Complex H, either spatially or kinematically, it appears
likely that the M31 Cloud and Complex H are unrelated.

The positional coincidence of the M31 Cloud with M31 
is striking, however, as is the blending
in velocity.  Not only is the M31 Cloud  nearly centered on 
M31 itself, its position angle on the sky as well as its inclination (if
the cloud is disk-like) are both similar to those of M31.  These
morphological similarities make it reasonable to ask if the cloud is
somehow associated with the galaxy.  It seems unlikely that the M31
cloud is part of a very extended gaseous disk of M31, because the cloud
has a less extreme systemic velocity than the galaxy proper, and
extends well beyond the minor axis on one side of the galaxy, which
would not be the case if it were in normal galactic rotation.
Furthermore, there is no evidence for the cloud in maps made in the
velocities between $-300$ and $-400$ \kms, corresponding to velocities
from the approaching, southeast part of the M31 disk.  Although
the M31 Cloud is superimposed on M31 on the sky, there is no clear
evidence for an {\it interaction} with the galaxy.  Nevertheless the
location and morphology of the M31 Cloud remain striking.

\markcite{Davies75}Davies (1975) found a compact HVC in close
positional proximity to M31 which he argued is likely to be associated
with that galaxy, despite its different velocity.  The Davies
cloud, with an angular extent of only 0\degper5, is within about
1\degper5 of the center of M31, but has an LSR velocity of $-447$ \kms,
as close to the systemic velocity of the galaxy as the M31 Cloud.  At
the distance of M31, the Davies cloud would have a mass of 4.7 $\times
10^6$ \msun~and a diameter of 5 kpc.

\subsubsection{HVCs in the Direction of M33}
\label{sec:m33}

\begin{figure}[htpb] 
\centerline{\psfig{figure=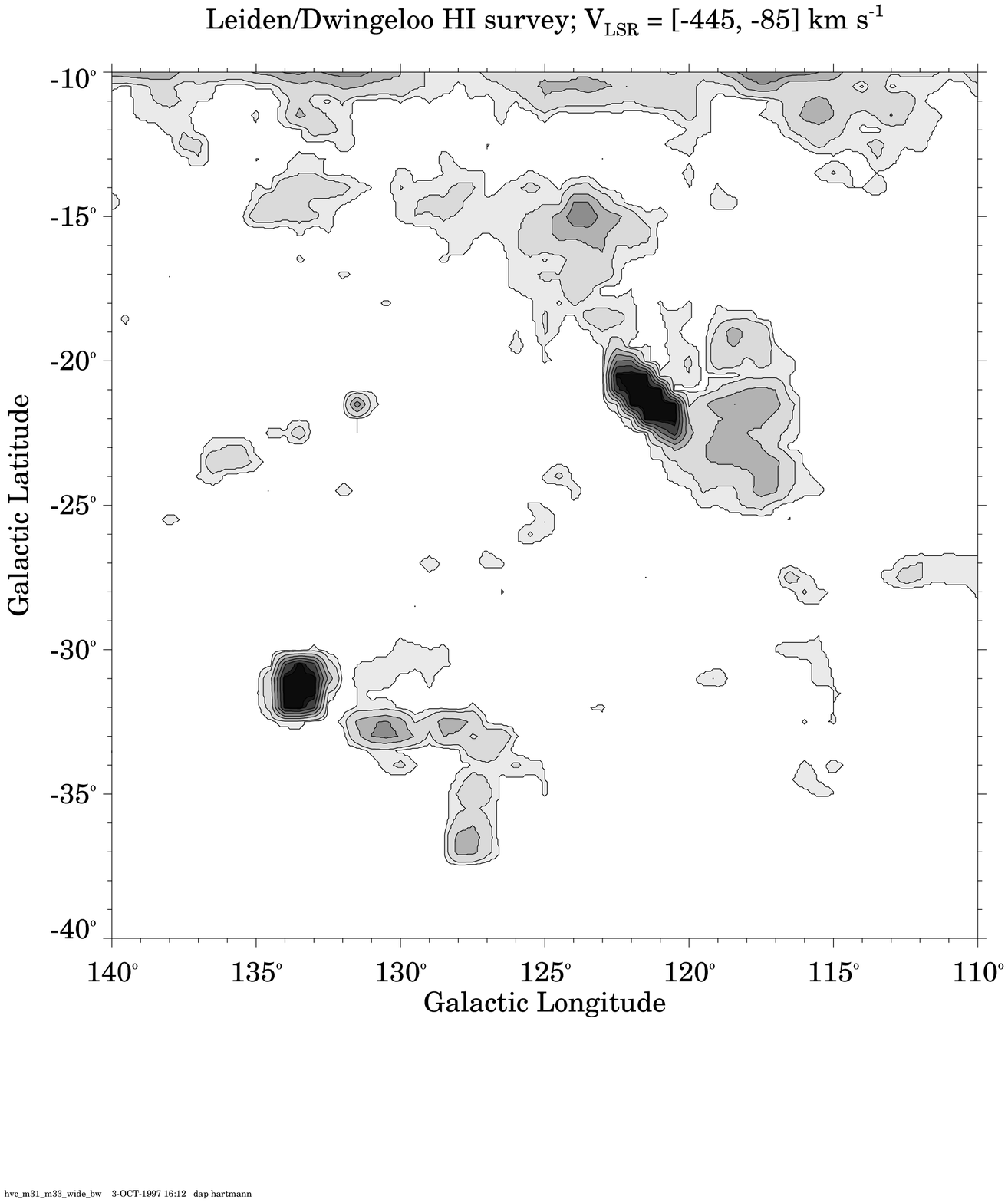,height=6in,angle=0,silent=1}}
\vskip -0.2in
%\plotone{m33m31.ps}
\caption{HI emission integrated over the broad velocity range $-445
\leq~v_{\rm LSR} \leq -85$ \kmse in the general vicinity of M31 and
M33.  Wright's cloud is the extended hook--shaped HVC patch centered near
$l = 128\deg$, $b = -33\deg$.  M31 and M33 both appear larger here than
in Figure~\ref{fig:m31} because the larger velocity range incorporates
more of the gas from the disks of these galaxies; similarly, several
additional HVCs are seen in this figure but not in
Figure~\ref{fig:m31}.}
\label{fig:m33_m31}
\end{figure}

Close to M33, \markcite{Wright74,Wright79}Wright (1974, 1979) found an
HVC in the velocity range $-440 < v_{\rm LSR} < -320 $ \kms; the cloud
is shown in Figure~\ref{fig:m33_m31} as it appears in the LD survey.  
M33 itself is the
bright object at $l = 133\deg$, $b = -31\degper5$; Wright's cloud
is the extended hook--shaped object to the right of, and slightly
below, that galaxy. The detailed maps which Wright made of this cloud
did not allow him to ascertain whether there is, or is not, any
connection between M33 and this very--high--velocity cloud of HI.  At the
distance of M33 the cloud would have an HI mass of $1.6 \times 10^8$
\msun~and a diameter of 70 kpc.

Figure~\ref{fig:m33_m31} also shows several other HVC patches that 
appear as a broken chain of emission along the line between M31 and
M33, suggesting some sort of tidal streamer.  These patches emit most
strongly in the velocity range $-110$ to $-130$ \kms, that is, within
about 50 to 70 \kmse of the systemic velocity of M33, $-179$ \kms. The
characteristic velocity of Wright's cloud differs from the systemic
velocity of M33 by about the same amount that the velocity of the M31
Cloud differs from the systemic velocity of M31.

\subsection{The Local Group Ensemble of HVCs}
\label{sec:localcomplex} 

\subsubsection{Spatial Distribution} 
\label{sec:lgspatial}

\begin{figure}[htpb] 
\vskip -0.2in
\centerline{\psfig{figure=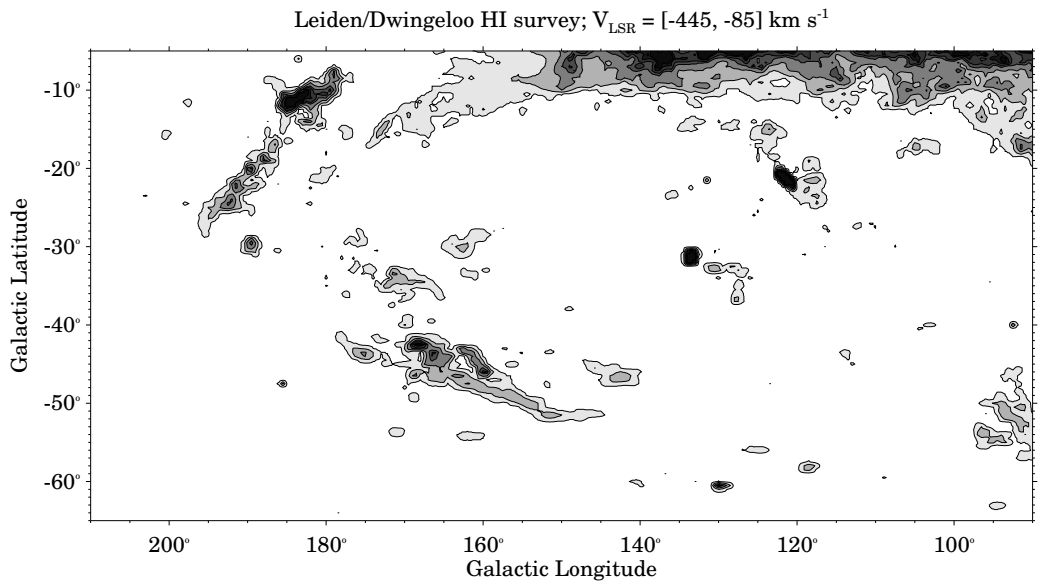,width=6in,angle=90,silent=1}}
\vskip -0.2in
%\plotone{hvcbarybw.ps angle=90}
\caption{HVCs seen over an extended range of latitude and longitude,
approximately centered on the direction of the barycenter of the Local
Group, and emitting in the same velocity range as represented in
Figure~\ref{fig:m33_m31}.  The band running along the top of the figure
to $l \simeq 160\dege$ represents conventional--velocity HI emission
from the gaseous disk of the Milky Way.  The entire region is replete
with HVCs: the WvW91 compilation tabulates some 100 individual clouds
within the boundaries of this figure.}
\label{fig:hvcbary}
\end{figure}

We have so far established that at least one HVC is extragalactic, and
that there are a number of HVCs plausibly associated with the most
massive members of the Local Group.  All of the emission seen in 
Figure~\ref{fig:m33_m31} at $b < -15\deg$,
with the exception of that from the
disks of M31 and M33, is contributed by HVCs, suggesting that the
entire region is replete with these objects. Figure~\ref{fig:hvcbary}
shows the high--velocity emission from a larger region of the LD
survey, $ 210\deg > l > 90\deg$, $-5\deg > b >-65\deg$, centered
approximately on the direction of the barycenter of the Local Group.
Emission from the gaseous disk of the Milky Way is seen as the band
running along the top of the image from $ l = 90\deg$ to $160\deg$.
All of the remaining emission is due to HVCs.  Some of the smaller HVCs
pervading this larger region are not catalogued by WvW91 because of the
more complete spatial sampling of the LD survey.

\begin{figure}[htpb]
%\plotone{aitoff.ps angle=-90}
\vskip -0.2in
\centerline{\psfig{figure=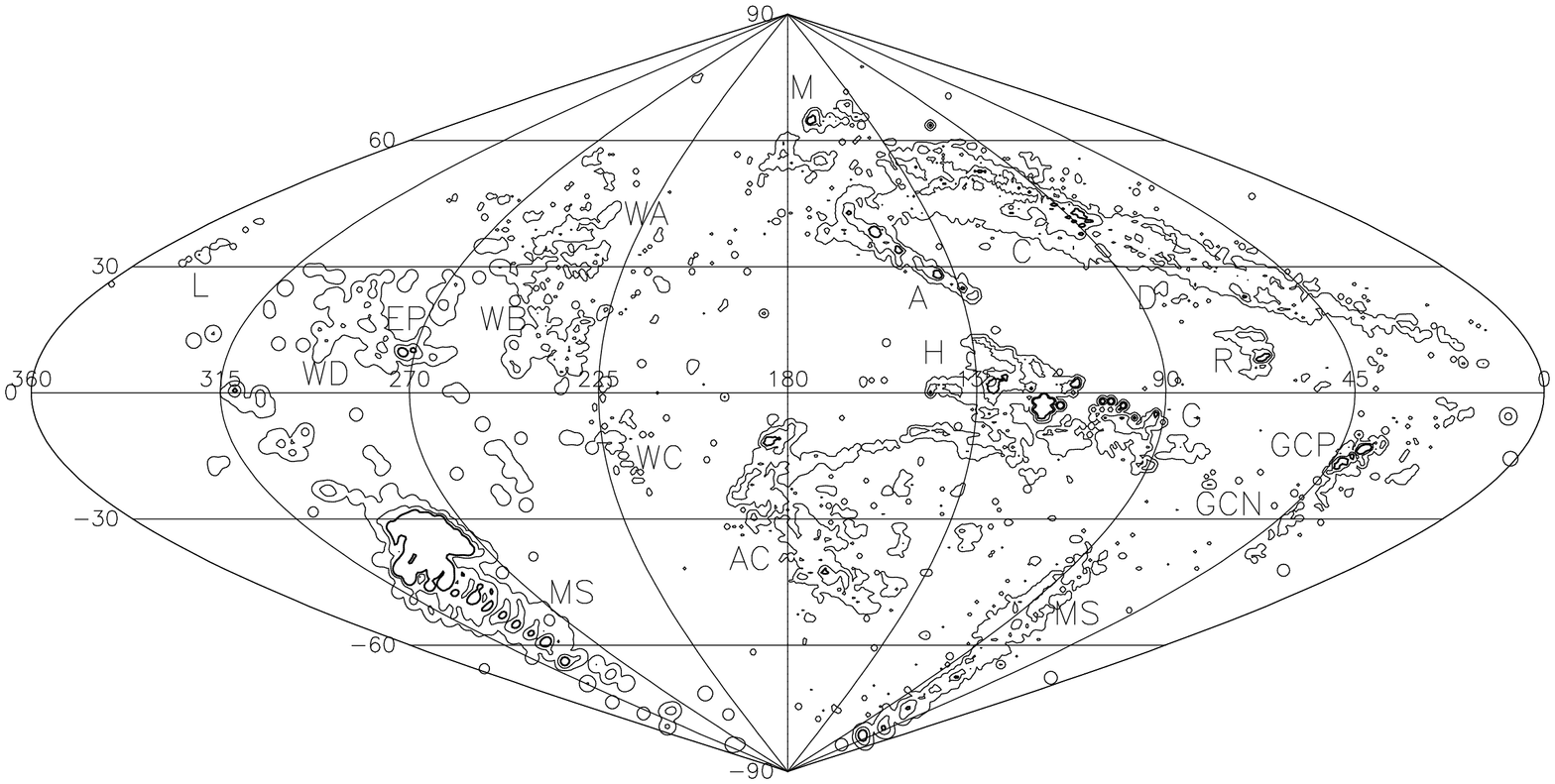,width=6in,angle=-90,silent=1}}
\vskip -0.2in
\caption{Reproduction of Wakker's (1991) Figure 2, in a somewhat
different projection, showing the locations of HVCs on the sky.
Individual HVC complexes are contoured by column density and labeled
according to Wakker's nomenclature; the principal groupings are
described in the text in the context of the Local--Group hypothesis.
In general, clouds with $l \leq~ 180\deg$ have negative LSR velocities;
those with $l \geq~ 180\deg$, positive LSR velocities.}
\label{fig:aitoff}
\end{figure}

HVCs are not uniformly distributed on the sky, and
\markcite{Wakker91}Wakker's (1991) Figure 2, (reproduced here as 
Figure~\ref{fig:aitoff}), demonstrates this well--known, important point.

The figure shows HI column--density contours of the HVCs catalogued
through 1990.  Wakker identifies 10 distinct populations but, for the
purposes of our discussion, identification of four major HVC groupings
suffices to encompass most of the HVCs over the entire sky.  These are:
(1) the Magellanic Stream (indicated by MS in Figure~\ref{fig:aitoff});
(2) the Northern Hemisphere Clouds ($210\deg > l >60\deg$; $70\deg > b >
30\deg$) comprising Complexes A, C, and M); (3) the Local--Group
Barycenter Clouds ($210\deg > l > 90\deg$; $10\deg >b > -60\deg$)
comprising Complexes G, H, and ACHV; and (4) the Local--Group
Antibarycenter Clouds ($310\deg > l >210\deg$; $50\deg > b > 10\deg$)
comprising the HVC Complexes WA, WB, WC, and WD discovered by
\markcite{Wannier72a}Wannier \& Wrixon (1972) and
\markcite{Wannier72b}Wannier, Wrixon, \& Wilson (1972) at moderate
positive--velocity deviation values, and the EP clouds (see references
in WvW91) at extreme positive--velocity deviations.

Each of these major groupings of HVCs has a distinct kinematic
signature.  The Magellanic Stream is a narrow streamer running
through the South Galactic Pole and the Magellanic Clouds, and is
contiguous over hundreds of degrees on the sky over a wide range of
velocities.  The Northern Hemisphere Clouds are also contiguous in
space and velocity, but the clouds have a narrower velocity extent
than the MS. These clouds, which are the most extensively studied of
all HVC groupings, all have negative velocities, with velocities
traceable into the IVC regime or even into the regime of conventional
Galactic--disk kinematics.  The Local--Group Barycenter Clouds comprise
the clouds seen in the general direction of the massive Local Group
galaxies, and all have negative velocities relative to the LSR.  The
Local--Group Antibarycenter Clouds all have positive velocities
relative to the LSR.  The clouds constituting this grouping are
situated approximately opposite to the direction of the barycenter of the
Local Group.  They do not form well--defined streamers:  the individual
members of this grouping are spatially distinct, relatively small in
extent, and with modest kinematic gradients.  Unlike the Northern
Hemisphere Clouds, both the Barycenter and the Antibarycenter groupings
are characterized by clouds with radial velocities that vary greatly
from cloud to cloud over the spatial extent of the entire
grouping.  In what follows, we eliminate the Magellanic Stream clouds
from the statistics of the HVCs, because their origin is known and is
evidently distinct from that of the other clouds.

\subsubsection {Kinematics of the Individual Groupings} 
\label{sec:lgkinematics}

We consider first the Local--Group Barycenter Clouds.  An 
estimate of the positional centroid of this grouping from 
Figures~\ref{fig:hvcbary} and \ref{fig:aitoff} yields $l = -143\deg$, $b =
-23\deg$. This centroid is remarkably close to the projected barycenter
of the Local Group calculated by \markcite{Einasto82}Einasto \&
Lynden-Bell (1982) to be at $l = 147\deg$, $b = -25\deg$, and well within the
error ellipse of their estimate. The angular extent of the Barycenter
grouping is approximately a steradian, roughly what one would expect if
the volume of the Local Group were filled with these clouds. But if
these clouds are related to the Local Group, they must then share its
kinematics. The mean velocity, velocity extent, and the velocity
centroid of the ensemble of HVCs provide a good test of which frame
of reference is most relevant.

Imagine a frame of reference moving at constant velocity relative to
the barycenter of a group of objects such as the HVCs.  In this moving
frame, the mean velocity of the ensemble will be shifted, but the
velocity dispersion will remain unchanged.  On the other hand, in a
reference frame that is rotating with respect to the barycenter of the
ensemble and offset from it, the observed velocity dispersion will be
increased; there may be also be a shift in the mean velocity, depending
on the distribution of the ensemble on the sky, although such a shift
would tend toward zero for full--sky coverage.  Consider, for example,
the radial velocities of globular clusters relative to the local
standard of rest and relative to the Galactic--center standard of rest (GSR).
Clearly, the velocity dispersion of the ensemble of globular clusters should be
smaller when measured relative to the GSR, and there would be no shift
in the mean velocity.  Transformation from LSR to GSR coordinates is
achieved by subtracting the motion of the LSR, 220 sin $l$ cos $b$
\kms, from the LSR velocity of each globular cluster.  Using the
compilation of \markcite{Harris96}Harris (1996), we find that the
globular clusters have a dispersion of 134 \kmse and a mean velocity of
9 $\pm$ 12 \kmse relative to the LSR; relative to the GSR, 
the dispersion is 119 \kmse with a mean of 3 $\pm$ 11 \kms, as
expected.

\begin{figure}[htpb]
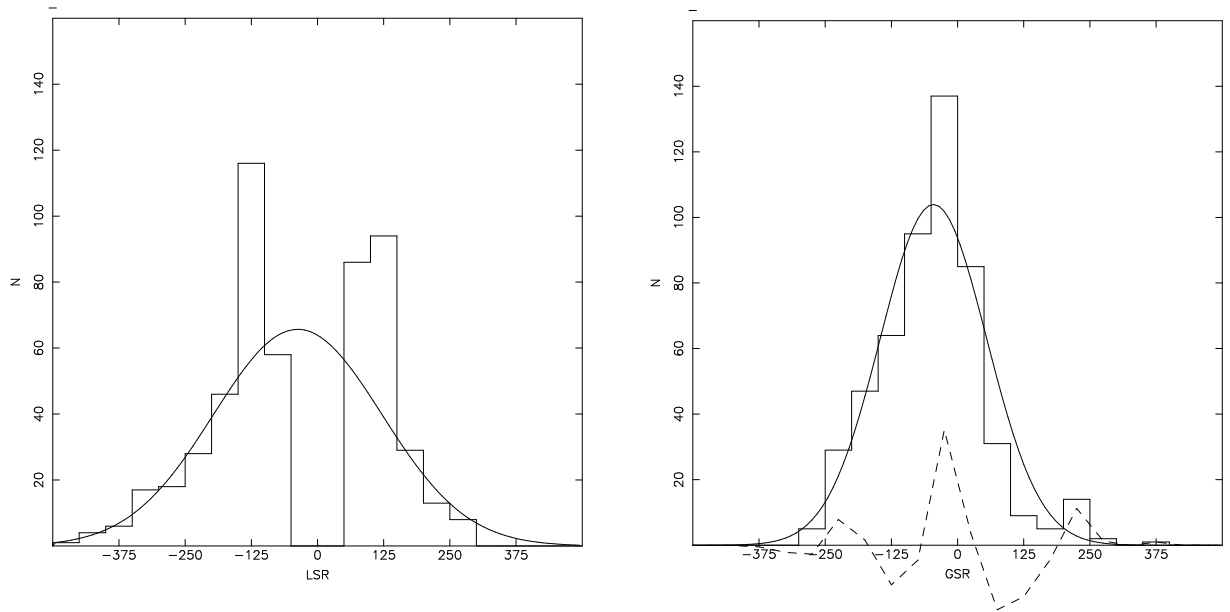

%\plottwo{hvclsr.ps}{hvcgsr.ps}
\hskip -1.7in
\centerline{\psfig{figure=hvclsr.ps,width=3.0in,angle=0,silent=1}}
\vskip -3.0in
\hskip 3.4in
{\psfig{figure=hvcgsr.ps,width=3.0in,angle=0,silent=1}}
\vskip -0.2in
\caption{ {\it Left:} Histogram of the distribution of HVC velocities
relative to the LSR.  HVCs which might in fact have $v_{\rm LSR}$ near
zero are not plotted, because such clouds would not be separable from
conventional--velocity Galactic emission. {\it Right:} Distribution of
HVC velocities relative to the GSR.  A Gaussian profile was fit to the
wings of both histograms: the $v_{\rm GSR}$ distribution of the HVCs is
more narrowly confined than the $v_{\rm LSR}$ one, suggesting, as
discussed in the text, that the GSR system is the more relevant
reference frame.  The data in both panels are from the WvW91 catalogue
of HVCs.}
\label{fig:hvclsr}
\end{figure}
If we now consider the system of HVCs in the same way, the change in
the velocity dispersion is even more dramatic;
the results are shown in Figure~\ref{fig:hvclsr}. The dispersion of the
complete HVC ensemble is 159 \kmse relative to the LSR; relative to the GSR,
it falls to 101 \kms.  The mean velocity in both cases is, however, 
significantly different from zero: $-37 \pm 7$ \kms, measured with respect to 
the LSR, and $-46 \pm 4$ \kmse measured with respect to the GSR.  These values 
suggest that the GSR inertial frame is more appropriate than the LSR 
frame, and that both the GSR and LSR frames are moving at constant 
velocity relative to the barycenter of the HVCs.  This result is at variance 
with a Galactic origin for the HVCs. If HVCs originated in the Galactic disk, 
then the velocity dispersion of the HVC ensemble would be lower in the LSR 
frame because the gas would conserve angular momentum.  If the HVCs originated 
in the Galactic center, the mean velocity of the ensemble would be zero.  
Evidently, the entire Milky Way is moving toward the barycenter of the HVCs.

If so, then the velocity centroid of the Local Group Barycenter
Clouds should reflect the motion of the Milky Way toward the
Local Group barycenter, which is given by \markcite{Einasto82}Einasto
\& Lynden-Bell (1982) as $-82$ \kmse relative to the LSR.  We define
the Barycenter Clouds as all of the HVCs within an ellipse centered on
$l= -143\deg$, $b = -23\deg$, with a major and minor axis of 60\dege
and 30\deg, respectively, and where the major axis is tilted by 30\dege
counterclockwise from \be = 0\deg. Some 96 of the clouds catalogued by
WvW91 fall within this ellipse.  Relative to the LSR, the mean velocity
of these clouds is $-173 \pm 10$ \kms; relative to the GSR, the mean
velocity is $-88 \pm 11$ \kms.  If we now subtract the motion of the
GSR relative to the Local--Group Standard of Rest (LGSR), the mean
velocity of the Local--Group Barycenter Clouds becomes $-28 \pm 10$
\kms, suggesting that the LGSR is a reference frame more appropriate to
the Local--Group Barycenter Clouds than either the GSR or the LSR
reference frames.

We note, however, that although the change of coordinate systems
produces a considerable lowering of the mean velocity of the
Local--Group Barycenter HVCs, the mean is still marginally negative.
Furthermore, the Antibarycenter Clouds, which are on the opposite
side of the Milky Way from M31, also have a negative mean velocity
relative to the GSR and to the LGSR of $-57 \pm 12$ \kmse and $-124\pm
12$ \kms, respectively.   It is reasonable to ask why reduction to the
LGSR does not produce a zero mean velocity if it is the
non--translating inertial reference frame, and, furthermore, why the
clouds that are approximately opposite on the sky to the Barycenter
Clouds are moving more rapidly toward the barycenter of the Local Group
than the Barycenter Clouds themselves.  We note in this regard that a
non--zero mean velocity of the HVCs relative to the barycenter would
occur if the entire cloud ensemble were either expanding or
contracting;  a negative mean velocity implies infall, as first
accounted for in the model of \markcite{Bajaja87}Bajaja et al. (1987).
Second, if the Milky Way is itself falling toward the barycenter of the
Local Group as postulated by \markcite{Kahn59}Kahn \& Woltjer (1959),
because the Milky Way lies between the Antibarycenter Clouds and the
barycenter, the Antibarycenter Clouds would be accelerated both toward
the Milky Way $and$ toward the barycenter of the Local Group, with the
sign and relative magnitude consistent with the observed mean radial
velocities.  In \S~\ref{sec:dynamics}, we discuss a model for the
dynamics of the HVCs in the Local Group that reproduces the observed
negative mean LGSR velocities in both the Barycenter and Antibarycenter
cloud groupings.  In this model, the HVCs are fragments remaining from
the (continuing) formation of the Local Group, falling towards its
center of gravity.

\subsubsection{Velocity Extrema}
\label{sec:extrema}

No cloud is catalogued in the WvW91 compilation with a positive
velocity higher than $v_{\rm LSR} = 297$ \kms, and no cloud with a
negative velocity lower than $-464$ \kms,  even though the useful
velocity range of the principal surveys contributing to that catalogue
is $-900$ to $+800$ \kms.  This limitation on the velocity extent of
the HVC ensemble suggests that the HVCs constitute a gravitationally
bound system of clouds rather than a collection of objects more or less
randomly distributed in extragalactic space, which would not show such
cutoff velocities.

One cannot use either the value of the velocity dispersion or the
values of the velocity extrema to differentiate between a Galactic and
an extragalactic origin, because both origins encompass the observed
values, but it is reasonable to ask what sort of mass would keep the
clouds bound in the Local Group hypothesis.  For a radius of 1.5 Mpc,
(see \S\ref{sec:dynamics} below) the measured one--dimensional velocity
dispersion of 106 \kmse gives a mass of $2.1 \times 10^{12}$ \msune for
a virialized ensemble of objects.  This mass is close to the mass of
$3.0 \times 10^{12}$ \msune obtained from the timing argument by
\markcite{Kahn59}Kahn \& Woltjer (1959).

\subsubsection{Angular--Size/Velocity Relation}
\label{sec:size_vel}

Imagine that all of the HVCs (again excluding the Magellanic Stream)
comprise a single class of extragalactic objects. If the clouds have a
uniform set of properties, such as a single power--law size
distribution independent of their location in the Local Group, then the
nearer clouds would have, on average, a larger angular extent than the
more distant ones. In that case, it should be possible to distinguish
kinematically the nearer clouds from the more distant ones. We note
first that most of the luminous mass in the Local Group is concentrated
in M31 and the Milky Way, and that these two galaxies are approaching
each other at a velocity of 123 \kms. Relative to the GSR, one would
then expect that the clouds closest to the Milky Way would have
velocities closest to 0 \kms.  If the HVCs are falling toward the Local
Group barycenter, then those clouds with the most negative velocities
with respect to the GSR would be the most distant; in the terms of the
present analysis, these would correspond to clouds on the far side of
M31, approaching both M31 $and$ the barycenter.  Increasing distance
would correspond roughly to increasingly negative velocity with respect
to the  GSR. We therefore would expect a significant correlation
between angular size and GSR velocity.

\begin{figure}[htpb] 
\centerline{\psfig{figure=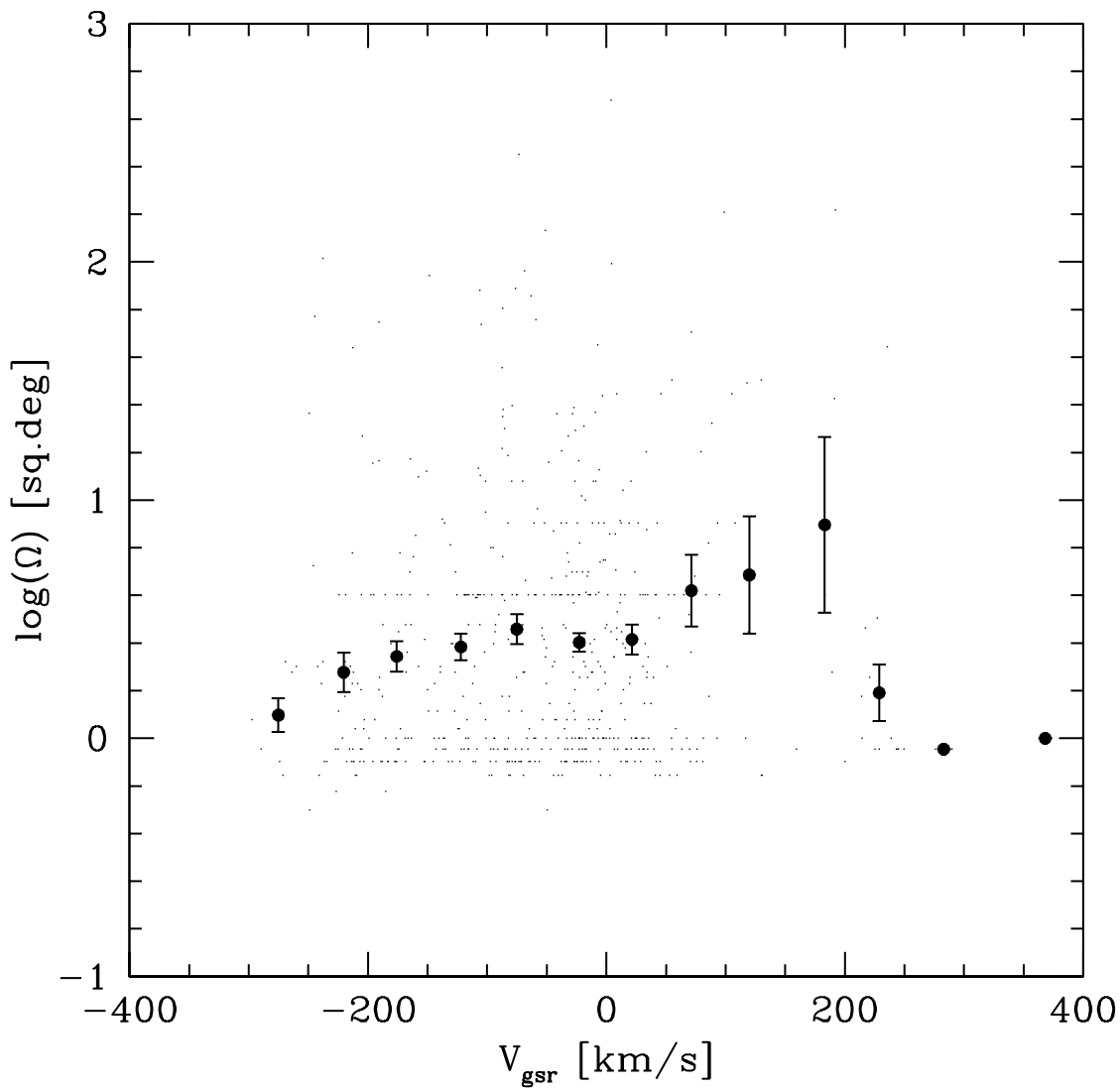,width=6in,angle=0,silent=1}}
\vskip -0.2in
%\plotone{sizevel.ps} 
\caption {Variation of angular size with respect to GSR velocity for
the HVCs in the WvW91 catalogue.  Individual clouds are shown as small
dots; the large dots represent the material averaged over bins 50 \kmse
wide. The error bars represent $\pm \sqrt{N-1}$ uncertainties; bins
without such bars represent only a single point. The apparent
quantification at low values of $\Omega$ reflects the finite sampling
of the data in WvW91. HVCs with very negative values of $v_{\rm GSR}$
tend to have smaller angular sizes.}
\label{fig:sizevel}
\end{figure}

The angular size, $\Omega$, tabulated by WvW91, plotted against
GSR velocity, with clouds collected in bins 50 \kmse wide,
is shown in Figure~\ref{fig:sizevel}. The error bars give the
statistical uncertainty of the mean in each bin. Except for the highest
positive velocities (which represent very few clouds in any case), there is
a good correlation between angular size and GSR velocity in just the sense
expected if the HVCs were associated with the Local Group. The slope of
the distribution depends in detail on the dynamics of the Local Group,
the distribution of HVC sizes, and the trajectories of the individual
clouds, as discussed in some additional detail below. It is difficult
to see how a Galactic origin for the HVCs could produce this
correlation; there is no such correlation in a plot of $\Omega$ versus
$v_{\rm LSR}$.

\section{Local Group Dynamics}
\label{sec:dynamics}

\subsection{Cosmological Background}
\label{sec:cosmobackground}

The continuing accretion of gas and dark matter onto galaxies and
groups is an inevitable prediction of all hierarchical models of the
formation of structure in the Universe.  According to these models,
gravitational fluctuations in the dark matter first collapse to form
small, bound objects. If the velocity dispersion in these
``mini--halos'' exceeds 10 \kms, then they are able to accumulate
baryons (\markcite{Ikeuchi86}Ikeuchi 1986; \markcite{Rees86}Rees 1986;
\markcite{Bond88}Bond, Szalay, \& Silk 1988; \markcite{Babul92}Babul \&
Rees 1992; \markcite{Miralda93}Miralda-Escude \& Rees 1993;
\markcite{Kepner97}Kepner, Babul, \& Spergel 1997).  These mini---halos
will collect into filaments; those nearby would then fall onto the
Local Group, resulting in an accretion shock at the edge of the Local
Group.

Outside this shock, the gas and dark matter in the mini--halos would
likely move together as the clouds fall onto the Milky Way.  It is not
clear whether the gas and dark matter will be separated at the
accretion shock or whether a cloud will survive ram--pressure stripping;
it is also not clear whether tidal forces will destroy the mini--halos.
The subsequent evolution of the clouds (once they are imbedded in a hot
intragroup medium) will depend on things such as the rate of 
evaporative heating (which
depends on unknowns such as the magnetic field structure). 
The fate of the dark matter will depend upon the mean
density in these small halos.

The location of most of the baryons in the local universe is not known,
but it is reasonable to suspect that they are associated with the
mini--halos which are collected into the filaments and sheets around
galaxy clusters.  Measurements of the deuterium abundance
(\markcite{Tytler96}Tytler, Fan, \& Burles 1996) suggest that $\Omega_{\rm b} h^2
\simeq 0.025$ at the epoch of nucleosynthesis.  Analyses of the
Lyman--$\alpha$ forest (\markcite{Rauch97}Rauch et al. 1997) infer a
lower limit on baryon abundance in cold gas of $\Omega_{\rm b} h^2 > 0.023$.
There is little cold gas seen in absorption at low redshifts, however,
and the local stellar density, $\Omega_*h^2$, is only 0.007
(\markcite{Fukugita97}Fukugita et al. 1997).  Numerical
simulations (\markcite{Cen95}Cen et al. 1995) suggest that much of the
gas is in filaments and sheets, where it has not yet been detected.

While most of the gas in the filaments would be ionized, some of it may
be neutral. The Lyman--$\alpha$ forest and the Lyman--limit systems
trace the distribution of this neutral gas on cosmological scales
(\markcite{McGill90}McGill 1990; \markcite{Cen94}Cen et al. 1994;
\markcite{Zhang95}Zhang, Annios, \& Norman
1995;\markcite{Hernquist96}Hernquist et al. 1996;
\markcite{Miralda96}Miralda-Escude et al. 1996; \markcite{Bi97}Bi \&
Davidsen 1997; \markcite{Bond97}Bond \& Wadsley 1997).  We suggest in
this paper that Galactic astronomers have been observing the same type
of gas clouds and identifying them as HVCs.

The Local Group is probably a dynamically young system, with the Milky
Way and M31 approaching each other for the first time. Hierarchical
models suggest that there is continuing accretion of gas onto the Local
Group though filaments.  While most of this gas is likely to be
ionized, there would be neutral gas that is able to cool when higher
densities are reached in dark--matter mini--halos.  In such a
situation, there would be small, neutral, gas clouds with velocity
dispersions of $\sim$ 10 \kmse embedded in the large, coherent,
velocity fields of the filaments. These filaments would contain 
mass comparable to the total mass in the Local Group.

The dynamics of the Local Group is probably rather simple: more than
98\% of the mass of the Local Group is contributed by M31 and its
satellites and by the Milky Way and its satellites
(\markcite{Raychaudhury89}Raychaudhury \& Lynden-Bell 1989, hereafter
RL). Thus the dynamics of the Local Group can be approximated as a
two--body problem, with M31 and the Milky Way approaching each other on
nearly radial orbits (\markcite{Einasto82}Einasto \& Lynden-Bell
1982).

The gravitational effects of neighboring galaxies complicates this
simple two-body interaction. The neighboring galaxies exert a net force
on the Local--Group barycenter that produces a bulk flow of the entire
Local Group and, in addition, exert a tidal force that compresses and
shears the Local Group. While the bulk flow is probably primarily due
to distant mass concentrations such as the Great Attractor and the Coma
Cluster, the tidal force is primarily due to neighboring galaxies. RL
estimate the tidal field from samples of nearby galaxies. The effect of
the Local Group neighbors is to compress the flow of material in one
dimension and to shear it along an axis that is not far from the line
joining M31 and the Milky Way.

\subsection{Simulating the Dynamics of the HVCs}
\label{sec:simulation}

We have done a simple simulation of the formation and evolution of the
Local Group. The dynamics are approximated as a modified restricted
three--body problem. The test particles in the simulation are subject
to the gravitational forces of M31 and the Milky Way, and to the
external tidal field of the neighboring galaxies. We use the model of
RL to simulate the external tidal field and its temporal evolution (see
their Table 5). For these parameters, M31 today has a tangential
velocity of 38 \kmse and the typical galaxy is assumed to have a
mass--to--light ratio of 66. (We also ran a simulation for a
mass--to--light ratio of 80, but the results do not differ
significantly from those discussed below.)

\begin{figure}[htpb] 
\centerline{\psfig{figure=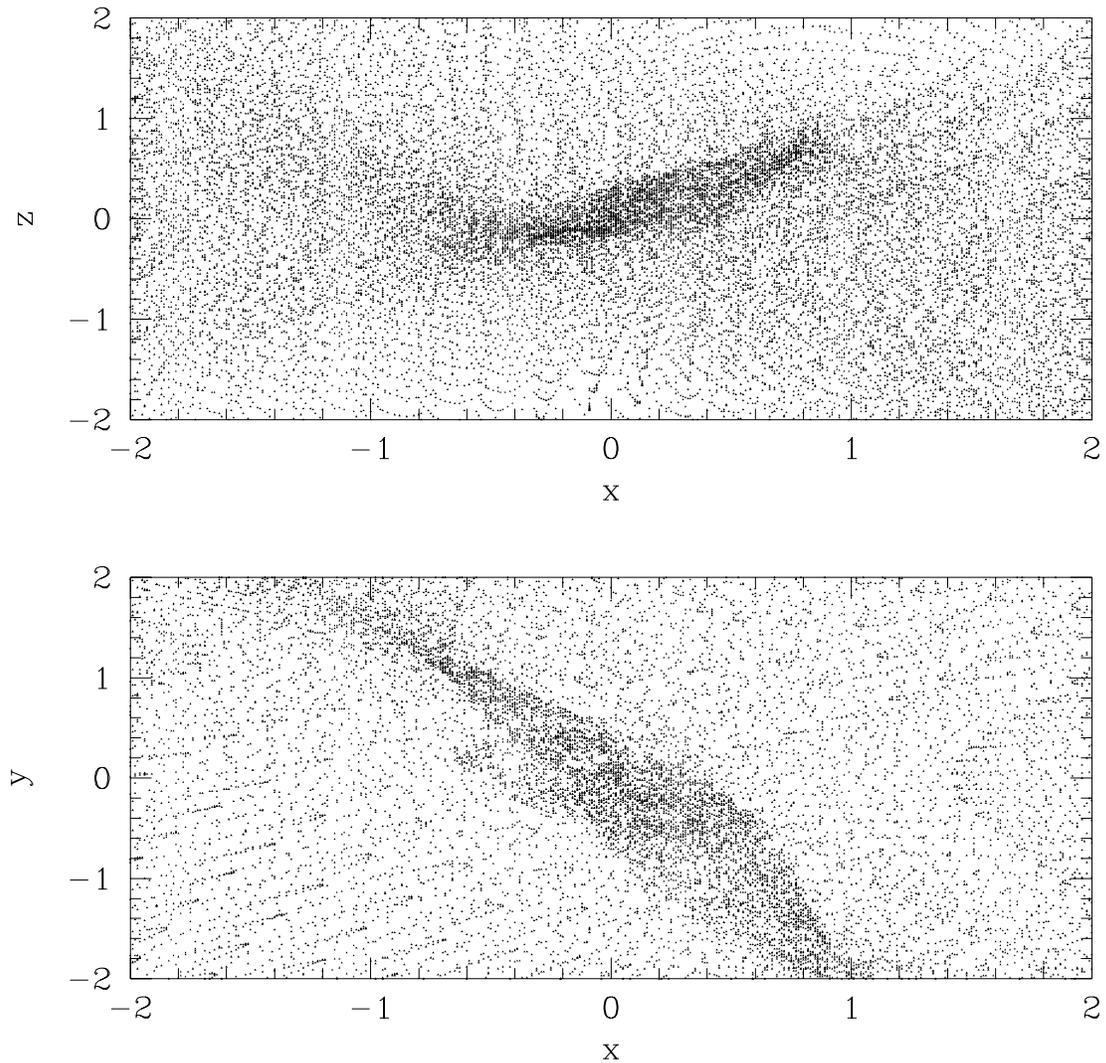,height=6in,angle=0,silent=1}}
\vskip -0.2in
\caption{Locations of the simulated HVC particles at the termination of
the simulation described in the text.  The upper panel shows the
$(x,z)$ projection; the lower panel shows the $(x,y)$ one.  Distances
are given in Mpc.  The Milky Way is located at (0.23, --0.38, 0.18); M31,
at (--0.12, 0.19, 0.09).  Any test particle that fell inwards towards the
two galaxies and passed within 100 co--moving kpc of their centers was
assumed to be accreted onto the galaxies and was excluded from this
plot and the subsequent discussion.} 
\label{fig:filament} 
\end{figure}

The simulation begins when the physical distance between M31 and the
Milky Way is 0.1 Mpc. We represent M31 and the Milky Way as ``sticky
particles", assigning one third of the Local Group mass to the Milky
Way and the remaining two thirds to M31. Any test particle that is
falling towards the two galaxies and passes within 100 co--moving kpc
of their centers is assumed to be accreted onto the galaxies.

\begin{figure}[htpb] 
\centerline{\psfig{figure=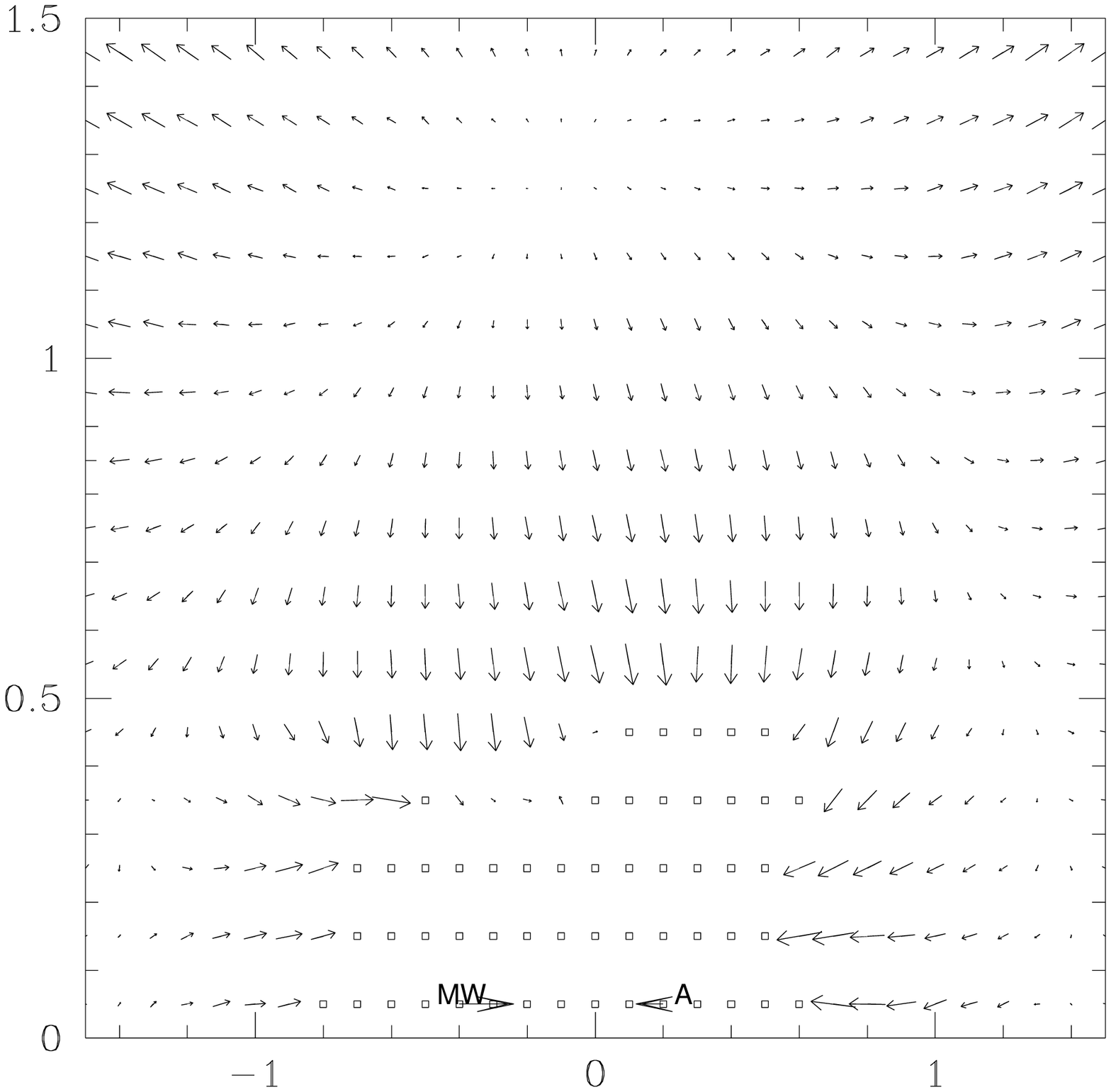,height=6in,angle=0,silent=1}} 
\vskip -0.2in
\caption{Simulated velocity flow in the Local--Group rest frame.  The
Milky Way and M31 (marked by MW and A, respectively) are moving towards
each other with relative velocity 124 \kms.  Their gravitational pull,
together with the external tidal field, produces flow towards a
filament (see the vectors above the plane).  Material in the filament
(and near the Local Group) flows along the filament as indicated;
material beyond the Local--Group accretion radius partakes in the
general Hubble flow.  The squares show regions where the gas random
velocity (computed with a 100 kpc smoothing length) exceeds (100
\kms)$^2$: this gas is likely shock heated. The distance at which the
Hubble flow is reversed is about 1.5 Mpc.}
\label{fig:arrowc}
\end{figure}

The net effect of the external tidal field and the gravitational pulls
of M31 and the Milky Way is to compress most of the test particles into
a filament. This can be seen in Figure~\ref{fig:filament}, which shows
the ($x,y$) and ($x,z$) projections of the test--particle density in the
simulation, and in Figure~\ref{fig:arrowc}, which shows the velocity
field.  We identify the test particles remaining at the end of the
simulation with the HVCs. Today, $\sim 25$\% of the mass in the
simulation would be bound to the Milky Way, and $\sim 50$\% to M31; the
remaining mass would still remain in the filament.  If the known HVCs
have distances similar to those implied by the simulation, the total
mass of the remaining neutral hydrogen is $\sim 10^{11}$ \msun.

The simulated velocity field shown in Figure~\ref{fig:arrowc}
suggests that the gas flow 
along the filament is smooth and likely cold to the ``left'' of the Milky Way 
and to the ``right'' of M31. In the region between M31 and the Galaxy, the 
streamlines cross; such a situation would likely produce shocks, suggesting 
that M31 and the Milky Way could be imbedded in a common halo of hot gas. In 
the simulation, we bin the particles into $0.1 \times 0.1 \times 0.1$ Mpc$^3$ 
cubes and compute density, velocity, and velocity moments in each cube. 
Whenever $<v^2>$ in a cube exceeds ($100$ \kms)$^2$, the gas in the cube is 
assumed to be hot.

\begin{figure}[htpb]
\vskip -1.1in
\centerline{\psfig{figure=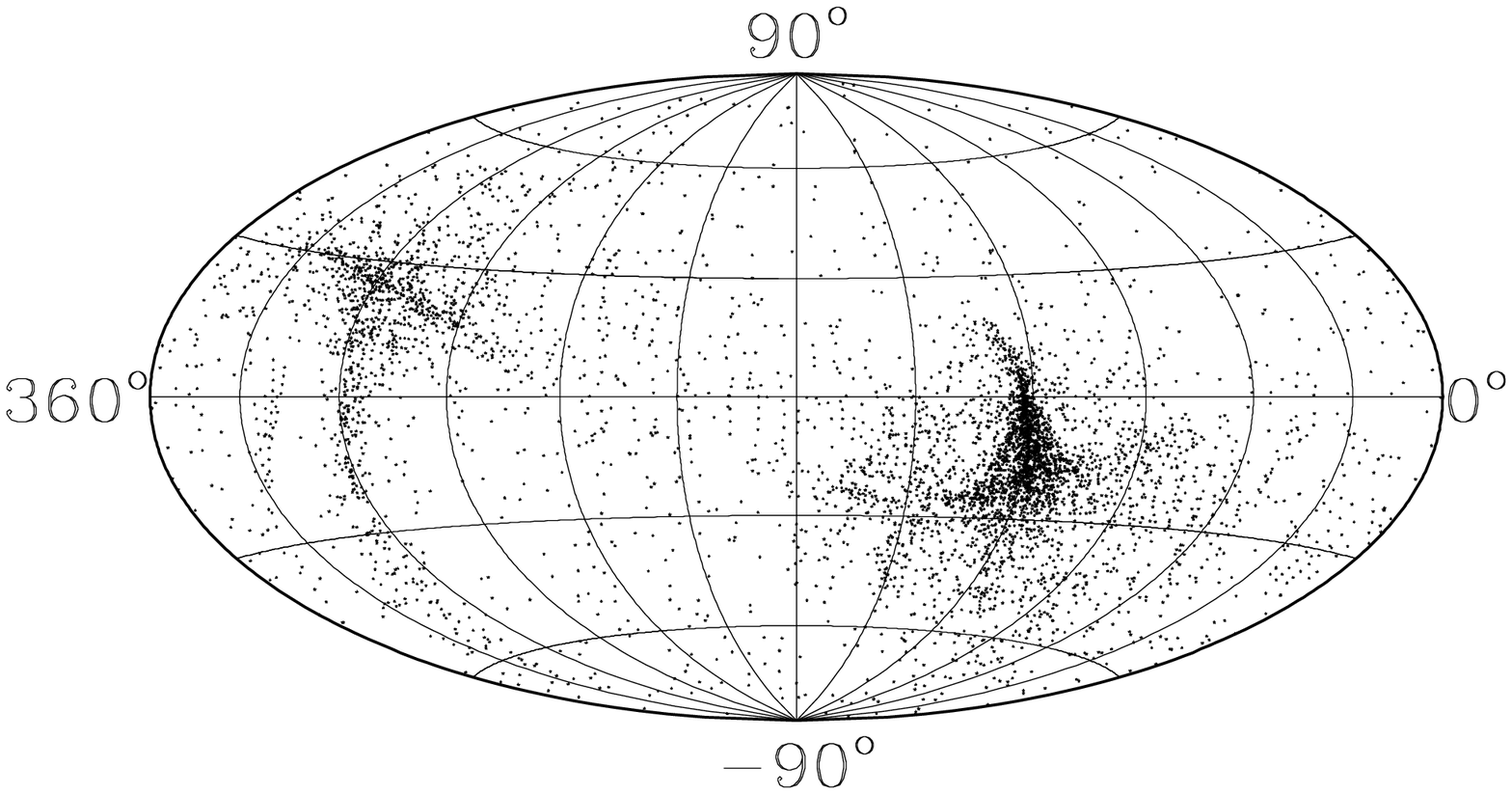,height=4in,angle=0,silent=1}}
\vskip -1.2in
\centerline{\psfig{figure=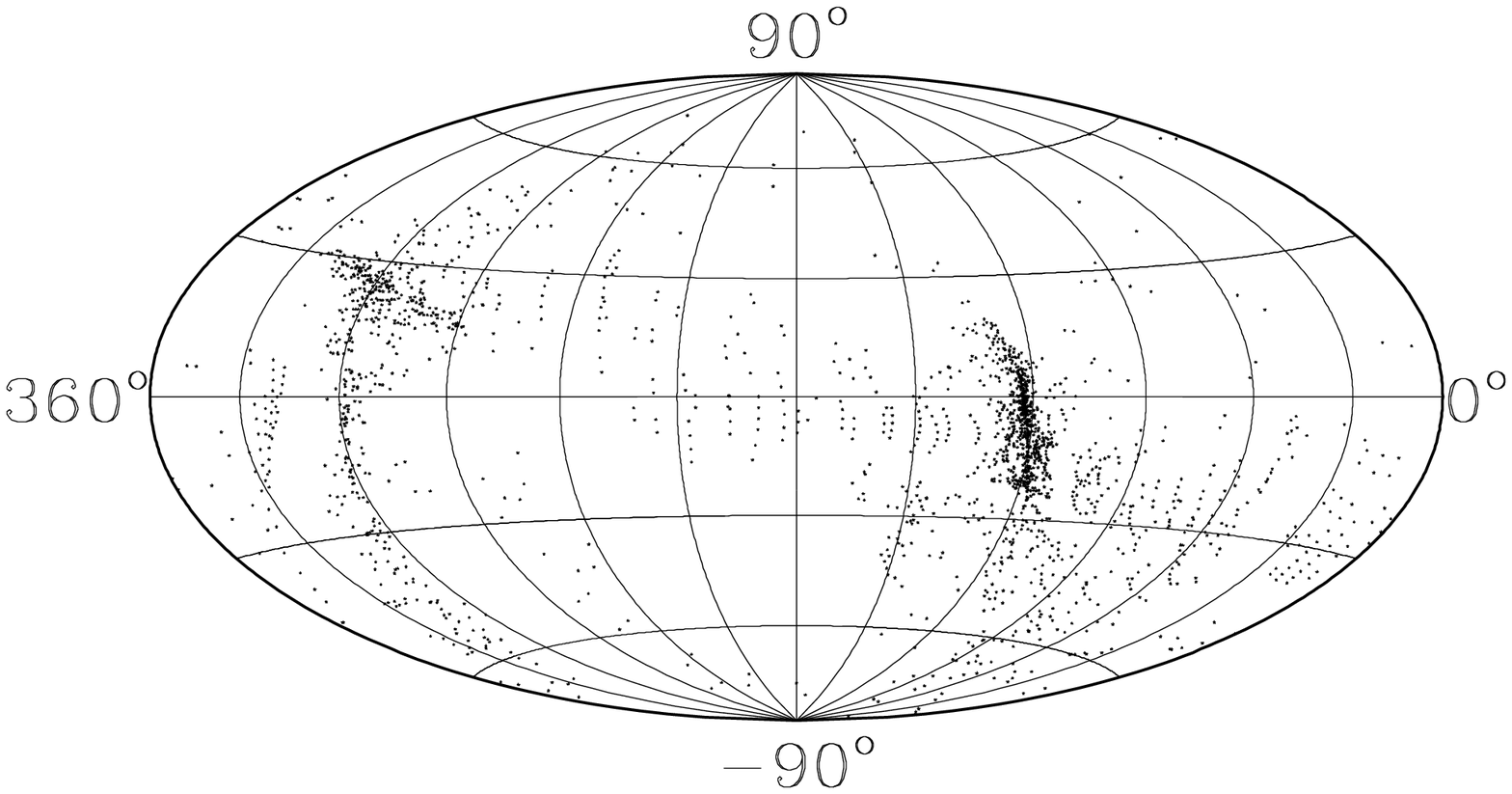,height=4in,angle=0,silent=1}}
\vskip -0.2in
\caption{Distributions on the sky of the simulated HVC particles,
calculated for two values of the velocity dispersion within the 100--kpc
smoothing box. {\it Upper:}  Projection showing all of the test
particles within 2 Mpc of the Milky Way, regardless of the velocity
dispersion. {\it Lower:} Same projection as in the upper panel but here
showing only those particles located in regions in which the gas random
velocity within the 100--kpc smoothing box is less than (100
\kms)$^2$.  The overall distribution on the sky is similar in both
cases.}
\label{fig:LGall}
\end{figure}

Figure~\ref{fig:LGall} shows that the tidal and gravitational fields
have compressed most of the gas into a large filament that stretches
from $l= 120\deg$, $b = -10\deg$ to $l = 300\deg$, $b = 10\deg$. This
orientation is due to two effects: the M31/Milky Way axis lies along a
line through $l = 122\deg$, $b=-21\deg$ and the tidal stretching by the
external galaxies lies along an axis pointing toward $l = 143\deg$, $b
= -23\deg$.  The lower panel in
Figure~\ref{fig:LGall} represents only the  cold gas.  The morphology
of the simulated particles does not depend sensitively on the removal
of the hot gas.

\subsection{Comparison with the HVC Observations}

\label{sec:obscomp}

\begin{figure}[htpb]
\vskip -1.5in
\centerline{\psfig{figure=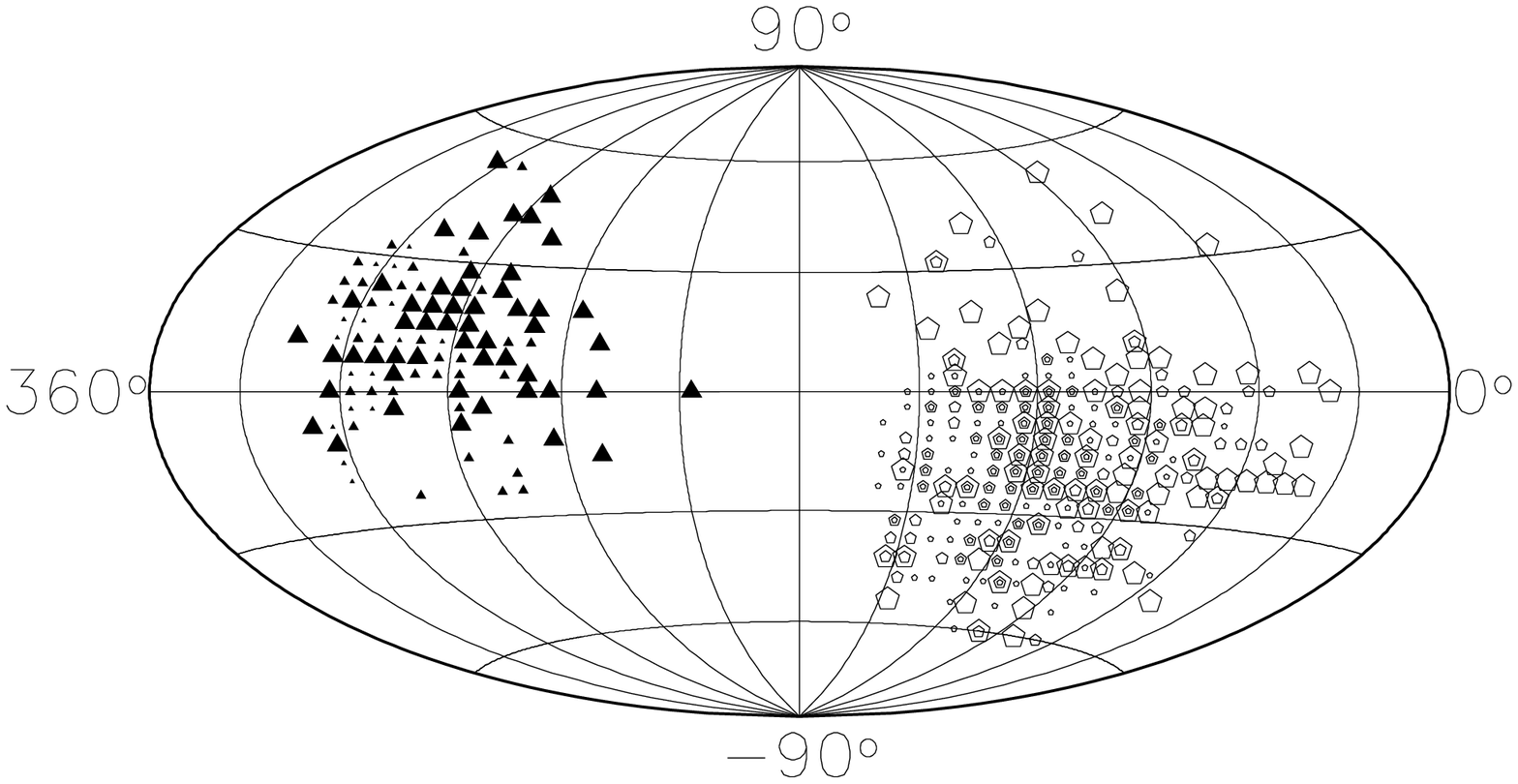,width=4in,angle=0,silent=1}}
\vskip -1.5in
\hskip 1.5in
{\psfig{figure=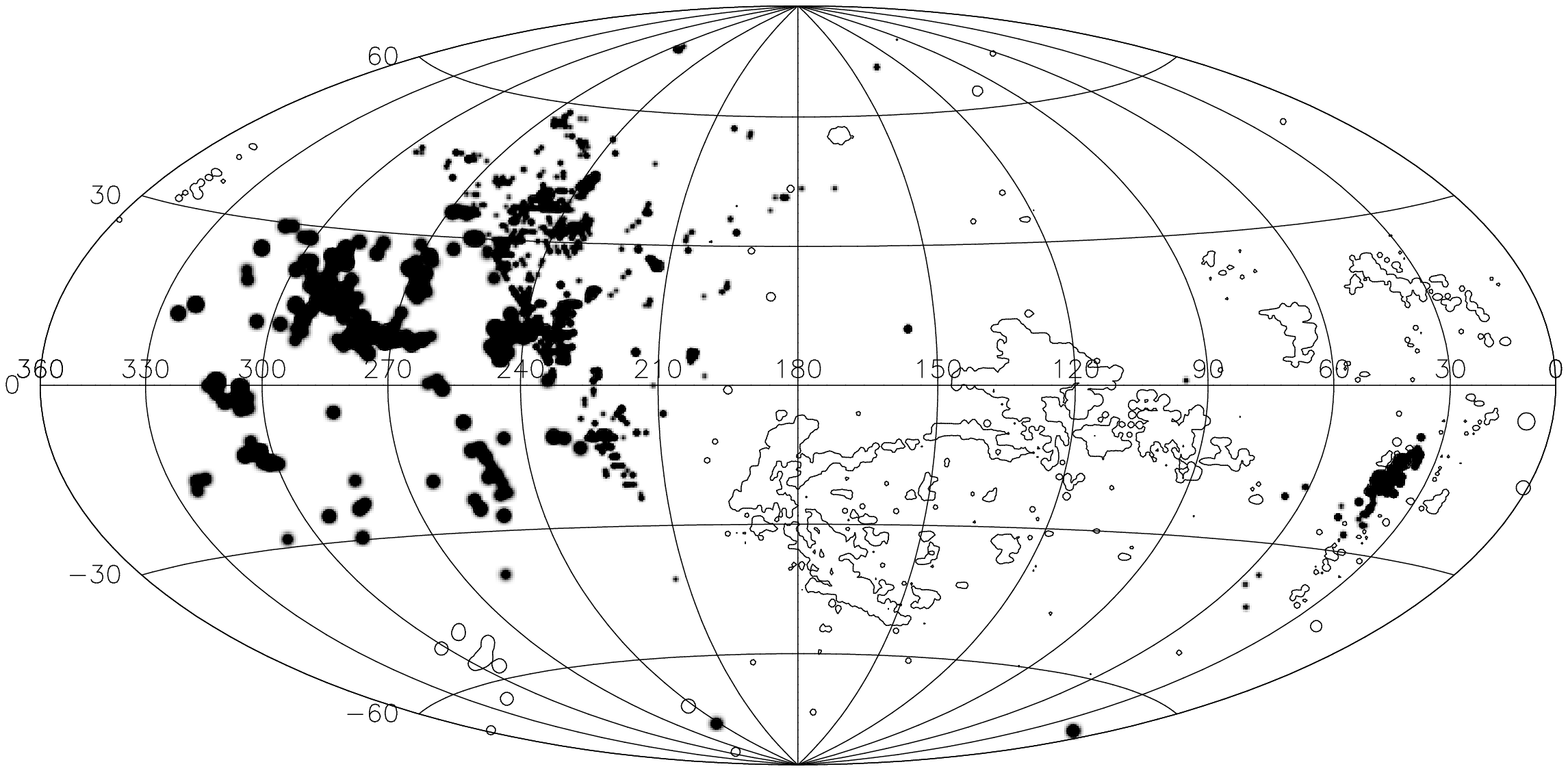,width=4in,angle=-90,silent=1}}
\vskip -0.2in
\caption{Comparison of simulated and observed sky-- and kinematic
distributions of the HVC ensemble. {\it Upper:} Distribution of all
simulated clouds having HI column densities greater than 3$\times
10^{18}$ cm$^{-2}$ and $|v_{\rm LSR}|$ greater than 200 \kms,
regardless of the dispersion within the smoothing box. The small,
medium, and large symbols denote, respectively, simulated clouds with
column densities between $3\times10^{18}$ and $1\times 10^{19}$,
between $1\times 10^{19}$ and $3 \times 10^{19}$, and greater than $3
\times 10^{19}$ cm$^{-2}$.  Strictly speaking, these simulated column
densities are total ones, i.e. including the dark--matter content.  The
triangles represent clouds with negative velocities;  the stars, clouds
with positive velocities.  This figure represents the distribution of
HVCs if the clouds have not been destroyed by passage through a hot
intergalactic medium and if collisions between HVCs are rare.  {\it
Lower:}  Distribution of observed HVCs, as in Figure~\ref{fig:aitoff}
in Aitoff projection, but excluding the Magellanic Stream and the
Northern Hemisphere Complexes A, C, and M, which are evidently
relatively nearby and thus unrepresentative of the angular size of
individual clouds in the Local--Group ensemble. Positive velocities are
denoted by filled contours, negative velocities by open contours. The
simulated spatial and kinematic distributions resemble, in essence, the
observed distributions.  The lower panel was kindly provided by Bart
Wakker.} 
\label{fig:lb} 
\end{figure}

We estimate the column density associated with the simulated Local--Group 
HVCs by binning the particles in $l,b,v$ space and then
integrating the total column at each grid point. We assume that
$\Omega_{\rm b}/\Omega_{\rm tot} = 0.1 $ and that all of the overdensity in the
Local Group region is already bound to M31 and to the Milky Way.
By restricting our analysis to gas that has not
yet  passed through the Local Group accretion shock, we exclude any
nearby gas. It is not clear that this is a valid restriction, as some
of the gas clouds may in fact survive passage through this shock.
However, the similarity in the spatial distributions shown in the two
panels of Figure~\ref{fig:LGall}\ suggests that inclusion or removal
of the hot gas should not strongly affect the comparison of the
simulation with the observations. Accordingly, the upper panel in
Figure~\ref{fig:lb}\ shows $all$ of the regions with the same velocity
and column density criteria, and thus represents the HVCs in the
simulation. The larger symbols correspond to regions of higher column
density in the simulation.

We now compare the simulated results shown in the upper panel of 
Figure~\ref{fig:lb}\ with the observations.  We first remove from the
observations the HVCs constituting the Magellanic Stream as well as the
grouping of Northern Hemisphere Clouds and plot the results in the
lower panel of Figure~\ref{fig:lb}; positive and negative LSR
velocities are indicated separately.  Direct distance measurements
(\markcite{Danly93}Danly et al. 1993; \markcite{van Woerden97}van Woerden
et al. 1997) indicate that the Northern Hemisphere Clouds are relatively
nearby and thus would distort the overall statistical comparison.
The agreement with the observations of the simulated spatial
distribution, velocity separation, and direction of the HVC velocities
is rather good, considering the simplicity of the model.

\begin{figure}[htpb]
\hbox{
%\hspace{-0.9in}
\psfig{figure=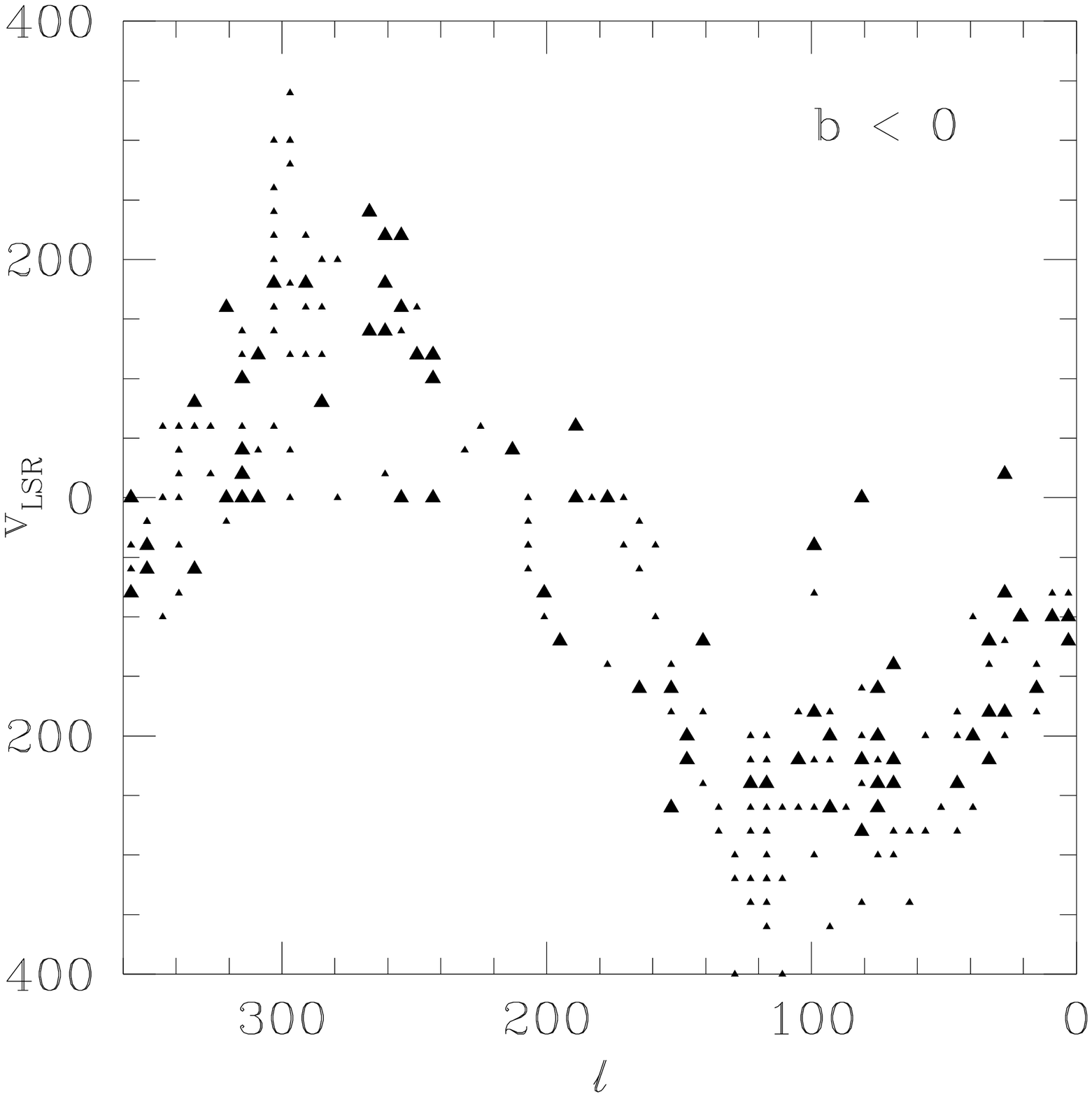,height=3in,angle=0,silent=1}
%\hspace{-1.1in}
\psfig{figure=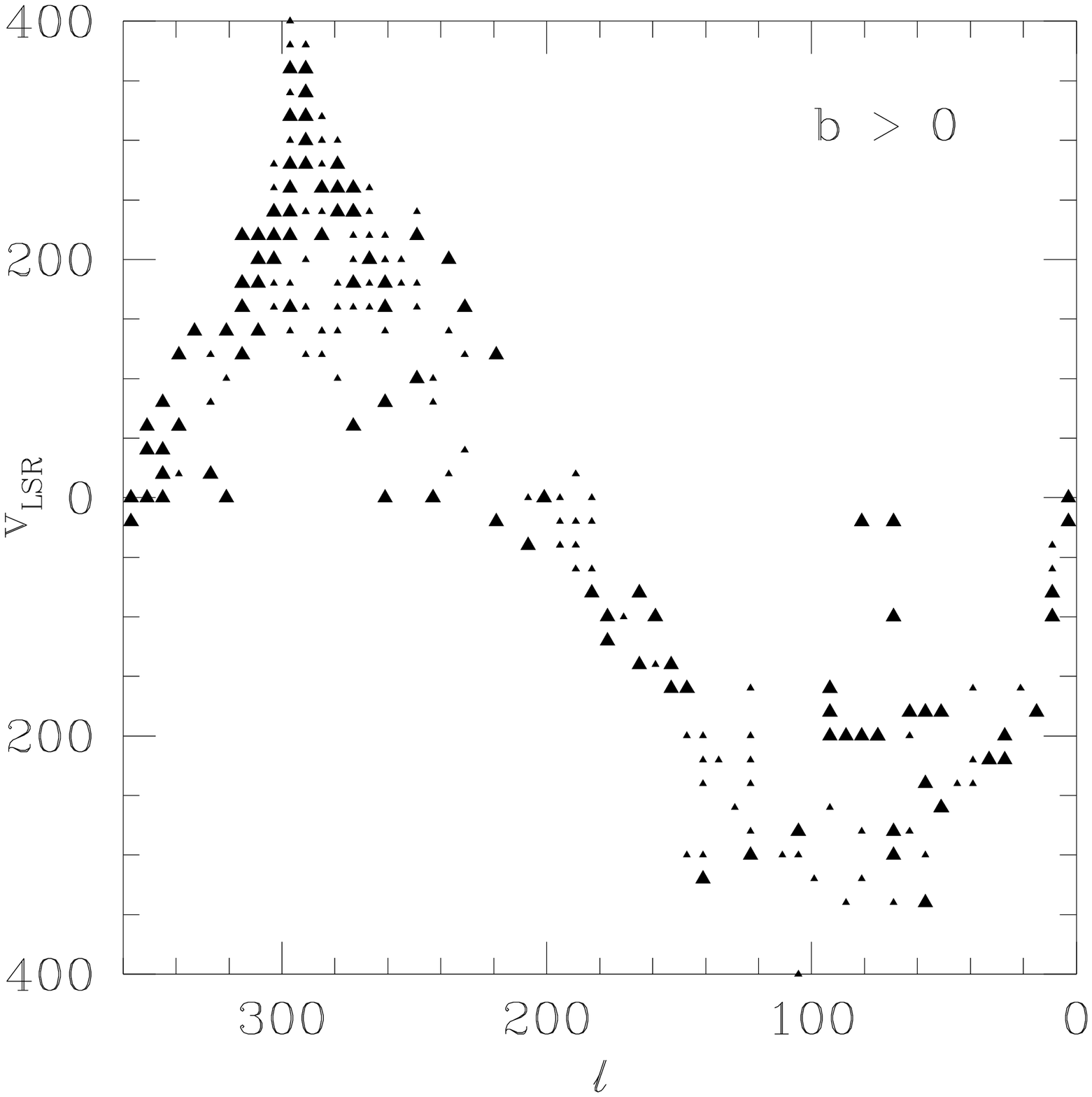,height=3in,angle=0,silent=1}}
\vbox{\vskip -0.1in}
\centerline{\psfig{figure=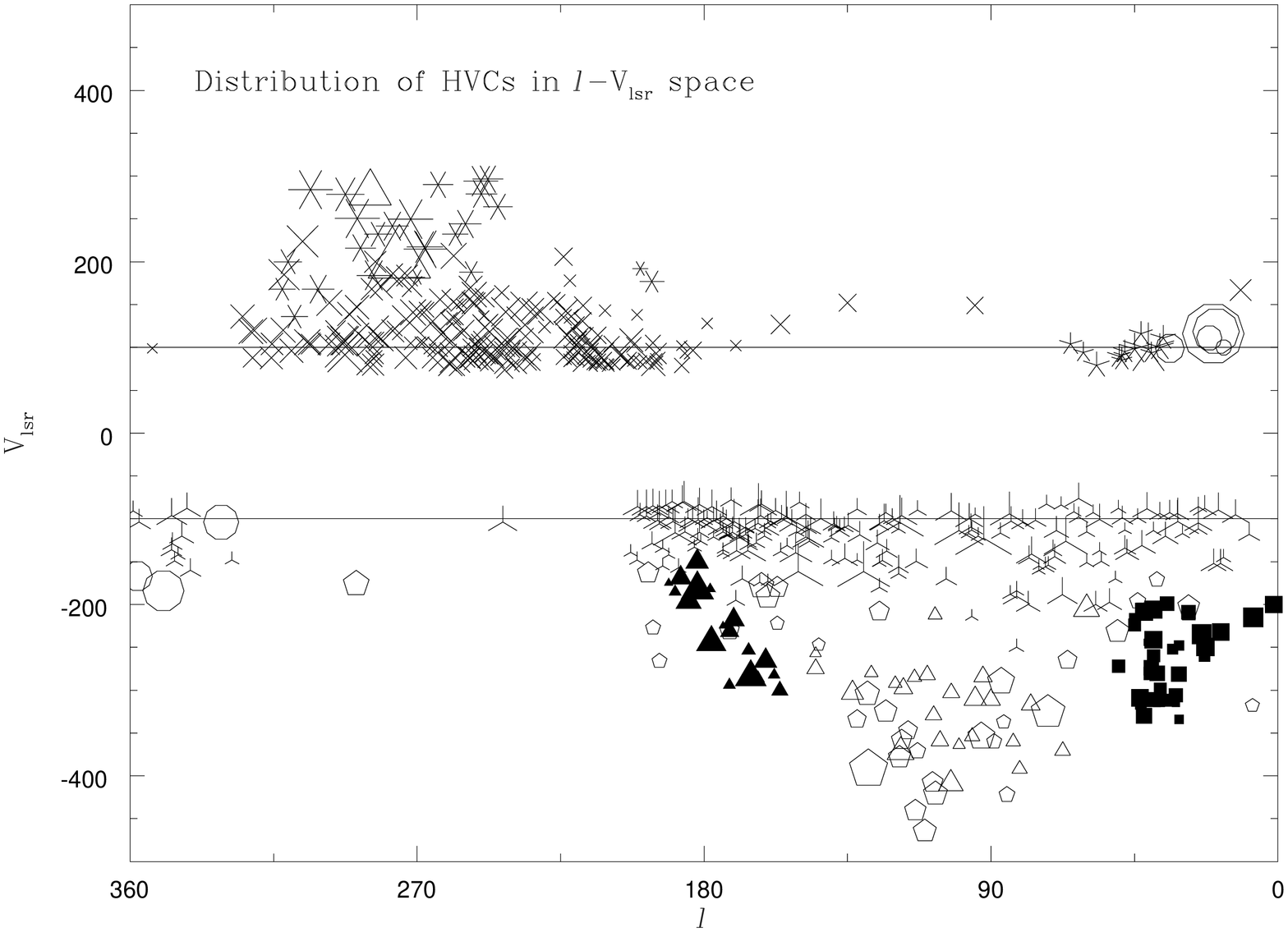,height=3in,angle=-90,silent=1}}
\vskip -0.2in
\caption{Comparison of the simulated longitude--velocity distribution
of the HVCs with the observed situation.  Radial velocities are
relative to the LSR.  {\it Upper:} Simulated kinematic distribution of
clouds with HI column densities greater than $3\times 10^{18}$
cm$^{-2}$, plotted separately for $b < 0 \deg$ and for  $b > 0 \deg$.  {\it
Lower:} Longitude--velocity diagram of the observed HVC ensemble, as
compiled by WvW91.  The symbols are proportional in size to the flux
from the individual clouds, and are keyed to the individual complexes
defined by Wakker (1991).  Clouds with LSR velocities $|v_{\rm LSR}| <
80$ \kmse are not considered here as HVCs, regardless of their
location.  The general features observed are accounted for by the
simulation. }
\label{fig:lvwakker}
\end{figure}

We next compare the observed and simulated longitude--velocity
distributions  of the HVCs in the LSR frame.  Figure
~\ref{fig:lvwakker} shows several features well represented in the
simulation.  The envelope of the observed velocities is approximately
sinusoidal, but is displaced from $v_{\rm LSR}$ = 0 \kmse by about
$-100$ \kms.  Both the functional form of the envelope and the
displacement from zero (due in part to the motion of the LSR toward the
barycenter of the Local Group) are reproduced in the simulation.
Furthermore, the amplitude of the envelope is reproduced.  Certain
details, such as the negative--velocity HVC gas between $180\deg > l >
205\deg$ and at $l > 340\deg$ are also accounted for by the simulation.

The Local--Group hypothesis thus is able to account for important
aspects of the spatial and of the kinematic distributions of HVCs.  The
model is simple, and rather insensitive to the tunable parameters.
It suggests identifying HVCs as the structures from which the Milky Way
and M31 have been built, and which fuel the continuing star formation in
both galaxies. It further suggests that the HVCs are among the first
structures to form in the Local Group, and that these clouds can be 
expected to be associated with copious dark matter.

\subsection{Lyman--$\alpha$ Clouds}
\label{sec:darkmatter}

The simulation suggests that the Local Group is similar to other galaxy
groups which have been numerically modeled. In numerical simulations of
the formation of large--scale structure, most galaxies and groups are
in filaments (e.g. \markcite{Hernquist96}Hernquist et al. 1996;
\markcite{Bond95}Bond, Kofman, \& Pososyan 1995).  Within these filaments,
hot gas is
associated with individual groups.  Outside the groups, the gas is
primarily cold. This cold filament gas seems to have properties similar
to those of the Lyman--$\alpha$ forest and Lyman--limit lines observed
in absorption toward distant quasars (see e.g.
\markcite{Hernquist96}Hernquist et al. 1996 and \markcite{Katz96}Katz
et al. 1996).

\begin{figure}[htpb]
\centerline{\psfig{figure=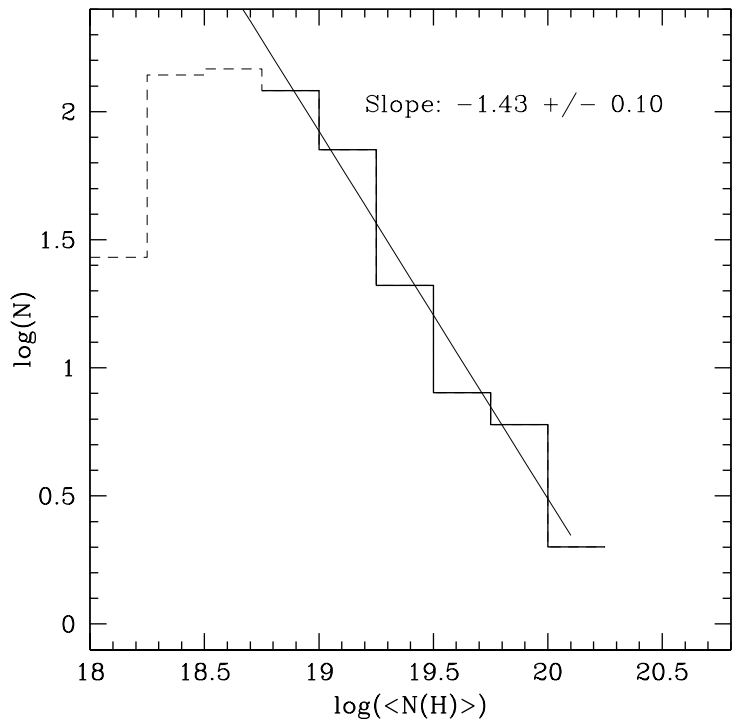,width=4in,angle=0,silent=1}}
\vskip -0.2in
\caption{Histogram of the mean column--density distribution of the HVCs from 
the WvW91 compilation.  The straight line fits the data at the higher column 
densities, $N_{\rm HI} > 3 \times 10^{18}$ \c2, for which the sample is likely 
complete.}
\label{fig:columndens}
\end{figure} 

The mini--halos may account for the core--halo structure seen in some
of the HVCs (\markcite{Giovanelli73}Giovanelli, 
\markcite{Verschuur73}Verschuur, \& Cram 1973;
\markcite{Cram76}Cram \& Giovanelli 1976; 
\markcite{Giovanelli77}Giovanelli \& Haynes 1977;
\markcite{WakkerS91}Wakker \& Schwarz 1991; \markcite{Wolfire95}Wolfire
et al. 1995). When the ionization rate is not high enough to balance
cooling, there is an instability that enables cold gas to accumulate in
the center of the halo (\markcite{Murakami90}Murakami \& Ikeuchi 1990;
\markcite{Kepner97}Kepner et al. 1997). Over a wide range
of parameters, this cold gas is primarily atomic rather than molecular.
In this picture, the HVCs are gravitationally bound by the dark matter
rather than pressure confined (see e.g. \markcite{Wolfire95}Wolfire
et al. 1995).  Because this instability may lead to some star formation
(\markcite{Murakami90}Murakami \& Ikeuchi 1990), it would be intriguing
to see if there are ultra--low--surface--brightness galaxies associated
with some of the HVCs.

If the formation of the Local Group is typical of the formation of
small groups, and if HVCs are indeed the leftover building blocks of
local galaxy formation and evolution, similar objects would be expected
in other galaxy groups.  We speculate that the Ly--$\alpha$ clouds may
be such systems and discuss them below; we discuss in
\S\ref{sec:exgalhvcs} HVCs in emission toward
other galaxies. The HVCs correspond to Lyman--limit systems in the
column density range $2 \times 10^{17}$ \c2 $<N_{\rm HI} < 2 \times
10^{20}$ \c2.  As can be seen from Figure~\ref{fig:columndens}, the
HVCs span the middle of this range. Lower column densities correspond
to clouds in the Ly--$\alpha$ forest; higher column densities
correspond to damped Ly--$\alpha$ clouds.

\markcite{Wolfe93}Wolfe (1993) has plotted the frequency distribution
of the full range of column densities detected in the Ly--$\alpha$
absorbers, showing that this distribution follows a power law with an
index of $-1.25$ over the entire range of column densities and with an
index of $-1.67$ for log \thinspace $N_{\rm HI} > 20$ \c2.  That the
power--law index of $-1.4 \pm 0.1$ shown in 
Figure~\ref{fig:columndens} is consistent with the index characterizing the
Ly--$\alpha$ absorbers suggests that the HVCs may be manifestations of
the same phenomenon.  From a purely theoretical standpoint, the gas
column densities associated with the dark matter mini--halos are
expected to scale as $N_{\rm HI}^{-5/3}$ (\markcite{Rees88}Rees 1988;
\markcite{Milgrom88}Milgrom 1988), close to the value of the slope,
--1.4, seen
in Figure~\ref{fig:columndens}. Thus, one of the expectations of our
hypothesis is the existence of additional HVCs at lower column
densities along the M31/Milky Way axis; these clouds may, however, be
largely ionized.

We may roughly estimate the probability of detecting an HVC as a
Ly--$\alpha$ absorber for an observer located external to the Local
Group.  The covering fraction of HVCs seen against a background quasar
depends on column density and can be estimated from various
observations.  If the WvW91 catalogue is incomplete by a factor of 2 down
to $N_{\rm HI} = 1 \times 10^{18}$ \c2, then using the mean cloud
properties within a Hubble--flow turnaround radius of 1.5 Mpc gives a
covering fraction of 0.1, which is a value similar to that obtained by
WvW91 if the Outer Arm Complex, Magellanic Stream, and Complex C are
not included.  \markcite{Bowen95}Bowen, Blades, \& Pettini (1995)
estimate a covering fraction on the sky of 0.14 for $N_{\rm HI} > 1
\times 10^{17}$ \c2 from HST GHRS spectra, a fraction which would be
approximately doubled for an external observer.
\markcite{Murphy95}Murphy, Lockman, \& Savage (1995) obtain a higher
value of the covering fraction,  0.37, for 
$N_{\rm HI} > 7 \times 10^{17}$ \c2, but make no
correction for the large clouds, which, to be consistent with the Bowen
et al. estimate, would reduce the covering fraction to about 0.09, or
0.18 for an external observer, to the Murphy et al. sensitivity limit.

If HVCs are not randomly distributed within the Local Group, but show
some concentration toward M31 and the Milky Way, then, extending the
considerations to very distant systems, the probability of detection
would be increased for a quasar within one or two hundred kpc of a host
galaxy.  One therefore expects a probability of about 0.3 for $N_{\rm
HI}> 1 \times 10^{17}$ \c2 at an impact parameter of 1.5 Mpc, and
higher values at smaller impact parameters, depending on the degree of
concentration at zero redshift. This probability might increase further
for lower column densities, especially if the frequency distribution of
low--column--density HVCs is similar to that of the Ly--$\alpha$
absorbers summarized by \markcite{Wolfe93}Wolfe (1993).  Several
ionized clouds without associated HI have, in fact, been detected in
optical absorption lines toward BL Lac, as well as toward four
extragalactic supernovae at velocities between $-260$ and $+263$ \kmse
\markcite{WvW97}(WvW97).  One would expect a higher detection
probability at higher redshifts, if the evolution of the HVC system is
at all well described by our simulation (see \S\ref{sec:simulation}).
Hoffman et al. (1998) present Ly--$\alpha$ absorption data towards the quasar
3C~273, and show that the lowest--redshift absorber is located only $\sim$200
kpc distant from the galaxy MCG+00-32-16, and at a velocity separation of only
94 \kms.  They interpret this feature as due to a ``failed dwarf" member of a
poor galaxy group, i.e. as an HI cloud which has not formed stars; the
properties of this cloud are compatible with those suggested here for
HVCs.

\section{Discussion}
\label{sec:discussion}

\subsection{Distances} 
\label{sec:distances}
\subsubsection{Absorption--Line Distances}
\label{sec:absdist}

The most obvious direct test of whether the HVCs are Galactic or
extragalactic has involved searching for optical or UV absorption lines
at the velocity of the HI emission toward sources at known distances.
All but two of the attempts to find optical absorption lines toward
stars in the halo of the Milky Way have returned negative results.
\markcite{Danly93}Danly et al. (1993) and \markcite{vanWoerden98}van
Woerden et al. (1998) obtained distances toward Complexes M and A,
respectively, of $1.7 < d < 5$ kpc and $4 < d < 10$ kpc.  These
complexes have distances within the range expected for the Northern
Hemisphere Clouds from the tidal considerations discussed in
\S\ref{sec:bighvcs}. In the Local--Group hypothesis, such distances are
not representative of the HVC ensemble as a whole because the two
complexes are likely to be the nearest HVCs.

The HVC simulation and the cloud kinematics suggests
that the mean HVC distance is $\sim$ 1
Mpc, and that those clouds with the most negative radial velocities
relative to the GSR in the general direction of M31 
will be the most distant.  We expect that
it will be possible to obtain distances via stellar absorption lines
only for the three HVC groupings with the largest angular sizes, namely for
the Northern Hemisphere Clouds, the Magellanic Stream, and the Outer
Arm Complex. The largest remaining clouds, including Complex H, might
also be relatively nearby, compared to Local Group distances, with
distances $<$ 100 kpc. We would not, however, expect the small isolated
clouds in the Barycenter or Antibarycenter groupings, with angular
diameters $\leq$ 60 sq deg, to yield stellar absorption lines against
stars in the Milky Way.

\subsubsection{H$\alpha$ Distances}
\label{sec:ionization}

High--velocity clouds will be bathed by the 
interstellar radiation field leaking into the halo from the disk if
they are Galactic, or by the intergalactic radiation field if they
pervade the volume of the Local Group.  In either case, the radiation
field will ionize the outer envelope of neutral gas.

Recent observations of  H$\alpha$ emission from HVCs place some limits
on their distances (\markcite{Kutyrev89}Kutyrev \& Reynolds 1989;
\markcite{Songaila89}Songaila, Byrant, \& Cowie 1989; 
\markcite{Tufte96}Tufte et al. 1996;
\markcite{Weiner96}Weiner \& Williams 1996; \markcite{Tufte98}Tufte, 
Reynolds, \& Haffner 1998; 
\markcite{Weiner98} Weiner 1998).  \markcite{Weiner96}Weiner
\& Williams (1996) detected H$\alpha$ emission toward substructures in
the Magellanic Stream at levels of 370, 210, and 200 mR (corresponding
to emission measures, EM, of 0.5 -- 1.0 cm$^{-6}$ pc), and attributed 
this emission
to shock ionization because of the morphology of the emission.  More
recently, \markcite{Tufte98}Tufte et al. (1998) detected H$\alpha$
emission along 13 lines of sight toward complexes A, C, and M at intensities
ranging from 60 to 200 mR (EM = 0.15 to 0.5 cm$^{-6}$ pc).  They
attribute this emission to either ram--pressure shocks caused by the
passage of the clouds through low density hot halo gas or to ionizing
radiation leaking from the Galactic disk.  The Magellanic Stream is at
a distance of about 50 kpc and thus would be expected to have lower
H$\alpha$ intensities than the nearer A, C, and M complexes, if both sets
of clouds are photoionized.  However, leakage radiation would depend
weakly on distance from the plane as long as the Milky Way
subtends a large angle as seen from the clouds, and would likely be
non-uniform in the halo, which might explain the range and variation of
observed intensities.  On the other hand, if shock ionization is
responsible for the emission from the A, C, and M complexes, the smaller
H$\alpha$ intensitites may be the result of lower velocities relative
to the ambient halo gas than is the case for the Magellanic--Stream clouds.
In any event, we expect most of the remaining HVCs to exhibit lower--intensity
H$\alpha$ emission than either set of detections because nearly all of
them would be either farther away from the source of leakage radiation
or in a much lower--density portion of the halo.  Observations
of a number of other HVCs show that the detections or upper limits are
reasonably consistent with this picture.  \markcite{Weiner98}Weiner
(1998) has observed 4 HVCs over a range of
longitudes and has obtained upper limits of 60 mR toward all of them.
\markcite{Kutyrev89}Kutyrev \& Reynolds (1989) detected H$\alpha$
toward an HVC in Cetus at a level of 81 mR, only 20\% of the intensity
of the weakest of the \markcite{Weiner96}Weiner \& Williams (1996)
detections, and comparable to the weakest of the Tufte et al. (1998) 
detections.
Thus, at least four of these five clouds appear to be farther than Complexes
A, C, and M and the Magellanic Stream, regardless of the source of
ionization.

\markcite{Bland97}Bland-Hawthorn \& Maloney (1997) have 
modelled the ionizing photons leaking from the Galactic disk
in order to derive an expression for the mean EM of an HI cloud as
a function of distance from the Galactic plane.  For their preferred
value of the optical depth of the disk material to radiation at the
Lyman limit, 2.8, which best fits the Weiner \& Williams detections,
they obtain $ {\rm EM} = 6.7 \times 10^2~{r_{\rm kpc}}^{-2}~{\rm
cm}^{-6}~{\rm pc}$;  for $\tau = 2$, the coefficient rises to $1.8
\times 10^3$. An EM of 6.7 $\times 10^2$ cm$^{-6}$ pc corresponds to an
intensity of about 270 R, three orders of magnitude greater than the
Magellanic Stream detections.  Even at a distance of 10 kpc from the
plane, an EM of at least 7 ${\rm cm}^{-6}~{\rm pc}$ is expected,
easily within the range of past and present observations.
\markcite{Bregman86}Bregman \& Harrington (1986) also estimated the
Lyman continuum flux leaking into the halo and obtain values similar to
those of Bland-Hawthorne \& Maloney.  If these models are approximately
correct, one would also expect that a Galactic HVC with a vertical
distance of only a few kpc would be evident on optical photographs such
as the POSS images, which have an emission  measure sensitivity of
$\sim$ 100~cm$^{-6}$ pc.  Furthermore, given the detection of shock--excited
H$\alpha$ in the Magellanic Stream, one would expect higher
intensities from all of the HVCs with larger radial velocities and from most
with lower velocities because of the much higher density of ambient gas
in the lower Galactic halo if the HVCs are Galactic.  
Given the level of detections in the
Magellanic Stream and in Complexes A, C, and M, essentially all HVCs located
in the Milky Way halo should be easily detectable in deep H$\alpha$
surveys such as the WHAM survey (\markcite{Reynold96}Reynolds 1996).

At large distances from the Milky Way and M31, HVCs should be
detectable from the ionization expected from the diffuse ionizing
background radiation, for which there is currently  a 2--$\sigma$ upper
limit to the flux of 20 mR (\markcite{Vogel95}Vogel et al. 1995). 
This ionizing flux would
lead to an EM of $4 \times 10^{-2}$~cm$^{-6}$ pc.  On the
other hand, if the background ionization is as low as $6\times
10^{-24}$~erg \c2 s$^{-1}$ Hz$^{-1}$, as suggested by
\markcite{Kulkarni93}Kulkarni \& Fall (1993), the EM from
extragalactic HVCs could be as low as $2\times 10^{-3}$~cm$^{-6}~{\rm
pc}$; this would be the minimum EM expected from an HVC,
Galactic or extragalactic.  It is unclear at present whether the
diffuse background flux or the ionizing flux leaking from the Milky Way
and M31 dominates the Lyman--limit flux absorbed by the HVCs.

\subsection{HVC Metallicities}
\label{sec:chemistry}
 
Metallicity determinations should provide one of the clearest tests of
whether the HVCs are Galactic or extragalactic.  If the HVCs are
Galactic, then they would have abundances that are at least solar. In a
Galactic--fountain model, for example, gas ejected from the inner
Galaxy would have abundances greater than solar because the Galactic
metallicity gradient implies high metallicity in the inner Galaxy. Gas
that is simply ejected vertically and falls back vertically does not
attain high velocities relative to the LSR.  Furthermore, if the source
of the fountain is gas ejected into the corona by massive stars and
supernovae, this gas should be metal enriched, even at the solar
circle.

In the Local--Group infall model, on the other hand, we expect that HVCs would
have metallicities typical of the intergalactic medium.  This
characteristic metallicity is poorly known, but is probably
significantly greater than the primordial abundance.  In poor groups,
X-ray observations imply metallicities $\sim0.1$ solar
(\markcite{Davis96}Davis et al. 1996).  This non--primordial abundance
likely represents the chemical pollution of the intergalactic gas.
ASCA observations of rich clusters find metal abundances  of roughly
half the solar abundance (\markcite{Mushotzky96}Mushotzky et al. 1996).
Based upon the large metal abundances seen in intercluster gas,
\markcite{Renzini97}Renzini (1997) suggested that the metal abundance of
the intergalactic medium has today reached 1/3 of the solar value.
Thus, even though we identify the HVCs as gas clouds which are falling
into the Local Group for the first time, their phenomenological
association with Ly--$\alpha$ absorbers suggests that their
metallicities would be subsolar, with values of $\leq$ 0.1 to 0.3
solar, but not primordial.

Abundance measurements from absorption--line studies show, on the other
hand, that HVCs always have metallicities significantly less than
solar, generally $\leq 0.1$ solar (\markcite{Savage93}Savage et al. 1993;
\markcite{Lu94}Lu, Savage, \& Sembach 1994; \markcite{Sembach95}Sembach
et al. 1995; \markcite{Sembach96}Sembach \& Savage 1996;
\markcite{Lu97}Lu et al. 1997; see \markcite{WvW97}WvW97 for a review).
However, because some lines are saturated, and because observed lines
may not always be the dominant ionization stage, it is not always
possible to obtain reliable metallicities from the observations.
Furthermore, abundances of some species may be depleted onto grains,
further complicating metallicity determinations. However, analyses of
the IRAS and COBE data bases (\markcite{Wakker86}Wakker \& Boulanger
1986; \markcite{Schlegel97}Schlegel, Finkbeiner, \& Davis 1997) suggest
that the dust abundances in the Northern Hemisphere Complex are at
least three times lower than the locally determined value.  Also, because
the HVCs are observed primarily at high Galactic latitudes, if
they were Galactic then the HVCs should be more like the warm diffuse
clouds where the gas--phase depletions are considerably smaller.  Thus
depletion onto dust grains would be minimal because there is little
dust on which to deplete the gas, and because the HVCs would, in any event,
correspond to interstellar clouds where the depletion is already
relatively minimal.  Alternatively, the IRAS non-detections  
of HVCs may result from dust temperatures lower than those typically found
in the Galactic plane; in that case, heating by the interstellar
radiation field implies distances of at least 10 kpc from the Galactic
plane (\markcite{Wakker86}Wakker \& Boulanger 1986), a value consistent
with the Local Group hypothesis.

One HVC was measured to have an S/H ratio of 0.25 solar
(\markcite{lu97}Lu et al. 1997), which should represent its true
metallicity, because sulfur is not readily depleted onto grains and because
the SII transition observed should be the dominant ionization stage of the
ion.  Lu et al. conclude that this metallicity suggests a Magellanic Cloud
origin, even though the HVC in question is not part of the Magellanic
Stream.  The metallicity is reasonably consistent with the Local--Group
hypothesis, but inconsistent with a Galactic origin.  Complex C in the
Northern Hemisphere grouping exhibits a MgII line with an abundance of
0.10 solar and another, higher--velocity component (associated with
cloud 84 in the WvW91 catalogue and presumably not part of the Northern
Hemishere Cloud), with an abundance of 0.06 solar
(\markcite{Bowen93}Bowen \& Blades 1993; \markcite{Bowen95}Bowen
et al. 1995; \markcite{WvW97}WvW97).   
\markcite{Sembach96}Sembach \& Savage (1996)
give somewhat different abundances, an order of magnitude lower in the
case of cloud 84, but always well below solar values.  For MgII the
ionization correction should not cause significant uncertainty, although
Sembach \& Savage point out that the results may be affected by dust
depletion.  The expected depletion for magnesium in warm halo
gas is, however, only about a factor of three on average
(\markcite{Sembach96}Sembach \& Savage 1996); these two clouds may be
reasonably considered to be metal deficient even for halo gas, contrary
to expectations if the gas originated in a Galactic fountain.  We
therefore find that of two lines of sight with three individual HVCs,
there are metal deficiencies of as much as a factor of twenty, and that no
line of sight has abundances that even approach solar values.

\subsection{Extragalactic HVC Searches}
\label{sec:exgalhvcs}

If HVCs are indeed characteristically extragalactic, and are the leftover 
building blocks from which the Local Group formed, similar entities would be 
expected to be observed toward other galaxy groups. Because the details of the 
distribution would depend on the dynamics of the particular galaxy group being 
observed, the detailed spatial and kinematic patterns of the Local 
Group would not be preserved.  The general properties of the HVC phenomenon 
would, however, be preserved out to the Hubble--flow turnaround radius of the 
galaxy group in question, and might be observable with radio interferometers 
or, in some nearby groups, with single dishes.  As we discuss below, 
such clouds have apparently been detected.

We first consider the sensitivity of the VLA and of single dishes for a search 
for extragalactic HVCs.  For the VLA, the brightness sensitivity is greatest in 
the most compact configuration, the D--Array; maximum sensitivity is attained 
when the velocity channels are the broadest.  If we assume that the channels 
are 20 \kmse wide, equal to the mean linewidths of the HVCs in Table 1, and 
that all 27 antennas are used for an integration time of 8 hours (a single 
earth--rotation synthesis), we obtain an rms brightness sensitivity of 
0.12 K.  A detection with this many pixels and a single velocity channel 
requires a minimum signal-to-noise ratio of 5 $\sigma$, or an equivalent mean 
column 
density of $2.2 \times 10^{19}$ \c2 in a synthesized beam with a diameter of 
44\asec.  This estimate is a lower limit to the detectable column density 
because it assumes observing at the zenith, 100\% system efficiency, and 
natural weighting.

We now further assume that we are trying to detect an HVC system identical to 
the one we associate with the Local Group, with a turnaround radius of 1.5 Mpc 
and a mean HVC diameter of 28 kpc.  The primary beam (i.e. field of view) is 
30\amin; a cluster of Local--Group HVCs would have to be at a distance of 340 
Mpc to fill this primary beam.  The synthesized beam (resolution element) at
this distance is 73 kpc, diluting the signal by a factor of 6.7 and raising
the minimum detectable column density by the same factor.  Thus, one would
detect HVCs with a mean  $N_{\rm HI} > 1.5 \times 10^{20}$ \c2; 
Figure~\ref{fig:columndens} shows that 
there are only two Galactic HVCs with such large column densities.  The
detection threshold would increase if the clouds just fill the synthesized
beam, which would occur for a system at a distance of 130 Mpc, but, at that
distance, one would decrease the total number of clouds in the primary beam
by a factor of 6.7.  Given the frequency distribution of column densities
shown in Figure~\ref{fig:columndens}, 
we find that only 5.5\% of the HVCs have $N_{\rm HI} > 2 \times 10^{19}$ \c2, 
and only 2.0\% have $N_{\rm HI} > 3 \times 10^{19}$ \c2, a more realistic 
detection limit.  If another system at a distance of 130 Mpc has the same 
number of clouds as the Local Group, some 550, one would expect to detect only 
5 clouds at a threshold column density of $ 2 \times 10^{19}$ \c2, and only 2 
clouds at a threshold of $3 \times 10^{19}$ \c2.

All extragalactic searches have, however, focused on galaxies with
distances less than 130 Mpc.  In such nearby searches, the significance of a 
detection in a single pixel is not substantially increased unless a large 
fraction of the surface area of an HVC is at a column density much above the 
mean. In that case, one might be able to detect the higher--column--density 
parts of an 
HVC with lower sensitivity observations, yielding a size smaller than 
that of the cloud as a whole.  Spatial 
averaging could also increase the significance of a detection.  On the other 
hand, the smaller area covered by the primary beam means that there is a 
decreasing probability of intersecting a single cloud ($\propto d^{-2}$), such 
that the probability of detecting a cloud in a cluster at a distance of 13 Mpc, 
for example, is 0.05 at a column density of $2 \times 10^{19}$ \c2, if the 
clouds are uniformly distributed within the turnaround radius.  The detection 
probability increases with the square of the distance and is further increased 
if the distribution of HVCs is not random but shows some concentration in the 
deeper parts of the potential well, close to the most massive galaxies in the 
cluster.  In that case, a good strategy would involve targeted searches near 
massive galaxies in poor groups.  The detection probability is also enhanced to 
the degree that the WvW91 compilation is incomplete.  For most searches to 
date, we estimate that the probability of a detection with a radio 
interferometer of a single HVC like those in the Local Group is generally less 
than unity, but probably more than $\sim$ 0.1, depending on the
galaxy--cluster environment and its distance, suggesting that some
extragalactic HVCs might already have been detected.  In any event, we expect
that interferometric searches would rarely detect more than one cloud in the
primary beam for nearby objects, and no more than a few clouds at large
distances. 

For single--dish searches, the brightness sensitivity is much greater
than with an interferometer, but at distances greater than 3 Mpc an HVC
would be unresolved with a 25--m dish.  A 5 $\sigma$ detection of a
cloud with a mean column density of $1 \times 10^{18}$ \c2 at 3 Mpc
requires about 1/2 hour of on--source integration time with a sensitive
system, or a total of about 1 hour in real time.  Such an observation
would detect any cloud in the WvW91 compilation.  However, the surface
filling fraction of those clouds is only 5\%; about 20 hours of
observing would be necessary to find one cloud at a distance of 3 Mpc.
Most small groups of galaxies lie at greater distances, and although
the probability of intersecting a cloud goes up as the square of the
distance, the signal from a single cloud is decreased by the same
factor, and one detects only the higher--column--density HVCs for a fixed
integration time.  As with interferometers, the detection probability
depends on the unknown degree of clustering of the HVCs, and on the
incompleteness of the WvW91 compilation. With a small (25--m to 50--m)
antenna, one expects that a detection would be manifested as an
asymmetric wing on an HI line profile.  \markcite{Bates96}Bates \&
Maddalena (1996) searched for high--velocity wings using the NRAO 43--m
telescope and found such wings toward 8 of the 23 galaxies observed.
Further analysis and interferometric observations are required to
determine whether this gas is like the Local--Group HVC ensemble, but
their observations suggest that single dishes might be able to
detect extragalactic HVC candidates.  The newly--upgraded 300--m Arecibo
telescope affords approximately the same column--density sensitivity as
smaller dishes, but has a much smaller
beam which would resolve HVCs out to distances of
about 30 Mpc.  However, one still expects only a few detections in 20
hours of observation because of the expected small surface--filling
fraction of the HVCs. We therefore conclude that detection of HVC
analogues in other systems should be possible with a good search
strategy and a sufficient investment of observing time.

Several published HI searches have turned up clouds which
resemble what we expect for extragalactic HVCs.  Observations by
\markcite{vanderHulst88}van der Hulst \& Sancisi (1988) toward the
giant Sc galaxy M101 showed two clouds superimposed on the disk with 
radial--velocity differences of 130 and 160 \kmse from normal galactic
rotation.  The masses of those clouds are $1.6 \times 10^8$ and $1.2
\times 10^7$ \msune and the diameters are 16 kpc and 5 kpc,
respectively, for H$_0$ = 75 \kmse Mpc$^{-1}$. Van der Hulst \& Sancisi
argued that the M101 HVCs must be completely different from even the
largest and most massive of the Milky Way HVCs, if the latter are
objects in the Galactic halo at distances less than 10 kpc. The very
large kinetic energy ($\sim 10^{55}$ erg) of the M101 clouds relative
to the host galaxy implies that they are being accreted by the galaxy
rather than being expelled.  The mass and diameter of the larger M101
cloud is comparable to the mean value given in Table 3 for the Local
Group HVCs, and is thus consistent with its being an HVC similar to
those inferred in this paper.  Furthermore, the similarity in the
velocities of the two clouds discovered by van de Hulst \& Sancisi
suggest that they might be fragments of a single object falling onto
M101.

\markcite{Kamphuis92}Kamphuis \& Briggs (1992) found two HVCs toward
the galaxy NGC~628 with masses of $7.9 \times 10^7$ \msun~ and $9.5
\times 10^7$ \msune (for 
H$_0$ = 75 \kmse Mpc$^{-1}$) in the outer parts of the NGC~628 disk
which they, too, attribute to accretion.  The mean diameters of these
clouds are 38 kpc and 47 kpc, respectively; the respective peak HI
column densities are $1 \times 10^{20}$ and $3.7 \times 10^{19}$ \c2,
values within the range of what is observed for the Local--Group HVCs.
A third, smaller HVC is argued to be possibly related to one of the
larger clouds.

More recently, \markcite{Schulman96}Schulman et al. (1996) detected and
mapped an extragalactic HVC toward NGC 5668 with the VLA.  Using a
simple rotational model for the HI in NGC 5668, the authors identified
HI emission that is kinematically distinct from the galaxy itself and
found a cloud beyond the optical disk of the galaxy, extending in fact
even beyond the outer edge of the HI disk.  Although the cloud blends
with the emission from the disk of the galaxy, the diameter of the
cloud is estimated to be about 4\amin.  Schulman et al. conclude that
this feature is distinct from the HI in the disk and cannot be due to a
galactic fountain.  At a distance of 21 Mpc (for H$_0$ = 75 \kmse
Mpc$^{-1}$), the cloud has a diameter of about 25 kpc, and an HI mass
of $1 \times 10^8$ \msun.  The mass follows from the total amount of
kinematically distinct gas determined by Schulman et al. and adjusted for
their different value of H$_0$.

HVC analogues have also been found by \markcite{Taylor95}Taylor et al.
(1995), who made a VLA search for companion objects to HII galaxies.
They found six HI clouds without optical counterparts, in the fields of
21 galaxies.  These intergalactic HI clouds have masses in the range
0.6 to $1.7 \times 10^8$ \msun, and diameters ranging from 8 to 16 kpc.
Further, \markcite{Hunter94}Hunter, van Woerden, \& Gallagher (1994)
found a cloud with a mass of $6 \times 10^7$ \msune and a diameter of
$\sim$ 7 kpc toward NGC 1800.

Clearly, numerous extragalactic HI clouds have already been found, with
properties similar to those we infer for the HVCs associated with the
Local Group.  Several of these (in particular those of
\markcite{Taylor95}Taylor et al. 1995) show no direct spatial or
kinematic connection with the target galaxy. The others may be
extragalactic analogs of Complex C, which we argued above is probably
being tidally disrupted as it is being accreted by the Milky Way.  We
predict that observations in fields adjacent to other massive galaxies
will also turn up HI clouds with properties similar to those given in
Table 3, at about the rate determined by the detectability criteria
above.

\subsection{Re-examining Arguments Against the 
Extragalactic Origin of HVCs}
\label{sec:counterarguments}

Several cogent arguments have been offered against the extragalactic
origin of HVCs (see \markcite{WvW97}WvW97 and references there,
especially \markcite{Giovanelli77}Giovanelli 1977 and
\markcite{Giovanelli81}1981, and 
\markcite{Verschuur75}Verschuur 1975).  These arguments include
the following:
(1) the kinematics of galaxies in the Local Group do not match the HVC
kinematics; (2) if the HVCs are bound, then either they extend beyond
the Local Group or else some 90\% of their mass is in a form other than
neutral hydrogen; (3) no explanation is available for the multi--phase
nature of the HVCs, in particular, for the cold gas seen in the HVC
cores; (4) the small velocity gradients seen in the clouds remain
enigmatic; and (5) analogous systems of clouds are not detected near
external galaxies. These criticisms were made when the observational
situation was much less mature than it is now.  Specifically, the sky
north of $\delta = -30\deg$ is now quite uniformly surveyed in the HI
line, over the velocity range which encompasses the HVC
phenomenon. There is observational evidence for anomalous--velocity
structures in the vicinity of external galaxies, and the astrophysical
context now encompasses the idea of galaxy growth through accretion of
relatively modest clouds which may contribute dark and ionized matter
as well as the observed neutral gas.

It seems that the earlier arguments against the extragalactic HVC
origin do not pertain for the Local Group infall hypothesis which we
have discussed here.

$\bullet $ We showed in \S5 and \S6 that the kinematics of infalling
gas does, in fact, match the observed HVC kinematics.

$\bullet $ We indicated in \S4 that there is roughly 10 times more dark
matter than luminous gas associated with each HVC, the same ratio of
dark matter to matter accounted for as in galaxies and on cosmological
scales.

$\bullet $ We argued in \S6 that gas in dark matter halos with velocity
dispersions of 10 to  30 \kmse will form a two--phase structure as the
denser gas in the dark--matter cores cools and is shielded from the
intergalactic ionization field.

$\bullet $ We showed in \S6 that the small velocity gradients occur
because the HVCs are gravitationally focused into large--scale
filaments.

$\bullet $ We pointed out in \S6 that quasar absorption--line studies
do, in fact, reveal clouds with HI column densities of $10^{18}$ to
$10^{19}$ \c2, namely the Lyman--limit systems associated with galaxy
groups, and in \S7 that HI analogues have been detected toward a number
of other galaxies through 21--cm aperture synthesis.

\section{Implications for the Evolution of the Milky Way}
\label{sec:implications}

\subsection{Infall Rates and Implications for Star Formation}
\label{sec:accretion}

\begin{figure}[htpb]
\centerline{\psfig{figure=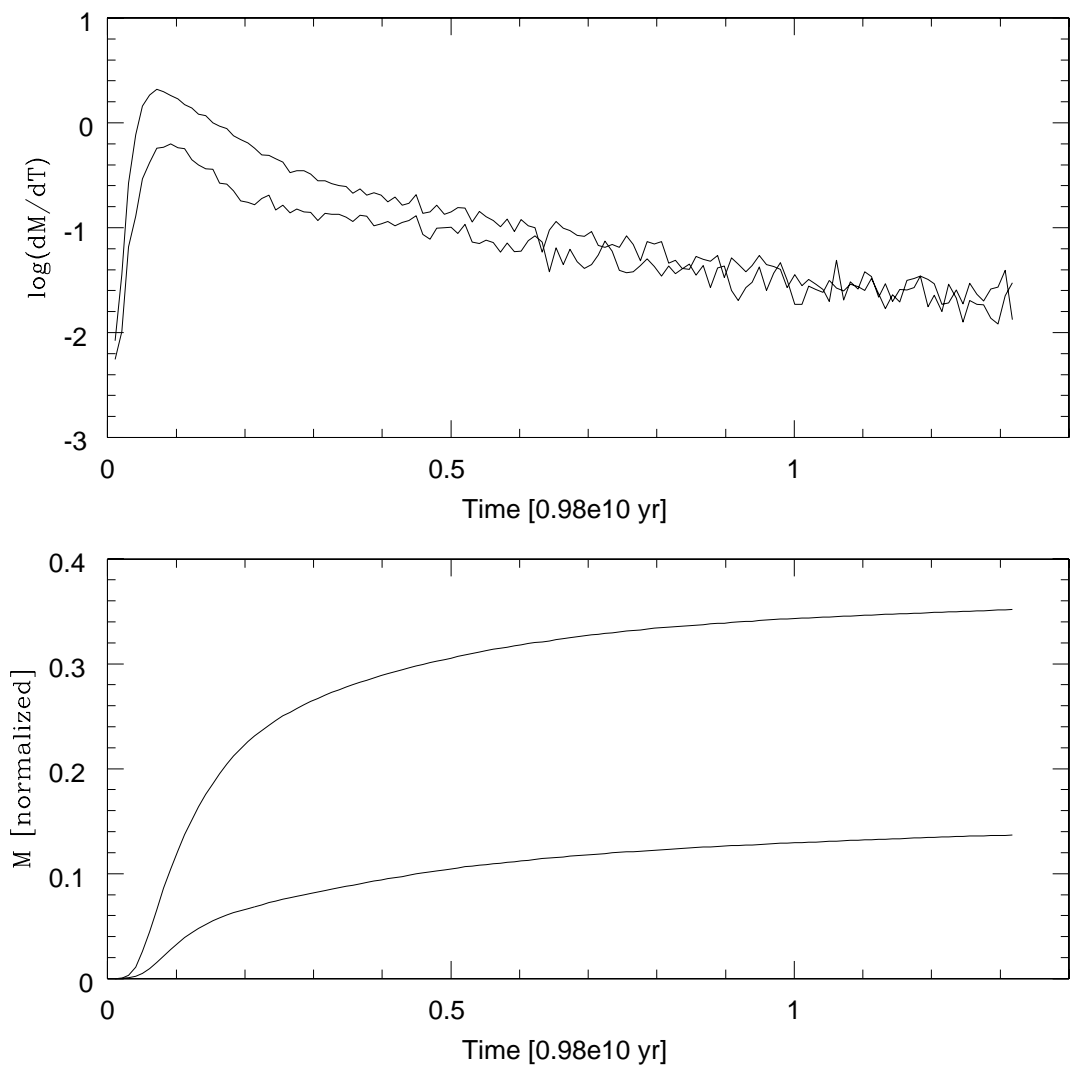,width=6in,angle=0,silent=1}}
\vskip -0.2in 
\caption{{\it Upper:} Normalized rate of simulated accretion of
clouds for the Milky Way (lower line) and for M31.
The M31 accretion rate is typically about twice that of the Milky Way.
After about 3 billion years the accretion rate becomes nearly
exponential with an e--folding time of about $5 \times 10^9$ y.  At the
current epoch, the accretion rate is flattening out, and is equivalent
to about 7.5 \msune y$^{-1}$ for the Milky Way.~~~{\it Lower:} Normalized
accreted mass for the Milky Way (lower line) and
M31.  The plot shows that most of the mass is accreted at early times
and that additional mass is being added to both galaxies quite slowly
at the current epoch.  } 
\label{fig:accrete2} 
\end{figure}

If the Milky Way and M31 are accreting material in the form of HVCs, 
it is possible to estimate the infall rate from the simulation discussed
in \S\ref{sec:simulation}, normalized by the total HVC mass currently 
observed.  The
results of this estimate are shown in Figure~\ref{fig:accrete2}.  The
accretion rate is determined by assuming that any cloud in the
simulation that passes within 100 co--moving kpc of the center of
either the Milky Way or M31 is ultimately absorbed by that galaxy.
This is a gross simplication of the accretion process, but it is
nonetheless useful to explore the accretion history in the simulation.

The upper panel of Figure~\ref{fig:accrete2} shows that accretion is
rapid early on, reaching a peak at about 30 times the current rate for
the Milky Way within a billion years after the beginning of the
simulation.  Although the accretion rate for M31 is about twice that of
the Milky Way in the first few billion years, the rates at the present
epoch are nearly equal.  About half of the mass accreted by the
Milky Way falls onto it during the first 2 billion years.  Using the
direct normalization from the simulation, the present--day accretion
rate shown in Figure~\ref{fig:accrete2} for the Milky Way is 7.5
\msune y$^{-1}$, corresponding to a neutral--gas accretion rate of
about 1.2 \msune y$^{-1}$.

Another way to estimate the accretion rate is by comparison with the
observed HVCs. Figure~\ref{fig:accrete2} shows that roughly 30\% of
the total mass accreted by the Milky Way has been acquired during the past
9 billion years.  As mentioned in \S6, this mass is approximately
equal to the mass currently observed in HVCs, and is about 25\% of the
original inventory.  There are 518 HVCs in the WvW91 compilation that
are not part of the Magellanic Stream.  Comparison of the compilation
with maps made from the LD survey revealed a number of small clouds not
in the WvW91 catalogue because they lie between the grid points of the
coarser surveys; the incompleteness, however, probably changes the
total number by less than a factor of two.  Thus the total mass in HVCs at the
present time is contributed by $\sim$1000 clouds with a mean mass of $3
\times 10^8$ \msun, or $3 \times 10^{11}$ \msun.
Roughly 50\% of the total accreted over the past 12 billion years
corresponds to an average rate of 12.5 \msune y$^{-1}$.
Figure~\ref{fig:accrete2} indicates that there have been about 2
e--folding times in the past 12 billion years. The present--day total
accretion rate should be about 1/e of the average rate of about 4.6
\msune y$^{-1}$, or a neutral--gas accretion rate of $\sim$ 0.8 \msune
y$^{-1}$, in reasonable agreement with the accretion rate given above.

We should not take these numbers as more than rough estimates, in view
of the uncertainties and simplifications present in the simulation.
The gas accretion rate, for example, would be significantly larger if a
large quantity of ionized gas were associated with the HVCs.  It is
worth noting in this context that \markcite{Lacey85}Lacey \& Fall
(1985) and \markcite{Blitz97}Blitz (1997) have argued that the short (2
to 5 $\times 10^8$ y) molecular--gas depletion time for the Milky Way
could be compensated with gas infall equal to the net rate of gas
conversion into stars, some 1 to 3 \msune y$^{-1}$.  The gas accretion
rate we infer for the HVCs is within this range, and could provide the
fuel for the continuing star formation in the Milky Way and in spiral
galaxies in general.

\subsection{The Galactic Fountain}
\label{sec:galfountain}

The most frequently discussed origin for the HVC phenomenon is the
``galactic fountain'' proposed by \markcite{Shapiro76}Shapiro \& Field
(1976) and elaborated by \markcite{Bregman80}Bregman (1980) and others.
In the galactic--fountain model, gas is heated by supernova explosions
in the disk to temperatures of $\sim 10^6$ K, convects to the Galactic
corona where it radiatively cools, and then falls back to the disk as
high--velocity gas. To explain the high velocities observed at latitudes
away from the zenith, Bregman requires a radial outflow which conserves
angular momentum; this outflow cannot account for  radial velocities
larger than about $\pm 200$ \kms, yet many HVCs are observed with
higher velocities.  The galactic--fountain model makes several other
predictions about the nature of the HVCs (see \markcite{Wakker90}Wakker
\& Bregman 1990): the HVCs would be metal rich because the gas is
largely ejected from the inner Galaxy ($Z > Z$\sun); their
characteristic distances would be between 0 and 10 kpc from the
Galactic plane; and their vertical velocities would be less than 70 to
100 \kms.

The predictions of the galactic fountain model are not consistent with
many of the observations of the ensemble of HVCs.  The HVCs observed to
date all have substantially subsolar abundances and at least two have
subsolar metallicities (see \S\ref{sec:chemistry}).  
If the heavy elements in the clouds were bound
into dust grains in clouds at distances less than 10 kpc, then the HVCs
would have been detected in the IRAS 60 \microne band, but they were
not detected (\markcite{Wakker86}Wakker \& Boulanger 1986).
Absorption--line observations towards AGNs imply that the filling
fraction for HVCs at column densities as low as $1 \times 10^{17}$ \c2
is a factor of 1.5 to 2.0 greater than that found in emission in WvW91
(\markcite{Murphy95}Murphy et al. 1995; \markcite{Bowen95}Bowen et al. 1995).
Excluding the
Magellanic Stream and the Outer Arm Complex, this implies a surface
filling fraction of 20\% down to these low column densities, yet only
two lines of sight have yielded positive absorption--line detections of
the dozens (or more) stars towards which high--velocity absorption has
been sought.  This result is difficult to understand if the
characteristic distance to an HVC is only a few kpc. Finally, many HVCs
have vertical velocities that exceed 100 \kms.  

%new paragraph here 1/16
% WBB, Jan. 24:  I haven't changed anything in this new paragraph, but I
%think that it hangs rather loosely here, and in any case the implied
%picture of the ISM dynamics is more violent, and hotter, than the obs.
%suggest, and that this has been shown fairly nicely by a number of recent
%ISM simulations, among them Bregman's own.  Is the para. necessary?
%Or maybe the wording could be such to give the fountain a blessing
%regarding some aspects of the dynamics of the ISM, but not relevant to HVCs.

One of the attractions of the galactic fountain model is that it seems
naturally to fit into a coherent picture for the dynamics of the
interstellar medium.  Supernovae eject significant amounts of mass into
the ISM and drive gas upwards through chimneys
(\markcite{Heiles90}Heiles 1990).  \markcite{McKee93}McKee (1993)
estimates an outflow rate of several solar mass per year.  Mass
balance seems to require that this gas is somehow returned to the
disk.  

\begin{figure}[htpb]
\centerline{\psfig{figure=fountain.ps,width=8in,angle=+90,silent=1}}
\vskip -0.2in
%\special{psfile=fountain.ps angle=90}
\caption{Latitude-velocity plots of all of the individual velocity
components in the LD survey biased by velocity width to deemphasize
high--velocity emission.  Each component from each spectrum is plotted
as a dot in the figure.
}
\label{fig:fountain}
\end{figure}

Is there evidence  for the existence of a Galactic fountain? We carried
out a decomposition of the individual spectra of the LD survey into
velocity components, identified from intensity maxima in each spectrum,
and selected according to a minimum velocity width. This selection
tends to bias the resulting component list against high--velocity
emission.  Figure~\ref{fig:fountain} summarizes the results of the
decomposition and was produced by marking with a dot
each velocity component in the list at a given Galactic latitude,
independent of its longitude. The plot is therefore a
latitude--velocity plot of HI velocities which accentuates the emission
observed at low velocities.  The dominant feature in 
Figure~\ref{fig:fountain} is the nearly
vertical band of emission, but note that the emission near 0 \kmse
toward both Galactic poles is negative,
smoothly increasing to slightly
positive values near the Galactic equator.  Gas near zero \kmse is
overwhelmingly local, but in a static HI layer, the velocities
would be precisely zero at latitudes away from the Galactic equator.
(At $|b| \sim 0\deg$, the Galactic rotation along the long lines of
sight intercepted would introduce some non--zero velocities, slightly
positive when averaged over the longitude range of the LD survey.).  An
error in the determination of the LSR would have a different signature,
resulting in deviations of equal magnitude but opposite signs in the
two Galactic hemispheres.  Rather, Figure~\ref{fig:fountain} suggests
that the gas is falling toward the plane in both hemispheres.  The
predominance of negative velocities toward the poles has been known for
many years (see e.g. \markcite{Weaver74}Weaver 1974;
\markcite{Kulkarni85}Kulkarni \& Fich 1985; \markcite{Lockman91}Lockman
\& Gehman 1991). We note that the deviation from zero velocity is quite
symmetric with respect to the Galactic plane, implying that the gas
motions are both systematic and quite general.

\begin{figure}[htpb]
\centerline{\psfig{figure=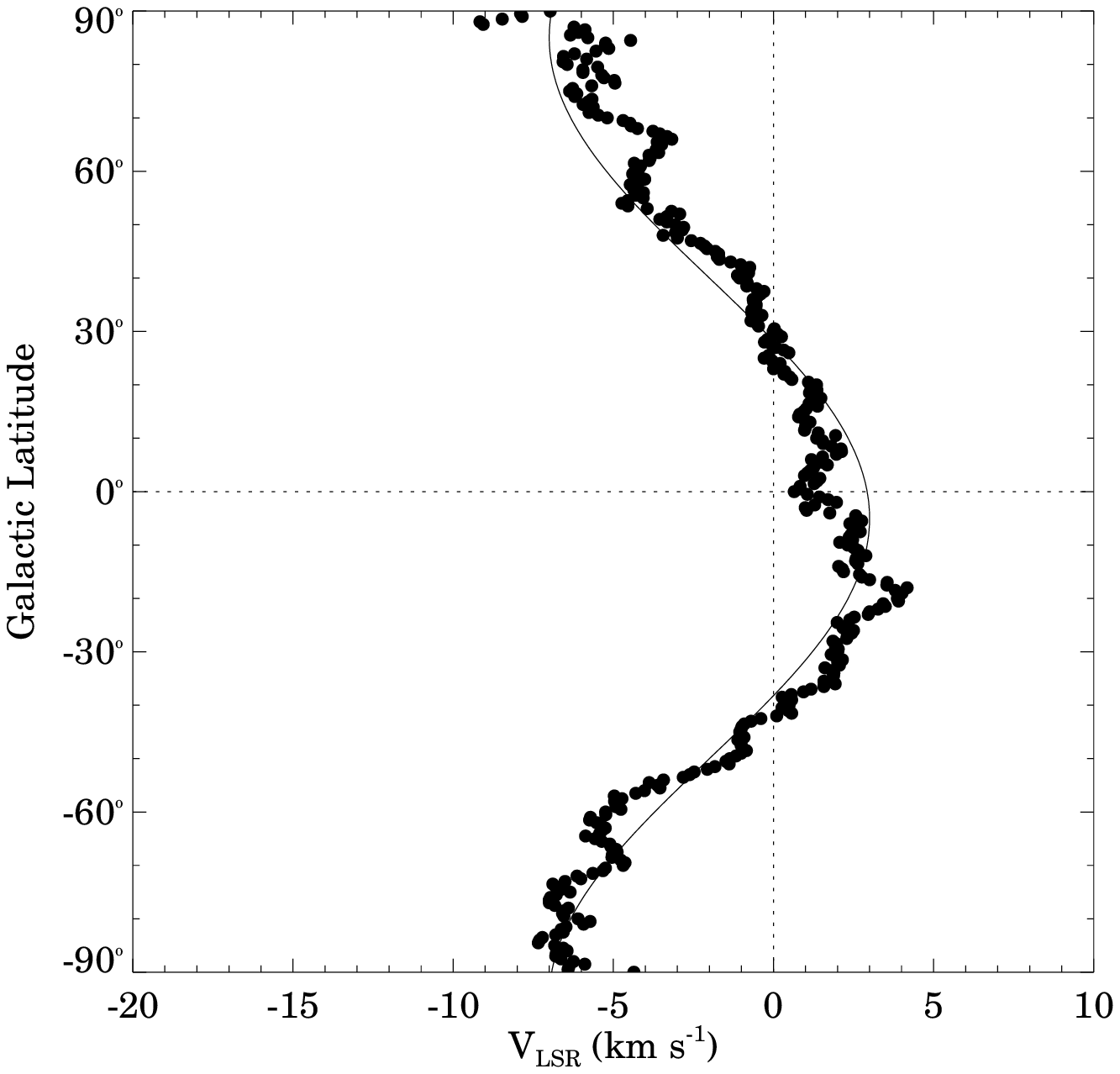,width=6in,angle=0,silent=1}}
\vskip -0.2in
\caption{Components from Figure~\ref{fig:fountain} averaged over
longitude at each 0\degper5 of latitude (the resolution of the LD
survey) for $|v_{lsr}| < 20$ \kms.  What is plotted is therefore
represents all of the local HI (out to 500 -- 1000 pc) within the
disk.  A disk in hydrostatic equilibrium would show all velocities at 0
\kms.  See text for the equation of the fitted sine curve.
}
\label{fig:fountfit}
\end{figure}

If the motions implied by Figure~\ref{fig:fountain} represent
vertical infall onto the Galactic plane, then the velocities should
show a sin\thinspace$b$ dependence which other systematic motions would
not show.  Figure~\ref{fig:fountfit} shows the mean of the velocity
components, averaged over all longitudes at each latitude of the LD
survey, for the local gas, which is obtained by restricting the
calculation of the mean to an LSR velocity range within $\pm$ 20 \kmse
of zero.  Superimposed on the data is the least--squares fit $v_{\rm
LSR} = -1.8 -4.5~{\rm sin}(2b -49)$ \kms. Deviations from the fit are
generally less than 1 \kms, showing that a sine curve is a good
description of the data; such a functional form implies vertical
infall.  The LD survey is not an all--sky survey, however, so we
checked the eventual effect of the incomplete sky coverage on the
Figure~\ref{fig:fountfit} curve by making latitude--velocity plots
separately in different longitude ranges. Although the LD coverage in
the first quadrant is complete in both longitude and latitude, and
quite incomplete in the third and fourth quadrants, the general shape
and magnitude of the curve shown in Figure~\ref{fig:fountfit}
is present in all quadrants, implying that the motion seen in the figure
is a general property of the local gas.  The results imply that the
entire HI layer is collapsing toward the equator at a velocity of 6.3
\kms, out to a distance of at least several hundred pc, i.e. the
radial distance probed at $b = 30\deg$. 
This is an average gas motion; although there may be local deviations,
the local HI layer is apparently not in hydrostatic equilibrium.

We propose that the inflow toward the plane of the HI layer is a
manifestation of the Galactic fountain; but rather than velocities of
tens up to 100 \kms, the returning flow evidently does not exceed the
sound speed of the material in which it is embedded.  Thus the
returning gas, once it has become neutral, does not have a vertical
velocity in excess of about 10 \kms. The infall velocity is regulated
by mass conservation at different heights above the plane and by the
requirement that the gas motions not be supersonic, at any $z$--height.
Furthermore, the data suggest that the motions are almost entirely in
$z$, as suggested originally by \markcite{Shapiro76}Shapiro \& Field
(1976), without a large radial component.

We estimate the mass infall rate, d$M$/d$t$, from ${\rm d}M/{\rm d}t =
\rho A v $, where $\rho$ is the mean density of the neutral gas, $v$
its mean infall velocity, and $A$ the area of the Milky Way onto which
the gas falls.  $N_{\rm HI}$ is well determined toward the Galactic
poles and has an average value of $1.7 \times 10^{20}$ \c2
(\markcite{Kulkarni85}Kulkarni \& Fich 1985). We take $\rho$ as $N_{\rm
HI}m_{\rm HI}/z_{\rm HI}$, where $z_{\rm HI}$ is the total thickness (i.e.
twice the scale height) of the HI layer, a loosely defined concept
unless the emission is deconvolved into separate components (see e.g.
\markcite{Falgarone73}Falgarone \& Lequeux 1973;
\markcite{Lockman91}Lockman \& Gehman 1991). If we take $z_{\rm HI}$ to
be twice the scale height in units of 1 kpc, and if we assume that what
we observe locally is representative of the infall out to a distance of
10 kpc from the Galactic center, then, correcting for helium, ${\rm
d}M/{\rm d}t = 5.3 z_{\rm HI}$ M\sune y$^{-1}$. If most of the HI seen
toward the galactic poles is the cold HI component, then $z_{\rm HI} =
400$ pc (\markcite{Falgarone73}Falgarone \& Lequeux 1973;
\markcite{Lockman91}Lockman \& Gehman 1991), and ${\rm d}M/{\rm d}t =
2.1$ M\sune y$^{-1}$.  If the Galactic fountain returns gas to radii as
large as 15 kpc, or about twice the Sun's distance from the center,
${\rm d}M/{\rm d}t$ increases by a factor of 2.3.  Thus for the
expected range of HI scale height and radius over which the Galactic
fountain operates, the expected infall rate from
Figure~\ref{fig:fountfit} is several M\sune y$^{-1}$.  However, unlike
the detailed fountain models proposed to date, the observations imply
that the return flows are subsonic and almost entirely normal to the
plane.

We now compare the measured mass infall rate with the estimated mass
injection rate into the Galactic halo by supernovae. Taking all of the
phases of HI into account, and extrapolating to a radius of 10 kpc, 
(\markcite{Heiles90}Heiles (1990) obtained a
mass injection rate of 1.7 M\sune y$^{-1}$, in agreement with our
estimated infall rate of 2.1 M\sune y$^{-1}$.
\markcite{McKee93}McKee (1993) found a similar mass injection rate of
several M\sune y$^{-1}$.  Thus even though the HI layer is evidently
not in hydrostatic equilibrium, the total gas layer in the disk and
corona appear to be in steady--state dynamical equilibrium.

\subsection{Chemical Evolution of the Disk}
\label{sec:disk}

The above analysis has implications for
understanding the evolution of the Galactic disk.  The constant rain of
high--velocity clouds implies that the disk experiences episodic
accretion of gas. The relatively weak Galactic fountain discussed above
implies slow radial mixing of metals.  The combination of these two
effects would lead to a chemically inhomogeneous interstellar medium, with
large radial metallicity gradients.

There is, in fact, evidence for significant chemical inhomogeneities in
the Galactic disk. \markcite{Edvardsson93}Edvardsson et al. (1993)
carried out a high--resolution study of disk stars, finding both a
remarkably small scatter in [$\alpha$/Fe] at fixed age and Galactic
distance, and a relatively large scatter in [Fe/H] at fixed age.  Their
study also confirmed the classic G--dwarf problem. They suggest that
these trends reflect the combination of episodic accretion and
relatively inefficient mixing.  \markcite{Friel93}Friel \& Janes (1993)
found a large spread in metallicity at fixed age, also consistent with
the combination of episodic accretion and slow mixing.  The
chemical--evolution models of \markcite{Pilyugin96}Pilyugin \& Edmunds
(1996) confirm that this combination accounts for the observed
metallicity trends.

Observations of chemical abundances in Orion also appear to confirm the
combination of episodic accretion plus local enrichment scenario.
\markcite{Meyer94}Meyer et al. (1994) argue that the low oxygen abundance
in Orion, some 40\% of the solar value, suggests recent infall. The
\markcite{Cunha92,Cunha94}Cunha \& Lambert (1992, 1994) studies of OB
associations found evidence for local self--enrichment of this gas.

\section{Conclusions and Predictions} 
\label{sec:finale}

Most cosmologists believe that galaxy formation is a hierarchical
process:  galaxies grow by accreting small clouds of gas and dark
matter. This process is a continuing one and we expect that galaxies
and groups are currently accreting new clouds. We simulated this process
for the Local Group and found that properties of the accreted clouds
are similar to certain properties of the high--velocity--cloud
phenomenon (excluding the Magellanic Stream HVCs):

$\bullet $ Most of the HVCs are located either near the general
direction of M31, towards the barycenter of the Local Group, or in the
antibarycenter direction, some 180\dege from the direction of M31 (see
Figure~\ref{fig:lb}).

$\bullet $ HVCs have chemical abundances similar to that of
intra--group gas, and different from the abundances characteristic of
the inner Galaxy. If HVCs were ejected from the inner Galaxy as part of
a Galactic fountain, then their metal abundance would exceed the solar
value, and this is not observed.

$\bullet $ HVCs have an angular--size/velocity relation that is
consistent with the clouds being nearly self--gravitating, and at a
distance of $\sim 300$ kpc.

If the Local--Group HVC hypothesis discussed in this paper is correct,
then studies of HVCs can directly probe the process of galaxy
formation. The validity of this hypothesis can be tested by a number of
future observations:

$\bullet $ Most observations of nearby galaxies would not have detected
the gas clouds that are equivalent to the HVCs. Moreover, many HI maps
of external galaxies extend just beyond the Holmberg radius. Our
discussion would have the typical HVC located nearly a Mpc from the
galactic center. It will be interesting to test our hypothesis with
deep HI observations of isolated groups and filaments, searching for HI
clouds associated with groups, rather than with individual galaxies.

$\bullet $ Lyman--limit clouds, which are seen in absorption towards
distant quasars, have column densities similar to those of the HVCs.
Observations of nearby Lyman--limit and Lyman--$\alpha $ systems show
that they are not all associated directly with individual galaxies, but
rather with groups of galaxies (\markcite{Oort81}Oort 1981;
\markcite{Stocke95}Stocke et al. 1995; \markcite{vanGorkom96}van Gorkom
et al. 1996; \markcite{Rauch96}Rauch, Weymann, \& Morris 1996). In the
scenario outlined in this paper, we expect that these clouds would
have properties similar to those of the local HVCs. Thus, it would be
interesting to use STIS to look for lower--column--density
high--velocity HI clouds, which would correspond to the Lyman--$\alpha
$ clouds.

$\bullet $ The simulations predict large amounts
of gas accreting onto M31 and the Local Group from the region of space
beyond M31, under the gravitational attraction of both M31 and our own
Galaxy. Because this gas is several Mpc away, the gas clouds are
expected to have small angular sizes and relatively low column densities.  
Deep HI observations in the
M31 direction should be able to detect this gas.

The hypothesis central to this paper, namely that HVCs are at distances 
of around 1 Mpc,
would be falsified by the detection of absorbtion in an HVC seen against
stars in the Milky Way halo in the direction of M31 or
in the anti--M31 direction.  On the other hand, further measurements
of low levels of H$\alpha$ emission towards these HVCs will
strengthen the case for their extragalactic nature.

%Connection to Intermediate Velocity Clouds
% Vincent de Heij is working in Leiden on the HVC/IVC connections, or rather 
%lack thereof, but I don't think the paper should wait for those resutls.
%Davis/Moore paper on Magellanic Stream

\section{Acknowledgements}
We have had useful conversations with many people about this work in
the last year and a half; it is difficult to do justice to all of
them.  We are particularly grateful to Bart Wakker for providing a
machine readable version of the WvW91 compilation, and for many
discussions via the Internet.  We are grateful for the comments and
insights provided by George Field, Jacqueline van Gorkom, Dave
Hollenbach, Buell Jannuzi, Jay Lockman, Dan McCammon, Chris McKee,
Blair Savage, and Amiel Sternberg.

\clearpage

% ------------------------ REFERENCES --------------------

\end{document}